%% file: main.tex
\documentclass[a4paper,11pt,bibliography=totoc,listof=totoc,headinclude=true,cleardoublepage=empty,oneside]{scrbook}
\usepackage[english,ngerman]{babel}
\usepackage[utf8]{inputenc}
\usepackage{ifthen}
\usepackage{color}
\usepackage{amsmath,amsthm,amssymb,amsfonts}
\usepackage{graphicx}
\usepackage{psfrag}
\usepackage{enumitem}
\usepackage{caption}
\usepackage{subcaption}
\usepackage{stmaryrd}
\usepackage{shadethm}
\usepackage{mwe}
\usepackage{float}
\usepackage{multicol}

\usepackage[unicode,colorlinks=true,pagebackref=false]{hyperref}

\KOMAoptions{footinclude=false} % Fusszeile wird nicht zu Satzspiegel gezaehlt
\KOMAoptions{headsepline=false} % Trennlinie zwischen Kopfzeile und Text
\KOMAoptions{DIV=12} % beeinflusst Satzspiegel
\KOMAoptions{BCOR=8mm} % Bindekorrektur
\recalctypearea % berechne Satzspiegel neu

\definecolor{change}{rgb}{0,.55,.55}

\def\doubleunderline#1{\underline{\underline{#1}}}
\newcommand\numberthis{\addtocounter{equation}{1}\tag{\theequation}}

\numberwithin{equation}{section}

\newshadetheorem{problem}{Problem} 
\definecolor{shadethmcolor}{rgb}{0.90, 0.90, 0.90}
\definecolor{shaderulecolor}{rgb}{0,0,0}
\newshadetheorem{theorem}{Definition}
\definecolor{shadethmcolor}{rgb}{0.90, 0.90, 0.90}
\definecolor{shaderulecolor}{rgb}{0,0,0}
\newshadetheorem{lemma}{Lemma} 
\definecolor{shadethmcolor}{rgb}{0.90, 0.90, 0.90}
\definecolor{shaderulecolor}{rgb}{0,0,0}
\newshadetheorem{definition}{Theorem} 
\definecolor{shadethmcolor}{rgb}{0.90, 0.90, 0.90}
\definecolor{shaderulecolor}{rgb}{0,0,0}
\begin{document}

\input{components/coverpage}

\cleardoublepage

\input{components/titlepage}

\input{components/disclaimer}

\input{components/acknowledgements}

\input{components/kurzfassung}

\input{components/abstract}

\input{components/tableofcontent}

\input{chapters/introduction}

\input{chapters/theory}

\input{chapters/turbulence}

\input{chapters/simulation}

\input{chapters/discretization}

\input{chapters/discretization2}

\input{chapters/results}

\input{chapters/conclusion}

\listoffigures
 
\listoftables

\renewcommand{\bibname}{References} 
\bibliographystyle{alpha} 
\bibliography{bibliography/literature}

\end{document}

%% file: components/coverpage.tex
% !TEX root = ../main.tex
% The titlepage.
% Included by MAIN.TEX

%--------------------------------------------------
% The title page
%--------------------------------------------------

\pagenumbering{Alph}
\selectlanguage{english}

\begin{titlepage}
  %\vspace*{-2cm}
  \begin{center}
    \includegraphics[width=0.3\textwidth]{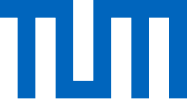}
    \vskip 1cm
    {\huge Technische Universität München}
    \vskip 5mm
    {\Large Department of Informatics}
    \vskip 4cm
    {\Large Master's Thesis in Computational Science and Engineering}
    \vskip 1cm
    {\huge\bfseries Exactly Divergence-free Hybrid Discontinuous Galerkin Method for \\[1ex] Incompressible Turbulent Flows}
    \vskip 1cm
    {\Large Xaver Mooslechner BSc}
    \vskip 4cm
    \includegraphics[width=0.3\textwidth]{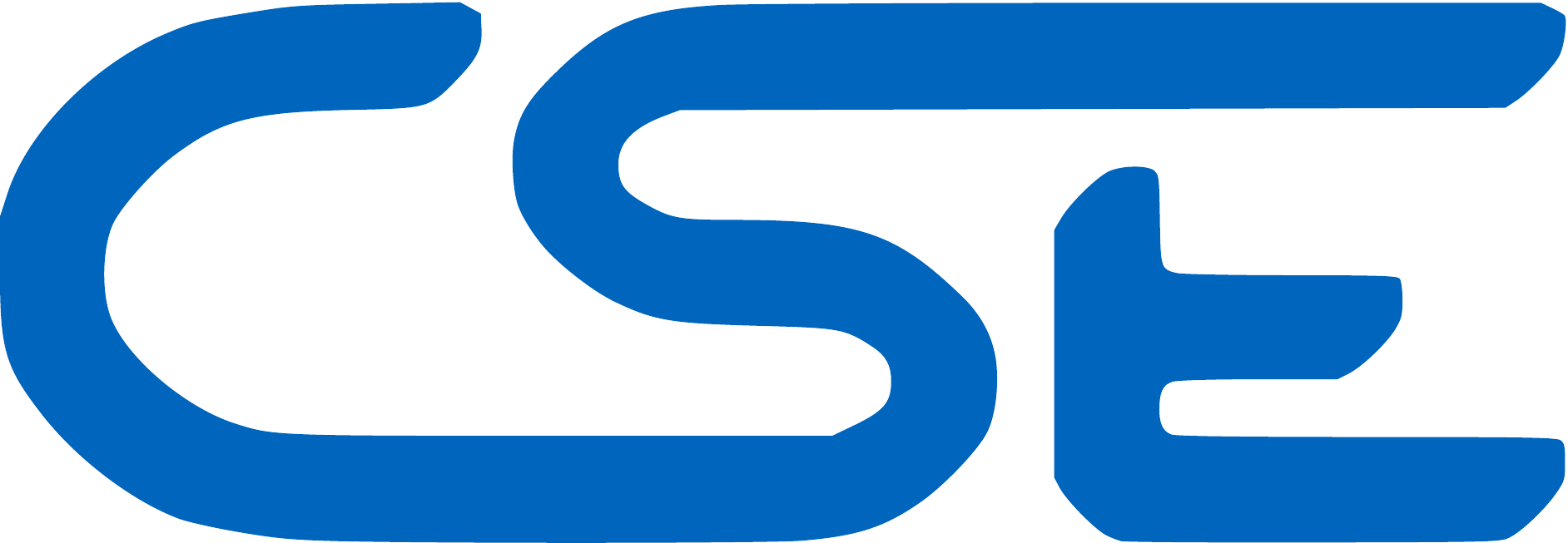}
  \end{center}
\end{titlepage}

\cleardoublepage

%% file: components/titlepage.tex
% !TEX root = ../main.tex
% The titlepage.
% Included by MAIN.TEX

%--------------------------------------------------
% The title page
%--------------------------------------------------

\pagenumbering{Alph}
\selectlanguage{english}

\begin{titlepage}
  %\vspace*{-2cm}
  \begin{center}
    \includegraphics[width=0.3\textwidth]{images/titlepage/tum_logo.pdf}
    \vskip 1cm
    {\huge Technische Universität München}
    \vskip 5mm
    {\Large Department of Informatics}
    \vskip 10mm
    {\Large Master's Thesis in Computational Science and Engineering}
    \vskip 1cm
    {\huge\bfseries Exakt divergenzfreie hybride diskontinuirliche Galerkin Verfahren für \\[1ex] inkompressible turbulente Strömungen}
    \vskip 10mm
    {\huge\bfseries Exactly Divergence-free Hybrid Discontinuous Galerkin Method for \\[1ex] Incompressible Turbulent Flows}
    \vskip 15mm
    \begin{tabular}{ll} 
    	{\large Author: Xaver Mooslechner BSc}\\[1ex]
    	{\large 1\textsuperscript{st} Examiner: Univ.Prof. Dr. Folkmar Bornemann}\\[1ex]
    	{\large 2\textsuperscript{nd} Examiner: Univ.Prof. Dr. Joachim Schöberl, Technische Universität Wien}\\[1ex]
    	{\large Assistant advisor: Dr. Philip Lederer, Technische Universität Wien}\\[1ex]
    	{\large Submission date: 22.04.2020}\\[1ex]
	\end{tabular}
  \vskip 15mm
  \includegraphics[width=0.3\textwidth]{images/titlepage/cse_logo.pdf}
  \end{center}
\end{titlepage}

\cleardoublepage

%% file: components/disclaimer.tex
% !TEX root = ../main.tex

\chapter*{Disclaimer}

\thispagestyle{empty}
\selectlanguage{english}

\vspace*{2cm}

I hereby declare that this thesis is entirely the result of my own work except where other-
wise indicated. I have only used the resources given in the list of references.

\vspace*{3cm}

\noindent
Vienna, 22.04.2020 %\today
\hfill 
\begin{minipage}[t]{5cm}
\centering
\underline{\hspace*{5cm}}\\
\end{minipage}

\cleardoublepage

%% file: components/acknowledgements.tex
% !TEX root = ../main.tex
\chapter*{Acknowledgement} 

\thispagestyle{empty}
\selectlanguage{english}

I wish to express my gratitude to my supervisor, Prof. Dr. Joachim Schöberl, for providing me the opportunity to write this master thesis within his work group at the technical university of Vienna. I want to thank you for the incredible amount he has taught me during this work. At the same time, I want to express my appreciation to Prof. Dr. Folkmar Bornemann for taking the official supervision of my thesis on him.
\\[2ex]
I also want to thank Dr. Philip Lederer for introducing and enhancing my knowledge in this topic and leading me to find solutions for arising problems.
\\[2ex]
Furthermore, I want to thank my girlfriend Caroline for all the support during my studies. 

\cleardoublepage

%% file: components/kurzfassung.tex
\chapter*{Kurzfassung}

\thispagestyle{empty}
\selectlanguage{english}

Diese Arbeit beschäftigt sich mit der Untersuchung eines $H(\mathrm{div})$-konformen hybriden diskontinuirlichen Galerkin Verfahrens für inkompressible turbulente Strömungen.
\\[1ex]
Die Diskretisierungsmethode liefert einige günstige physikalische und lösungs-orientierende Eigenschaften, welche von Vorteil sein können für das Auflösen von rechenintensiven turbulenten Strukturen. Eine herkömmliche Methode zur Diskretisierung der Navier-Stokes Gleichungen mit den bekannten Taylor-Hood Elementen ist auch gegeben, um einen Vergleich wiedergeben zu können. Die vier Hauptmethodiken zur Simulation von turbulenten Strömungen sind erläutert: die Reynolds-gemittelte Navier-Stokes Simulation, die Grobstruktursimulation, die Methode zur Mehrskalenvariationsrechnung und die direkte numerische Simulation. Die Grobstruktursimulation und Mehrskalenvariationsrechnung zeigen gute Ergebnisse in der Berechnung von traditionell schwierigen turbulenten Strömungs-fällen. Diese Genauigkeit kann nur durch direktes berechnen der Navier-Stokes Gleichungen übertroffen werden, was jedoch mit sehr hohen Kosten verbunden ist. Eine sehr verbreitete Herangehensweise ist die Reynolds-Mittelung, da diese sehr kostengünstig ist. Diese Prinzipien wurden an beiden Diskretisierungsverfahren angewendet und validiert anhand der turbulenten Kanalströmung.
\\[1ex]
Alle numerischen Simulationen wurden mit Hilfe der finiten Elementen Bibliothek Netgen/NGSolve durchgeführt. 
\cleardoublepage

%% file: components/abstract.tex
\chapter*{Abstract}

\thispagestyle{empty}
\selectlanguage{english}

This thesis deals with the investigation of a $H(\mathrm{div})$-conforming hybrid discontinuous Galerkin discretization for incompressible turbulent flows.
\\[1ex]
The discretization method provides many physical and solving-oriented properties, which may be advantageous for resolving computationally intensive turbulent structures. A standard continuous Galerkin discretization for the Navier-Stokes equations with the well-known Taylor-Hood elements is also introduced in order to provide a comparison. The four different main principles of simulating turbulent flows are explained: the Reynolds-averaged Navier-Stokes simulation, large eddy simulation, variational multiscale method and the direct numerical simulation. The large eddy simulation and variational multiscale have shown good promise in the computation of traditionally difficult turbulent cases. This accuracy can be only surpassed by directly solving the Navier-Stokes equations, but comes with excessively high computational costs. The very common strategy is the Reynolds-average approach, since it is the most cost-effective. Those modelling principles have been applied to the two discretization techniques and validated through the basic plane channel flow test case.
\\[1ex]
All numerical tests have been conducted with the finite element library Netgen/NGSolve.

\cleardoublepage

%% file: components/tableofcontent.tex
\pagenumbering{roman}
\selectlanguage{english}

\hypersetup{linkcolor=black}
\tableofcontents

\cleardoublepage
\pagenumbering{arabic}

%% file: chapters/introduction.tex
\chapter{Introduction}
\label{chapter:introduction}
%%%%%%%%%%%%%%%%%%%%%%%%%%%%%%%%%%%%%%%%%%%%%%%%%%%%%%%%%%%%%%%%%%%%%%%%%%%%%%%%%%%%%%%%%%%%%%%%%%%%%%%%%%%%%%
%%%%%%%%%%%%%%%%%%%%%%%%%%%%%%%%%%%%%%%%%%%%%%%%%%%%%%%%%%%%%%%%%%%%%%%%%%%%%%%%%%%%%%%%%%%%%%%%%%%%%%%%%%%%%%

{
\section{Motivation}
The phenomena of turbulence in fluid flow is one of the most impressing problems of classic mechanics. Since it has been initially observed and described by Leonardo da Vinci in the 16th century, a lot of effort was successively dedicated to understand the emergence of turbulence and their structures. The very chaotic, irregular and apparently unpredictable behavior of turbulent flows leads to a challenging subject of study. Nevertheless, this phenomena is by far not unknown and occurs very frequently in our daily life, from pouring milk in a cup of coffee to circulation of air in the atmosphere. Especially for many scientific and engineering purposes, turbulence plays a vital role and its prediction is of great interest.
\\[1ex]
In the past, due to the complexity of the governing equations of fluid motion, the analytical approach is highly restricted to simple flow cases and can by far not be applied to turbulence. As a result of this problem, the experimental analysis was the only method to deal with this phenomena. By the rapid growth of computational power and reduction of its operational cost, the numerical approach is becoming more and more advantageous compared to expensive experiments. Since the past few decades, this benefit drives the scientific community to establish methods for studying turbulent flows. The most important aspects of the development of methods in computational fluid dynamics (CFD) is accuracy of the solution and cost effective algorithms. In order to accurately simulate turbulence over an appropriate time period, different simulation techniques and discretization methods need to be developed.

\section{Outline of the thesis}
In the first chapter, the focus lies on the theory behind fluid dynamics and the derivation of the governing equations, the Navier-Stokes equation. This section should reacquaint the basics, which are indispensable for the upcoming chapters. \\ 
In chapter two, an overview of turbulence is given. The most relevant fields of interest are described in more detail and should provide sufficient knowledge. The turbulence section is divided in stochastic, spectral and theory description.
\\
The third chapter is dedicated to the different simulation principles and modelling. It provides the three main approaches how to numerically compute turbulent flow. In the last years, a rather new method has emerged in the field of CFD, which is described in this section as well. All of the four principles are analyzed and have been conducted with the different types of discretization techniques. 
\\
The two different spatial discretizations of the famous saddle-point problem are expressed in the fourth and fifth chapter of the thesis. One is the standard Continuous Galerkin (CG) method with the well-known elements from the Taylor-Hood family, which ensures only discrete divergence-free velocity property. The other technique is composed of a Hybrid Discontinuous Galerkin (HDG) method, developed by Christoph Lehrenfeld and Joachim Schöberl in \cite{LEHRENFELD2016339}. This mixed method guarantees an exact divergence-free flow and leads to an appropriate physical description.
\\
The last chapter briefly shows the results of the performed simulations and further on discusses the differences. 
\section{Implementation}
All numerical examples were implemented and tested in the finite element library Netgen/NGSolve, see \cite{netgen} and \cite{ngsolve}.
\cleardoublepage
%\section*{Notation}
%In this thesis we stick to the following notation:
%\\[1ex]
%\cleardoublepage
\section*{List of Abbreviations}
The following abbreviations are used in this thesis:
\\[1ex]
\newlist{abbrv}{itemize}{1}
\setlist[abbrv,1]{label=,labelwidth=1in,align=parleft,itemsep=0.1\baselineskip,leftmargin=!}
\begin{abbrv}
\item[CFD] computational fluid dynamics
\item[CG] continuous Galerkin
\item[DG] discontinuous Galerkin
\item[DNS] direct numerical simulation
\item[DOF] degree of freedom
\item[GS] grid scale
\item[HDG] hybrid discontinuous Galerkin
\item[IMEX] implicit-explicit
\item[LBB] Ladyshenskaja-Babuška-Brezzi
\item[LES] large eddy simulation
\item[PDE] partial differential equation
\item[RANS] Reynolds-averaged Navier-Stokes
\item[RST] Reynolds stress tensor
\item[RTT] Reynolds-transport-thoerem
\item[SGS] sub-grid scale
\item[VMS] variational multiscale
\end{abbrv}
}

%% file: chapters/theory.tex
\chapter{Derivation of the Navier-Stokes equations}
\label{chapter:theory}
%%%%%%%%%%%%%%%%%%%%%%%%%%%%%%%%%%%%%%%%%%%%%%%%%%%%%%%%%%%%%%%%%%%%%%%%%%%%%%%%%%%%%%%%%%%%%%%%%%%%%%%%%%%%%%
%%%%%%%%%%%%%%%%%%%%%%%%%%%%%%%%%%%%%%%%%%%%%%%%%%%%%%%%%%%%%%%%%%%%%%%%%%%%%%%%%%%%%%%%%%%%%%%%%%%%%%%%%%%%%%

{
A fluid element represents an accumulation of fluid molecules within an infinitesimal small volume $dV$, where its averaged motion over an infinitesimal time interval $dt$. Each element has its density $\rho$ and velocity $\underline{u}$. This macroscopic view leads to the continuum description, where the characteristics of the fluid element are prescribed via partial differential equations (PDE).
\\
In the following, we assume the physical domain $\Omega\subset\mathbb{R}^{3}$ and the fixed time interval denoted by $\lbrack0,T\rbrack$. Further on, we assume the fluid to be Newtonian, pure and viscous. The following relevant quantities are shown in Table~\ref{table:2.1}.

\begin{table}[!ht]
	\centering
	\begin{tabular}{|c|l|l|}
		\hline
		Description         & Quantity & Unit   \\ \hline
		velocity            & $\underline{u}(\underline{x},t)\in C^2(\Omega \times [0,T],\mathbb{R}^3)$   & $m\, s^{-1}$    \\
		density             & $\rho(\underline{x},t)\in C^1(\Omega \times [0,T],\mathbb{R})$ & $kg\, m^{-3}$   \\
		pressure            & $p(\underline{x},t)\in C^1(\Omega \times [0,T],\mathbb{R})$   & $kg\, m^{-1}\, s^{-2}$ \\
		force & $\underline{F}(\underline{x},t)\in C^0(\Omega \times [0,T],\mathbb{R}^3)$       & $kg\,m\,s^{-2} $    \\ 
		kinematic viscosity & $\nu\in\mathbb{R}\setminus \{ 0 \}$       & $m^{2}\, s^{-1}$    \\
		\hline
	\end{tabular}
	\caption{Physical quantities}
	\label{table:2.1}
\end{table}
We proceed with the following derivations as in \cite{oertel2015stroemungsmechanik} and \cite{kuhlmann2014stroemungsmechanik}.
\section{Lagrangian and Eulerian description}
In classical field theories, the kinematic of fluid flow can be basically described by two different point of views.
\\
The Lagrangian specification of the fluid field is one way of looking at a fixed fluid element as it moves through space and time. The observer of the fluid parcel monitors the change of its properties (e.g. velocity). At a specific time $t_{0}$, the fluid element has the position $\underline{\xi}(t_{0})=\underline{\xi_{0}}$. Since only one fluid parcel may stay at a certain position, the element is "labeled" $\underline{\xi_{0}}$. Therefore $\underline{\xi}(\underline{\xi}_{0};t)$ is the position vector and describes the trajectory of the fluid parcel 
\begin{equation}
	\underline{u}(\underline{\xi}_{0};t)=\frac{d\underline{\xi}(\underline{\xi}_{0};t)}{dt}.
	\label{eq1}
\end{equation}
The Eulerian description is a way of focusing at fluid flow on a specific location $\underline{x}$ in space as the fluid passes through time. As a result, the change of quantities at a fixed position in space depends on the change of each individual fluid element as well as the change of fluid elements at position $\underline{x}$.
\\[1ex]
The link between the Lagrangian and Eulerian specifications is given by the material derivative. Suppose the quantity $\phi(\underline{x},t)\in C^1(\Omega \times [0,T],\mathbb{R})$ from the Eulerian point of view and substitute $\underline{x}=\underline{\xi}(\underline{\xi}_{0};t)$ from the Lagrangian approach, it follows $\phi(\xi_{1},\xi_{2},\xi_{3},t)$. The total rate of change with respect of time of $\phi$ is
\begin{equation*}
	\frac{d\phi}{dt}=\frac{\partial\phi}{\partial t}+\frac{\partial\phi}{\partial x_{1}}\frac{\partial\xi_{1}}{\partial t}+\frac{\partial\phi}{\partial x_{2}}\frac{\partial\xi_{2}}{\partial t}+\frac{\partial\phi}{\partial x_{3}}\frac{\partial\xi_{3}}{\partial t}=\frac{\partial\phi}{\partial t}+\nabla\phi\,\cdot\,\frac{\partial\underline{\xi}}{\partial t}.
	\label{eq2}
\end{equation*}
Using Equation~\eqref{eq1}, it follows:
\begin{theorem}[\textbf{Material derivative}]
Let the quantitiy $\phi(\underline{x},t)\in C^1(\Omega \times [0,T],\mathbb{R})$ be an arbitrary function in a fluid transported with velocity \underline{u}. Then the material derivative is defined by
\begin{equation}
	\frac{D\phi}{Dt}=\frac{\partial\phi}{\partial t}+(\underline{u}\,\cdot\,\nabla)\phi.
	\label{eq3}
\end{equation}
\end{theorem}
So far, a single fluid element with infinitesimal expansion was considered. Now we extent our consideration to a dense pack of many fluid elements, a fluid volume. The relation between Lagrangian and Eulerian view for a fluid volume is given by the Reynolds-transport-theorem (RTT). 
\begin{theorem}[\textbf{Reynolds-transport-theorem}]
Let $V(t) \subset \Omega$ be an arbitrary volume with fixed number of fluid elements (fixed mass) and the quantitiy $\phi(\underline{x},t)\in C^1(\Omega \times [0,T],\mathbb{R})$ be an arbitrary function in a fluid transported with velocity \underline{u}. Then the Reynolds-transport-theorem is defined by
\begin{equation}
	\frac{D}{Dt}\int\limits_{V(t)} \phi(\underline{x},t)\:d\underline{x} = 
	\int\limits_{V(t)} \frac{\partial\phi(\underline{x},t)}{\partial t}\:d\underline{x}
	+\int\limits_{\partial V(t)} \phi(\underline{x},t)\underline{u}(\underline{x},t)\,\cdot\,\underline{n}\,d\underline{s},
	\label{eq4}
\end{equation}
while the surface $\partial V(t)$ defines the boundary of $V(t)$ and $\underline{n}$ the outward directed normal vector of the surface $\partial V(t)$.
\end{theorem}
Using Gauß' theorem for Equation~\eqref{eq4} yields
\begin{equation}
	\frac{D}{Dt}\int\limits_{V(t)} \phi(\underline{x},t)\:d\underline{x} = 
	\int\limits_{V(t)} \bigg(\frac{\partial\phi}{\partial t}
	+\nabla\,\cdot\,(\phi\underline{u})\bigg)\:d\underline{x}.
	\label{eq5}
\end{equation}

\section{Conservation of mass}
Again let define $V(t) \subset \Omega$ be an arbitrary volume with fixed mass and take density as our quantity, the total mass at time t is
\begin{equation*}
	m(t) = \int\limits_{V(t)} \rho(\underline{x},t)\:d\underline{x}.
	\label{eq6}
\end{equation*}
As the mass has has to be conserved, the rate of change with respect to time has to be
\begin{equation*}
	\frac{Dm}{Dt} = \frac{D}{Dt}\int\limits_{V(t)} \rho(\underline{x},t)\:d\underline{x} = 0.
	\label{eq7}
\end{equation*}
With use of Equation~\eqref{eq4}, the conservation of mass reads
\begin{equation}
	\frac{Dm}{Dt}= 
	\int\limits_{V(t)} \frac{\partial\rho}{\partial t}\:d\underline{x}
	+\int\limits_{\partial V(t)} \rho\underline{u}\,\cdot\,\underline{n}\,d\underline{s}
	=0.
	\label{eq8}
\end{equation}
The first term of Equation~\eqref{eq8} on the right-hand side represents the change in mass due to rate of change of density with respect to time and the second term gives balance of in- and outcoming mass flux. Further using Equation~\eqref{eq5} yields to the integral and differential form of the continuity equation 
\begin{equation}
	\int\limits_{V(t)} \bigg(\frac{\partial\rho}{\partial t}
	+\nabla\,\cdot\,(\rho\underline{u})\bigg)\:d\underline{x}
	=0,
	\label{eq9}
\end{equation}
\begin{equation}
	\frac{\partial\rho}{\partial t}
	+\nabla\,\cdot\,(\rho\underline{u})
	=0.
	\label{eq10}
\end{equation}
In this thesis, we strictly focus on incompressible fluid flow and $\rho=const$, therefore
\begin{equation}
	\frac{D\rho}{Dt}
	=0,
	\label{eq11}
\end{equation}
substituting Equation~\eqref{eq11} into Equation~\eqref{eq10} leads to the incompressibility constraint for the velocity
\begin{equation}
	\nabla\,\cdot\,\underline{u}=0.
	\label{eq12}
\end{equation}

\section{Conservation of momentum}
By using Newton's second law, the conservation of momentum equation can be deducted. It follows that the change of momentum of a fluid volume (with fixed mass) equals the resulting forces affecting the fluid volume. The momentum vector is given by
\begin{equation*}
	\underline{P}=\int\limits_{V(t)} \rho(\underline{x},t)\underline{u}(\underline{x},t)\:d\underline{x}.
	\label{eq13a}
\end{equation*}
Therefore
\begin{equation*}
	\frac{D\underline{P}}{Dt}=\frac{D}{Dt} \int\limits_{V(t)} \rho(\underline{x},t)\underline{u}(\underline{x},t)\:d\underline{x}
	=\sum_{m} \underline{F}_{m}.
	\label{eq13}
\end{equation*}
Possible forces are:
\begin{itemize}
\item Volume forces (e.g. gravity) affect the control volume $V(t)$ by
\begin{equation}
	\underline{F}_{V} = \int\limits_{V(t)} \rho\underline{f}\:d\underline{x},
	\label{eq14}
\end{equation}
$\underline{f}$ presents the acceleration in vector form.
\item Surface forces (e.g. shear stress) acting on the surface of the control volume $V(t)$ by
\begin{equation}
	\underline{F}_{S} = \int\limits_{S(t)} \doubleunderline{\sigma}\,\cdot\,\underline{n}\,d\underline{s},
	\label{eq15}
\end{equation}
$\doubleunderline{\sigma}$ is a symmetric tensor of second order
\begin{equation*}
	\doubleunderline{\sigma}=-p\doubleunderline{I}+\doubleunderline{\tau}
	= \left(\begin{matrix}-p+\tau_{11}&\tau_{12}&\tau_{13}\\\tau_{21}&-p+\tau_{22}&\tau_{23}\\\tau_{31}&\tau_{32}&-p+\tau_{33}\end{matrix}\right).
	\label{eq16}
\end{equation*}
The diagonal components of $\doubleunderline{\tau}$ present the normal stresses, while the other components describe the shear stresses. By the use of the Gauß' theorem, the term in Equation~\eqref{eq15} can be reforumlated to
\begin{equation}
	\int\limits_{S(t)} \doubleunderline{\sigma}\,\cdot\,\underline{n}\,d\underline{s}
	=\int\limits_{V(t)} \nabla\,\cdot\,\doubleunderline{\sigma}\:d\underline{x}.
	\label{eq17}
\end{equation}
The divergence of a matrix as shown in Equation~\eqref{eq17} is taken row-wise.
\end{itemize}

The left-hand side of the momentum equation is given 
\begin{equation*}
\begin{split}
	\frac{D}{Dt} \int\limits_{V(t)} \rho\underline{u}\:d\underline{x}
	& =\int\limits_{V(t)} \frac{\partial\rho\underline{u}}{\partial t}\:d\underline{x} + \int\limits_{S(t)} \rho\underline{u}(\underline{u}\,\cdot\,\underline{n})\,d\underline{s} \\
	& =\int\limits_{V(t)} \bigg( \frac{\partial\rho\underline{u}}{\partial t} + \nabla\,\cdot\,(\rho\underline{u}\otimes\underline{u})\bigg)\:d\underline{x}.
\end{split}
\label{eq18}
\end{equation*}
Here, $\underline{u}\otimes\underline{u}$ indicates the outer (dyadic) product of two vectors.\\
The right-hand side of the momentum equation consists of 
\begin{equation*}
	\int\limits_{V(t)} \bigg(\nabla\,\cdot\,\doubleunderline{\sigma} + \rho\underline{f}\bigg) \:d\underline{x},
	\label{eq19}
\end{equation*}
while $\doubleunderline{\sigma}$ may be decomposed to $-p\doubleunderline{I}+\doubleunderline{\tau}$, resulting in 
\begin{equation*}
	\int\limits_{V(t)} \bigg(-\nabla p + \nabla\,\cdot\,\doubleunderline{\tau} + \rho\underline{f}\bigg) \:d\underline{x}.
	\label{eq20}
\end{equation*}
The final momentum equation in integral and differential form is then
\begin{equation}
	\int\limits_{V(t)} \bigg( \frac{\partial\rho\underline{u}}{\partial t} + \nabla\,\cdot\,(\rho\underline{u}\otimes\underline{u})\bigg)\:d\underline{x}
	= \int\limits_{V(t)} \bigg(-\nabla p + \nabla\,\cdot\,\doubleunderline{\tau} + \rho\underline{f}\bigg) \:d\underline{x},
	\label{eq21}
\end{equation} 
\begin{equation}
	\frac{\partial\rho\underline{u}}{\partial t} + \nabla\,\cdot\,(\rho\underline{u}\otimes\underline{u})
	= -\nabla p + \nabla\,\cdot\,\doubleunderline{\tau} + \rho\underline{f}.
	\label{eq22}
\end{equation} 
Since we are assuming incompressible flows, the constraint from Equation~\eqref{eq11} and \eqref{eq12} further simplifies the conservation of momentum equation. Exploiting the identity
\begin{equation}
	\nabla\,\cdot\,(\underline{u}\otimes\underline{u}) = (\underline{u}\,\cdot\,\nabla)\underline{u} + \underline{u}(\nabla\,\cdot\,\underline{u}).
	\label{eq23}
\end{equation}
As a result, the incompressible momentum equation in differential form reads as follows
\begin{equation}
	\frac{\partial\underline{u}}{\partial t} + (\underline{u}\,\cdot\,\nabla)\underline{u}
	= -\frac{1}{\rho}\nabla p + \frac{1}{\rho}\nabla\,\cdot\,\doubleunderline{\tau} + \underline{f}.
	\label{eq24}
\end{equation}

\section{The Navier-Stokes equations}
In the following, further assumptions have to be made to generalize the stress tensor $\doubleunderline{\sigma}$. In this thesis, we assume that the fluid is a Stokes fluid. For this sake, the properties of such fluid is given: 
\begin{itemize}
\item Conservation of angular momentum, means that $\doubleunderline{\sigma}$ is symmetric $\doubleunderline{\sigma} = \doubleunderline{\sigma}^{T}$.
\item $\doubleunderline{\sigma}$ is isotropic.
\item If the fluid rests or moves due to rigid-body motion, it holds $\doubleunderline{\sigma} = -p\doubleunderline{I}$.
\end{itemize} 
The general ansatz for the stress tensor, assuming viscous Newtonian fluid and incompressible flow, is 
\begin{equation*}
	\doubleunderline{\tau}=\mu(\nabla\underline{u}+\nabla\underline{u}^{T}),
	\label{eq25}
\end{equation*}
while $\nabla\underline{u}$ denotes the vector gradient $\nabla\underline{u}=(\nabla\,\otimes\,\underline{u})^{T}$.\\[1ex]
The divergence of the stress tensor can be further simplified if $\mu=const$ to
\begin{equation*}
	\nabla\,\cdot\,\doubleunderline{\tau} = \mu\Delta\underline{u}.
	\label{eq26}
\end{equation*}
Additionally, for the Navier-Stokes model one needs suitable initial and boundary conditions for $\underline{u}$. The initial condition is $\underline{u}(\underline{x},0) = \underline{u}_{0}(\underline{x})$, which has to be physically correct as well as divergence free. The boundary conditions are distinguished between Dirichlet, Neumann or Robin boundary. Let $\partial\Omega$ be subdivided into three parts $\partial\Omega=\partial\Omega_{D}\cup\partial\Omega_{N}\cup\partial\Omega_{R}$ with $\partial\Omega_{D}\cap\partial\Omega_{N}\cap\partial\Omega_{R}=\emptyset$. Dirichlet conditions are of the form $\underline{u}(\underline{x},t)=\underline{u}_{D}(\underline{x})$ for $x\in\partial\Omega_{D}$ and in applications often used for inflow or no-slip boundary conditions. In order to describe outflow conditions, the Neumann boundary conditions of the form $\frac{\partial\underline{u}}{\partial\underline{n}}=\underline{g}(\underline{x})$ for $x\in\partial\Omega_{N}$ are appropriate. The Robin boundary condition is a linear combination of the Dirichlet and Neumann condition. Since it is not a major issue in this thesis, this third type boundary condition will be neglected.\\[1ex]
Finally, the Navier-Stokes problem is complete and can be read as:
\begin{problem}[\textbf{Incompressible Navier-Stokes equations}]
Let $\nu,\rho\in \mathbb{R}\setminus \{ 0 \}$, $\underline{f}\in C^0(\Omega,\mathbb{R}^{3})$, $\underline{u}_{D}\in C^0(\partial\Omega_{D},\mathbb{R}^{3})$, $\underline{g}\in C^0(\partial\Omega_{N},\mathbb{R}^{3})$ and $\underline{u}_{0}\in C^{2}(\Omega,\mathbb{R}^{3})$ find $\underline{u}\in C^2(\Omega \times [0,T],\mathbb{R}^3)$ and $p\in C^1(\Omega \times [0,T],\mathbb{R})$ such that
\begin{align*}
	& \nabla\,\cdot\,\underline{u}=0 \numberthis & \textnormal{in}\:\Omega\times[0,T],\label{NSE_C} & \\
	& \frac{\partial\underline{u}}{\partial t} + (\underline{u}\,\cdot\,\nabla)\underline{u}
	= -\frac{1}{\rho}\nabla p + \nu\Delta\underline{u} + \underline{f} \numberthis & \textnormal{in}\:\Omega\times[0,T], \label{NSE_M} & \\
	& \underline{u}=\underline{u}_{0} & \textnormal{in}\:\Omega, t=0, & \\
	& \underline{u}=\underline{u}_{D} & \textnormal{in}\:\partial\Omega_{D}\times[0,T], & \\
	& \frac{\partial\underline{u}}{\partial\underline{n}}=\underline{g} & \textnormal{in}\:\partial\Omega_{N}\times[0,T].
\end{align*}
\label{problem1}
\end{problem}
As it is of indispensible importance, a specfic dimensionaless parameter can be deducted via the dimensionaless Navier-Stokes equations. The following variables are:
\begin{align*}
	& \underline{u}^{*} = \frac{\underline{u}}{U}, & p^{*} = \frac{p}{\rho U^{2}}, \\ & \underline{x}^{*} = \frac{\underline{x}}{L}, & t^{*} = t\frac{U}{L},
\end{align*}
while $U$ and $L$ are the corresponding characteristic length and velocity scales. If we insert the dimensionless variables into the continuity and momentum equation from Problem~\ref{problem1} (neglecting the volume force term), we get the Navier-Stokes equations in dimensionless form
\begin{align}
	& \nabla\,\cdot\,\underline{u}^{*}=0, & \\
	& \frac{\partial\underline{u}^{*}}{\partial t^{*}} + (\underline{u}^{*}\,\cdot\,\nabla)\underline{u}^{*}
	= -\nabla p^{*} + \frac{1}{Re}\Delta\underline{u}^{*}.
\end{align}
The new parameter $Re=\frac{UL}{\nu}$ is called Reynolds number and gives the ratio between inertial and viscous forces and is a measure for turbulence.
}

%% file: chapters/turbulence.tex
\chapter{Turbulence phenomena}
\label{chapter:turbulence}
%%%%%%%%%%%%%%%%%%%%%%%%%%%%%%%%%%%%%%%%%%%%%%%%%%%%%%%%%%%%%%%%%%%%%%%%%%%%%%%%%%%%%%%%%%%%%%%%%%%%%%%%%%%%%%
%%%%%%%%%%%%%%%%%%%%%%%%%%%%%%%%%%%%%%%%%%%%%%%%%%%%%%%%%%%%%%%%%%%%%%%%%%%%%%%%%%%%%%%%%%%%%%%%%%%%%%%%%%%%%%

{
\section{Historical view}
Da Vinci's observations and drawings primarily describe that turbulent flow regime contains of eddying motion and structures of whirls in the 16th century.\\
The experiment of Osborne Reynolds was one of the first attempts to experimentally quantify turbulence and his work laid the foundation of turbulence theory. Reynolds showed that two flow regimes exist, laminar and turbulent, and introduced a parameter (the Reynolds number) to distinguish between those states of fluid flow. The transition of laminar to turbulent flow occurs only if a certain critical Reynolds number has been exceeded.\\
Lewis F. Richardson firstly expressed the concept of energy cascade in well-developed turbulence and captured his observations in his famous poem in the 1920s. In the regime of turbulent flow, a wide range of different length scales exists, from large whirls to smaller ones. The biggest eddies of size of almost the characteristic length scale contain the most of kinetic energy, while the smaller whirls get supplied by the larger ones. Evoking a cascading waterfall until a certain length scale is reached, where the remaining energy dissipates due to viscosity.\\
Around 20 years later, Andrei N. Kolmogorov extended the work of Richardson and evolved the theory of turbulence and the concept of energy spectrum. This spectrum gives the distribution of energy among turbulence vortices as function of vortex size. In his analysis, he found that these scales are well separated, where the intermediate subrange of scales are statistically isotropic and Kolmogorov hypothesized a universal form of the energy spectrum in this region. \\
Over the last sixty years since the publishing of Kolmogorov's theory, much progress has been made in the field of turbulence flow. Particularly, the defining and identifying of coherent turbulent structures, vortex structures which persist in the flow for a relatively long time, through experiments have given great insight. A significant contribution to this progress is due to the development of advanced numerical simulation methods. These new techniques have made available data, which could not been measured in any experimental investigation.

\section{Characteristics}
However, defining turbulence is by no means easy. It is common to describe it by listing its characteristics. Further detailed description may be found in \cite{hinze1975turbulence} and \cite{pope_2000}.

\begin{itemize}
\item Turbulent flows are \textit{chaotic}. In the sense of small disturbances in the initial field, which will be amplified and leading to an uncorrelated flow field. This makes a deterministic approach impractical and intractable to describe its motion in full details as function of space and time. The random fluctuations may have amplitudes of ten to thirty percent of its mean value.

\item Turbulent flows are \textit{unsteady}. In the regime of turbulence, a true stationary solution does not exist due to its high irregularity. Only stochastic steady form can be reached.

\item Turbulent flows are \textit{rotational}. It has been shown many times that turbulence only arises and persists in rotational flows, in the presence of shear. An initially irrotational flow may become rotational by laminar-turbulent transition. Such process may only happen when inertial forces dominate viscous effects (at high Reynolds numbers) and small perturbations are no longer be damped by molecular viscosity.

\item Turbulent flows are \textit{diffusive}. Such flows cause very rapid mixing of any transported quantity like momentum or heat. The turbulent diffusion allows much faster mixing of quantities than if only molecular diffusion processes were involved. Diffusivity may be important from a practical point of view, for instance, the turbulent drag on an airfoil.

\item Turbulence is still a phenomenon of \textit{continuum mechanics}. Even the smallest eddies occurring in turbulent flow are far greater than any single fluid element (molecular scale).
\end{itemize}

\section{Stochastic description}
\label{section:RANS}
In general, a detailed description of flow quantities in time and space is neither advisable nor desirable. In order to be able to compare two different turbulent flows, it only makes sense if initial and boundary conditions match, therefore an appropriate statistical description is desired. We usual refer to a statistical representation of the fluctuations. A turbulent fluid flow is called statistically steady, if the averaged quantities of two far separated time instants are equal. Osborne Reynolds firstly occupies this issue and introduced the Reynolds decomposition and the Reynolds operator. He decomposed the flow quantities in its mean value and fluctuating part
\begin{align}
	& \underline{u}(\underline{x},t)=\langle\underline{u}\rangle(\underline{x})+\underline{u}'(\underline{x},t), & p(\underline{x},t)=\langle p \rangle(\underline{x})+p'(\underline{x},t). &
	\label{eq27}
\end{align}
The mean value is defined via the time-average operator
\begin{equation}
	\langle\phi\rangle(\underline{x}) = \lim\limits_{T \rightarrow \infty}\frac{1}{T} \int\limits_{0}^{T} \phi (\underline{x},t) \,dt,
	\label{time_avg}
\end{equation}
which is a Reynolds operator.\\
An ensemble average (stochastic mean) of a random variable $\underline{u}(\underline{x},t)$ calculated from $N$ independent realizations of the same phenomenon is defined as
\begin{equation}
	\langle\underline{u}\rangle(\underline{x},t) = \lim\limits_{N \rightarrow \infty}\frac{1}{N} \sum_{n=1}^{N} \underline{u}^{(n)}(\underline{x},t).
\end{equation}
An ensemble $\{\underline{u}^{(n)}(\underline{x},t)\}^{N}_{n=1}$ is a collection of notionally identical experiments. Since the flow is turbulent, the fluid motion differs from each instance of the ensemble, because microscopically differences in the (experimental) setup become significant as time progresses.\\
Nevertheless, it can be shown that for a very long time period $T\rightarrow\infty$ each realization of the ensemble can be considered as a representative of all possible realizations of an ensemble. In equilibrium systems, time and ensemble averages of physical quantities are equivalent due to ergodic principles.\\
In the following, a Reynolds operator is defined by satisfiying certain properties.
\begin{theorem}[\textbf{Reynolds operator}]
Let $\phi,\psi\in C(\Omega \times [0,T],\mathbb{R})$ and $c\in\mathbb{R}$, then the following properties are satisfied by the Reynolds operator:
\begin{multicols}{2}
\begin{enumerate}[label=(\roman*)]
 \item $\langle\phi+\psi\rangle=\langle\phi\rangle+\langle\psi\rangle$
 \item $\langle c\phi\rangle=c\langle\phi\rangle$
 \item $\langle c\rangle=c$
 \item $\langle\langle\phi\rangle\psi\rangle=\langle\phi\rangle\langle\psi\rangle$
 \item $\langle\frac{\partial\phi}{\partial\underline{x}}\rangle=\frac{\partial\langle\phi\rangle}{\partial\underline{x}}$
 \item $\langle\frac{\partial\phi}{\partial t}\rangle=\frac{\partial\langle\phi\rangle}{\partial t}$
\end{enumerate}
\end{multicols}
Hence, these conditions ensue:
\begin{multicols}{3}
\begin{enumerate}[label=(\roman*)]
\setcounter{enumi}{6}
\item $\langle\langle\phi\rangle\rangle = \langle\phi\rangle$
\item $\langle\phi '\rangle=0,\: \phi '=\phi-\langle\phi\rangle$
\item $\langle\langle\phi\rangle\langle\psi\rangle\rangle=\langle\phi\rangle\langle\psi\rangle$
\end{enumerate}
\end{multicols}
\label{def3}
\end{theorem}
However, fluctuation moments of second or higher order are not
necessarily zero. The standard deviation is often called root-mean-square velocity in turbulence flow and is defined as
\begin{align}
& u_{i,rms} = \sqrt{\langle u_{i}'u_{i}' \rangle} = \sqrt{\langle(u_{i}-\langle u_{i}\rangle)^{2}\rangle},
& u_{rms} = \sqrt{\frac{1}{3}\sum_{i}\langle u_{i}'u_{i}' \rangle}, &
\end{align}
while $u_i$ for $i=1,2,3$ denotes each component of the velocity.\\
By inserting the Reynolds decomposition into the Navier-Stokes equation of the form of Equation~\eqref{NSE_C} and \eqref{NSE_M} and apply the Reynolds operator on the whole system of equations, we derive the Reynolds-averaged Navier-Stokes equations (RANS)
\begin{align*}
& \bigg\langle\nabla\,\cdot\,(\langle\underline{u}\rangle+\underline{u}')\bigg\rangle = 0, \\ &
\bigg\langle\frac{\partial(\langle\underline{u}\rangle+\underline{u}')}{\partial t} + \big((\langle\underline{u}\rangle+\underline{u}')\,\cdot\,\nabla\big)(\langle\underline{u}\rangle+\underline{u}')\bigg\rangle
	= \bigg\langle-\frac{1}{\rho}\nabla (\langle p\rangle + p') + \nu\Delta(\langle\underline{u}\rangle+\underline{u}') + \underline{f}\bigg\rangle. &
\end{align*}
Using the properties from Definition~\ref{def3}, it can be further simplified to
\begin{align*}
& \nabla\,\cdot\,\langle\underline{u}\rangle = 0, \\ &
\frac{\partial\langle\underline{u}\rangle}{\partial t} + \bigg\langle\big((\langle\underline{u}\rangle+\underline{u}')\,\cdot\,\nabla\big)(\langle\underline{u}\rangle+\underline{u}')\bigg\rangle
	= -\frac{1}{\rho}\nabla\langle p\rangle + \nu\Delta\langle\underline{u}\rangle + \langle\underline{f}\rangle. &
\end{align*}
The second term of the left-hand side of the momentum equation is the nonlinear convective term, which needs particular considerations. It follows
\begin{align*}
\bigg\langle\big((\langle\underline{u}\rangle+\underline{u}')\,\cdot\,\nabla\big)(\langle\underline{u}\rangle+\underline{u}')\bigg\rangle
&= \bigg\langle \nabla\,\cdot\,\big((\langle\underline{u}\rangle+\underline{u}')\otimes(\langle\underline{u}\rangle+\underline{u}')\big) \bigg\rangle \\
& = \nabla\,\cdot\,\bigg\langle \langle\underline{u}\rangle\otimes\langle\underline{u}\rangle + \langle\underline{u}\rangle\otimes \underline{u}' + \underline{u}'\otimes\langle\underline{u}\rangle + \underline{u}'\otimes \underline{u}' \bigg\rangle \\
& = \nabla\,\cdot\,(\langle\underline{u}\rangle\otimes\langle\underline{u}\rangle + \langle\underline{u}'\otimes\underline{u}'\rangle) \\
& = (\langle\underline{u}\rangle\,\cdot\,\nabla)\langle\underline{u}\rangle + \nabla\,\cdot\,\langle\underline{u}'\otimes\underline{u}'\rangle.
\end{align*}
Thus we obtain the incompressible RANS equations
\begin{align}
& \nabla\,\cdot\,\langle\underline{u}\rangle = 0, \label{RANS_C}\\ &
\frac{\partial\langle\underline{u}\rangle}{\partial t} + (\langle\underline{u}\rangle\,\cdot\,\nabla)\langle\underline{u}\rangle
	= -\frac{1}{\rho}\nabla\langle p\rangle + \nu\Delta\langle\underline{u}\rangle + \langle\underline{f}\rangle - \nabla\,\cdot\,\langle\underline{u}'\otimes\underline{u}'\rangle. \label{RANS_M}&
\end{align}
Note that the Navier-Stokes equations and the RANS equations do not formally differ a lot, except the additional term $\nabla\,\cdot\,\langle\underline{u}'\otimes\underline{u}'\rangle$. This term is the divergence of the so called Reynolds stress tensor (RST)
\begin{equation*}
	 \langle\underline{u}'\otimes\underline{u}'\rangle =
	 \left(\begin{matrix}\langle\underline{u}_{1}'\underline{u}_{1}'\rangle&\langle\underline{u}_{1}'\underline{u}_{2}'\rangle&\langle\underline{u}_{1}'\underline{u}_{3}'\rangle\\\langle\underline{u}_{2}'\underline{u}_{1}'\rangle&\langle\underline{u}_{2}'\underline{u}_{2}'\rangle&\langle\underline{u}_{2}'\underline{u}_{3}'\rangle\\\langle\underline{u}_{3}'\underline{u}_{1}'\rangle&\langle\underline{u}_{3}'\underline{u}_{2}'\rangle&\langle\underline{u}_{3}'\underline{u}_{3}'\rangle\end{matrix}\right).
\end{equation*}
The application of the RANS equations only makes sense for turbulent flows, where it is assumed that highly fluctuating quantities occur. In the laminar case, $\underline{u}'=0$ and therefore the Navier-Stokes equations are reobtained. From Equation~\eqref{RANS_M}, we see that the RST affects the flow in form of additional stresses. It can be shown that this system of equations is not closed. No matter how many manipulations we perform, there are always more unknowns than equations relating them. This is known as the closure problem of turbulence and it arises because of the nonlinearity of the Navier-Stokes equations.\\
The RST term needs to be modeled to circumvent this problem. In the field of RANS simulation, the focus is primarily on modelling the RST and solving the RANS equations. A detailed overview is given in the upcoming chapter.

\section{Spectral description}
The statistical moments defined in the previous subchapter are single point moments. That is, they contain only information about a variable at a point.\\
In homogeneous turbulent flow, it only makes sense to have some statistical measure of spatial information about the flow. For example, to draw conclusions about length scale information, two point statistics are needed. The autocorrelation function is the correlation between velocity components at two different times. The normalized correlation function (autocorrelation function) is defined as
\begin{equation}
D_{i}(\tau)=\frac{\langle u_{i}(t)u_{i}(t+\tau)\rangle}{\langle u_{i}^2(t) \rangle}.
\end{equation}
Note that $D_{i}(0)=1$. Also from Schwartz's inequality, $D_{i}(\tau)<1$ for all $\tau \ne 0$. The autocorrelation coefficient is often used to define an integral scale of turbulence
\begin{equation}
L_{i}=\int_{0}^{\infty} D_{i}(\tau)\,d\tau.
\end{equation}
The integral length scale gives an estimate of the time interval, over which the velocity component is correlated.\\
The spatial correlation tensor gives the correlation between velocity components at two different spatial locations and has an important interpretation in turbulent flows. It is defined by
\begin{equation}
\doubleunderline{R}(\underline{r})= \left(\begin{matrix}\langle u_{1}(\underline{x})u_{1}(\underline{x}+\underline{r})\rangle&\langle u_{1}(\underline{x})u_{2}(\underline{x}+\underline{r})\rangle&\langle u_{1}(\underline{x})u_{3}(\underline{x}+\underline{r})\rangle\\\langle u_{2}(\underline{x})u_{1}(\underline{x}+\underline{r})\rangle&\langle u_{2}(\underline{x})u_{2}(\underline{x}+\underline{r})\rangle&\langle u_{2}(\underline{x})u_{3}(\underline{x}+\underline{r})\rangle\\\langle u_{3}(\underline{x})u_{1}(\underline{x}+\underline{r})\rangle&\langle u_{3}(\underline{x})u_{2}(\underline{x}+\underline{r})\rangle&\langle u_{3}(\underline{x})u_{3}(\underline{x}+\underline{r})\rangle\end{matrix}\right).
\end{equation}
To describe the various scales of spatial motion in a turbulent flow, it is more instructive to work with the Fourier  transform of the correlation tensor rather than the correlation tensor itself.\\
We assume that the velocity field may be Fourier transformed under the certain requirements. The Fourier transform of $u_{i}(\underline{x},t)$ is
\begin{equation}
\hat{u}_{i}(\underline{k})=\int_{-\infty}^{\infty} u_{i}(\underline{x})e^{-j\underline{k}\cdot\underline{x}}\,d\underline{x},
\end{equation}
herein $\underline{k}$ is the wave number and $j$ the imaginary unit. The inverse Fourier transform is
\begin{equation}
u_{i}(\underline{x})=\frac{1}{(2\pi)^3}\int_{-\infty}^{\infty} \hat{u}_{i}(\underline{k})e^{j\underline{k}\cdot\underline{x}}\,d\underline{k}.
\end{equation}
The Fourier transformed spatial correlation tensor $\doubleunderline{\hat{R}}$ is appropriately called the spectrum tensor or spectral density as it represents the contribution of a wave number $\underline{k}$
\begin{align}
\doubleunderline{\hat{R}}(\underline{k}) &= \int_{-\infty}^{\infty} \doubleunderline{R}(\underline{r})e^{-j\underline{k}\cdot\underline{r}}\,d\underline{r} \\
&=\underline{\hat{u}}\otimes\underline{\hat{u}}^*,
\end{align}
where $*$ indicates the conjugate complex value.\\
In other words, $\doubleunderline{\hat{R}}(\underline{k})$ gives the wavenumber distribution of the correlation tensor. Each wave number $\underline{k}$ corresponds to a physical space structure with a wavelength of $\frac{2\pi}{\underline{k}}$.\\
Of particular significance is the sum of the diagonal components of $\doubleunderline{R}(\underline{r})$ for $\underline{r}=0$. For this case we have
\begin{align*}
tr(\doubleunderline{R}(0))&=\underline{u}\,\cdot\,\underline{u} \\&=2E,
\end{align*}
which is twice the kinetic energy $E$. The operator $tr()$ indicates the trace of a square matrix. In the spectral space, we have
\begin{align*}
tr(\doubleunderline{\hat{R}}(\underline{k}))&=\hat{\underline{u}}\,\cdot\,\hat{\underline{u}}^{*} \\
&= 2\hat{E}(\underline{k}).
\end{align*}
And so
\begin{align}
E&=\frac{1}{(2\pi)^3}\int_{-\infty}^{\infty} tr(\doubleunderline{\hat{R}}(\underline{k}))\,d\underline{k}.
\label{energy}
\end{align}
We are interested in the spectral kinetic energy in dependence of the magnitude of $\underline{k}$, named $k$. Therefore, we integrate the spectrum tensor over a spherical shell with radius $k$
\begin{equation}
\doubleunderline{\hat{R}}(k)=\int_{0}^{2\pi}\int_{0}^{\pi} \doubleunderline{\hat{R}}(k\,\cos\theta,k\,\sin\theta\,\cos\phi,k\,\sin\theta\,\cos\phi)\sin\theta\,d\theta\,d\phi.
\end{equation}
And its trace is
\begin{align*}
tr(\doubleunderline{\hat{R}}(k)) = \hat{E}(k).
\end{align*}
We can rewrite Equation~\eqref{energy} to
\begin{equation}
E=\frac{1}{(2\pi)^3}\int_{0}^{\infty} \hat{E}(k)\,dk,
\end{equation}
$\hat{E}(k)$ represents the contribution of the kinetic energy at a wavenumber of $k$, which is called the three dimensional energy spectrum.\\
Integration of $\hat{E}(k)$ over all $k$ gives the total kinetic energy. With an eddy of a particular size, associated with a wavenumber of certain magnitude, the energy spectrum can be interpreted to give the distribution of energy among the different eddy sizes. As discussed above, the contribution of a wavenumber $k$ corresponds to a structure with a wavelength of $\frac{2\pi}{k}$. A large portion of the theoretical work on turbulent flows (including modeling) is concerned with the description of energy in the wavenumber spectrum and the transfer of energy among the different wavenumbers and frequencies.

\subsection{Kolmogorov's hypothesis}
As shortly described in the beginning of this chapter, another process in turbulent flow is the breakdown of the large coherent structures into small eddies, while energy is transferred via the cascadic breakdown of such vortices. In comparison to the more structured parental vortices that sometimes seem to be "predictable" or of periodic behavior, the little noisy eddies are completely three-dimensional, very random and tend to be completely independent of the big coherent structures. The observed scale distribution of these whirls match quite well with the theory of Kolmogorov. He assumes that the energy spectrum can be split into three sections:

\begin{itemize}
\item The energy-containing, or also called integral scales. These are motions of permanent character, mainly dominate the flow regime over many periods. More important, those scales contain by far the most of kinetic energy and are responsible of introducing turbulent kinetic energy to the system.

\item In the second region, known as the inertial sub-range, the transitive scales are taking place. These scales obey Kolmogorov’s famous law. They are completely independent of the forcing scale, not dominated by inertial forces rather than viscosity. Their main action is to transfer energy from the large scales to the very small ones. In this section the energy spectrum only depends on $k$ and $\epsilon$, the dissipation rate. It can be derived using dimensional analysis
\begin{equation}
\hat{E}(k)=C_{K}\epsilon^{2/3}k^{-5/3},
\end{equation}
where $C_{K}$ is the Kolmogorov constant.
\item The dissipation region, which comprises the smallest scales, is the place where the kinetic energy is dissipated by the viscous effects. These are scales of motion which are smaller than the Kolmogorov scale $\eta$, the length at which viscosity starts to strongly damp the turbulent motion. The end of the curve is characterized by a rapid dropoff in energy content. The scale $\eta$ is defined as
\begin{equation}
\eta=\bigg(\frac{\nu^3}{\epsilon}\bigg)^{\frac{1}{4}}.
\end{equation}
\end{itemize}
\begin{figure}[h]
    \centering
    \includegraphics[width=0.65\textwidth]{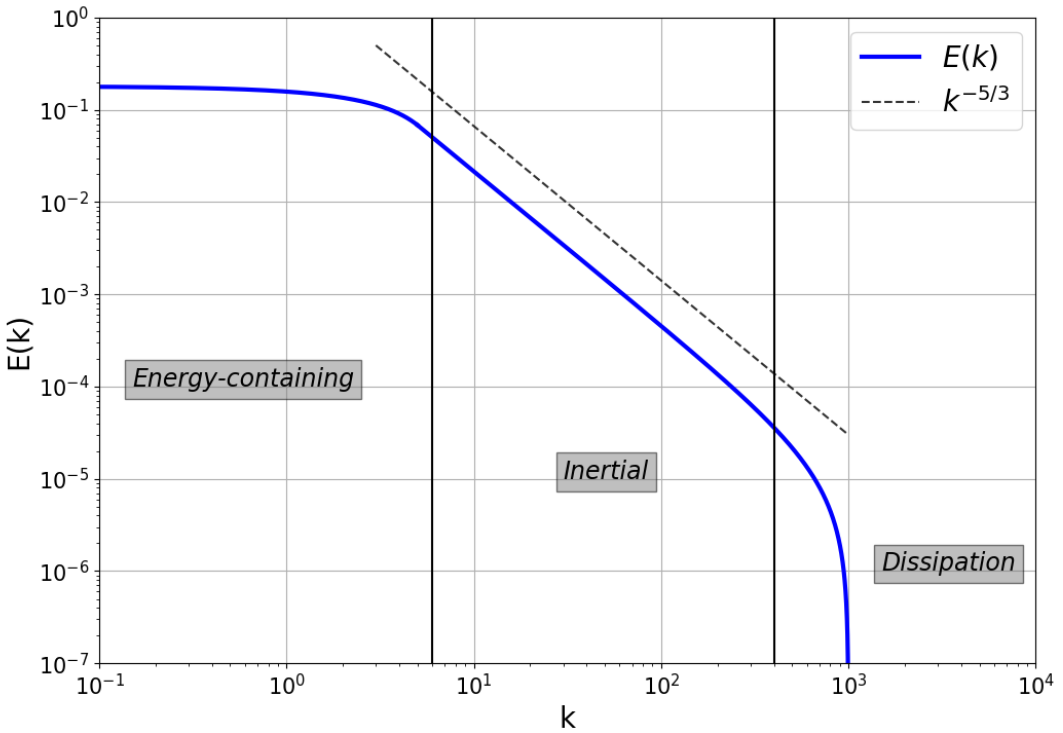}
    \caption{Energy spectrum over $k$.}
    \label{fig:energy_spectrum}
\end{figure}

\section{Turbulent channel flow}
The case of the fully developed turbulent plane channel flow is widely used in turbulence simulation for validation. Due to its geometric simplicity, many advantages are included as well as a lot of experimental and numerical data is available for comparison.\\
We assume a fully developed turbulent flow in a channel with a large extension in $x_1$- and $x_3$-direction. And so, the flow is assumed to be statistically stationary (homogeneous) in streamwise and spanwise direction. Under those assumptions, the following regularities may be deducted. In Figure~\ref{fig:channel_flow} the schematic representation of the turbulent channel flow may be seen.
\begin{figure}[h]
    \centering
    \includegraphics[width=0.85\textwidth]{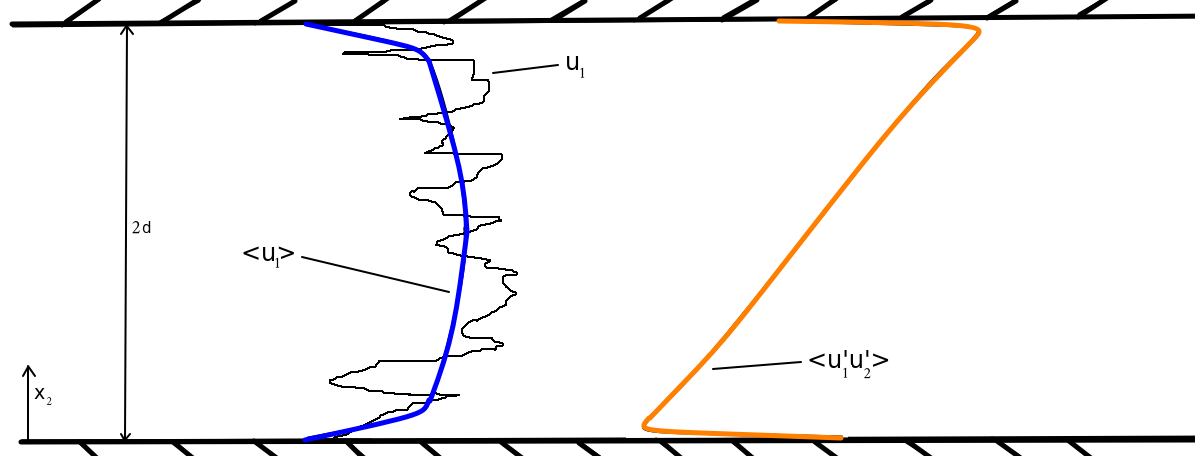}
    \caption{Schematic representation of the turbulent channel flow.}
    \label{fig:channel_flow}
\end{figure}
The bulk velocity and the corresponding bulk Reynolds number are defined as
\begin{align}
& U_{b}=\frac{1}{2\delta} \int_{0}^{2\delta} \langle u_1\rangle\,dx_{2}, & Re_{b}=\frac{U_{b}2\delta}{\nu}.
\end{align}
Due to symmetry in $x_{3}$, it follows that $\langle u_{3}\rangle = 0$. By using the continuity equation and homogeneity in $x_1$ and $x_3$, it follows that $\frac{\partial \langle u_2\rangle}{\partial x_2}=0$. The $x_1$-momentum equation of the RANS equations simplifies to
\begin{equation*}
0=-\frac{1}{\rho}\frac{d\langle p\rangle}{dx_1}+\nu\frac{\partial^2\langle u_1\rangle}{\partial x_2^2}-\frac{\partial\langle u'_1 u'_2\rangle}{\partial x_2}.
\end{equation*}
With the relation
\begin{equation}
\frac{\partial \tau_{ges}}{\partial x_2} = \frac{d\langle p\rangle}{dx_1},
\end{equation}
the total shear stress $\tau_{ges}$ is obtained
\begin{equation}
\tau_{ges} = \mu\frac{\partial\langle u_1\rangle}{\partial x_2}-\rho\langle u'_1u'_2\rangle.
\end{equation}
Due to the symmetry of the problem, the wall shear stress $\tau_{w}$ is defined as
\begin{equation}
\tau_{ges}(x_2=0)=\mu\frac{\partial\langle u_1\rangle}{\partial x_2}\bigg|_{x_2=0} = -\mu\frac{\partial\langle u_1\rangle}{\partial x_2}\bigg|_{x_2=2\delta} = \tau_{w}.
\end{equation}
Hence, the no-slip condition at the wall results in
\begin{equation}
-\frac{d\langle p\rangle}{dx_1}=\frac{\tau_{w}}{\delta},
\end{equation}
giving $\tau_{ges}$ as a linear function of $x_2$
\begin{equation}
\tau_{ges}=\tau_{w}(1-\frac{x_2}{\delta}).
\end{equation}
The profiles of the molecular and turbulent shear stress are shown in Figure~\ref{fig:stresses}.
\begin{figure}[h]
    \centering
    \includegraphics[width=0.65\textwidth]{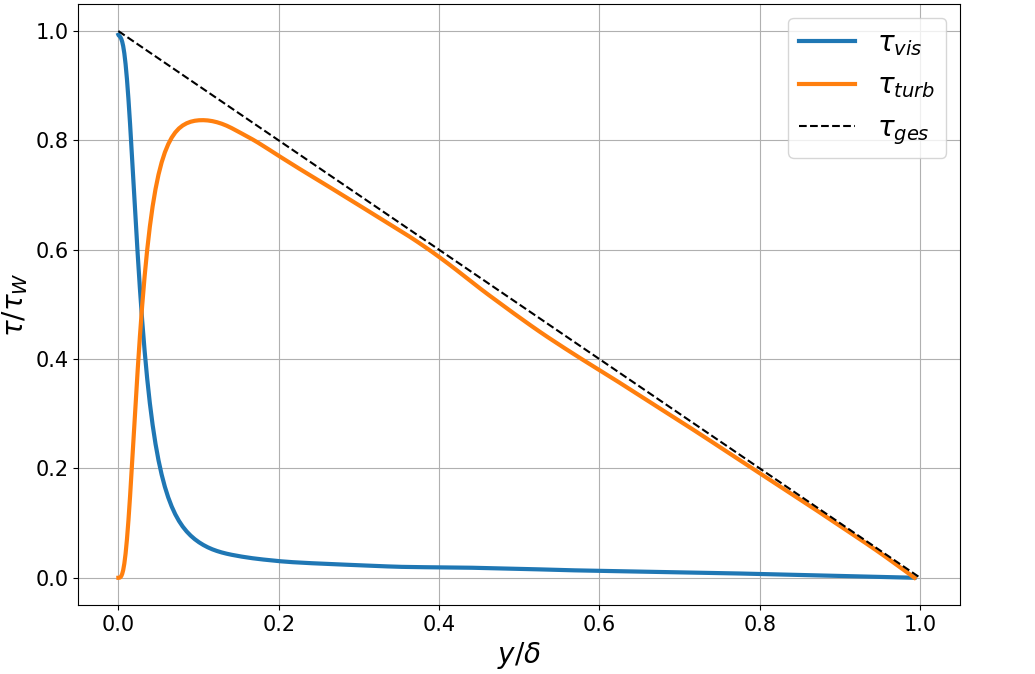}
    \caption{Normalized viscous shear stress profile and turbulent shear stress profile.}
    \label{fig:stresses}
\end{figure}
In vicinity of the wall, the viscous diffusion is clearly dominating. The turbulent diffusion is zero at the wall, but increases rapidly and dominates over almost the whole channel height.
\\[1ex]
The turbulent velocity profile is determined by $\nu,\tau_w,\rho$ and $\delta$. Due to the self similar behavior of the mean velocity profile, the following Reynolds number independent  parameters are introduced. The near-wall region is scaled by use of the friction velocity $u_{\tau}$ and the wall unit $l^{+}$
\begin{equation}
u_{\tau}=\sqrt{\frac{\tau_{w}}{\rho}},
\end{equation}
\begin{equation}
l^{+}=\frac{\nu}{u_{\tau}}.
\end{equation}
The corresponding friction Reynolds number $Re_{\tau}$ is defined as
\begin{equation}
Re_{\tau}=\frac{u_{\tau}\delta}{\nu}=\frac{\delta}{l^{+}}.
\end{equation}
The non-dimensional distance to the wall $y^+$ is scaled with the wall unit $l^{+}$
\begin{equation}
y^+=\frac{x_2}{l^{+}}=\frac{u_{\tau}}{\nu}x_2.
\end{equation}
The non-dimensional mean velocity is scaled with the friction velocity $u_{\tau}$ and reads
\begin{equation}
u^+=\frac{\langle u_1\rangle}{u_{\tau}}.
\end{equation}
\\[1ex]
The law of the wall for turbulent flows is derived by assuming that turbulence near the boundary is a function only of the flow conditions pertaining at that wall and is independent of the flow conditions further away.
The ansatz of so called universal velocity profile for turbulent flow near a wall is
\begin{equation*}
\frac{du^+}{dy^+}=\frac{1}{y^+}\Phi(y^+),
\end{equation*}
while $\Phi(y^+)$ is non-dimensional function of $y^+$. In order to approximate this ordinary differential equation, it may be divided into several sections.
\begin{itemize}
\item Viscous sublayer $y^+ < 5$: Since the turbulent fluctuations must go to zero at the wall, it follows that there is always a very small layer next to the wall, in which the flow is essentially laminar. Due to the no-slip condition at the wall, we get
\begin{equation*}
\tau_{w} = \mu\frac{\partial\langle u_1\rangle}{\partial x_2}\bigg|_{x_2=0} \Rightarrow \rho u_{\tau}^2=\rho\nu\frac{u_{\tau}}{l^+}\frac{du^+}{dy^+}\bigg|_{y^+=0}.
\end{equation*}
For small $y^+$ it follows
\begin{equation*}
\frac{du^+}{dy^+}\approx 1 \Rightarrow u^+ = y^+.
\end{equation*}
\item Logarithmic layer $y^+ > 30$, $y/\delta < 0.3$: Further from the wall, where the turbulent fluctuations dominate, the molecular diffusion may be neglected. In this region, $\Phi(y^+)=const$ since it does not depend on $\nu$ and therefore $y^+$ anymore. There it holds
\begin{equation*}
\frac{du^+}{dy^+}=\frac{1}{\kappa y^+}.
\end{equation*}
$\kappa$ indicates the Von Kármán constant. Integration over $y^+$ leads to the logarithmic law of the wall
\begin{equation*}
u^+=\frac{1}{\kappa}ln(y^+)+C_{log}.
\end{equation*}
For the channel flow, $\kappa=0.41$ and $C_{log}=5$ was determined via experimental and numerical tests.
\item Buffer layer $5<y^+<30$: In the buffer layer neither of the two laws hold.
\end{itemize}
This universal velocity profile is sketched in Figure~\ref{fig:law_of_wall} and compared with numerical data.
\begin{figure}[h]
    \centering
    \includegraphics[width=0.8\textwidth]{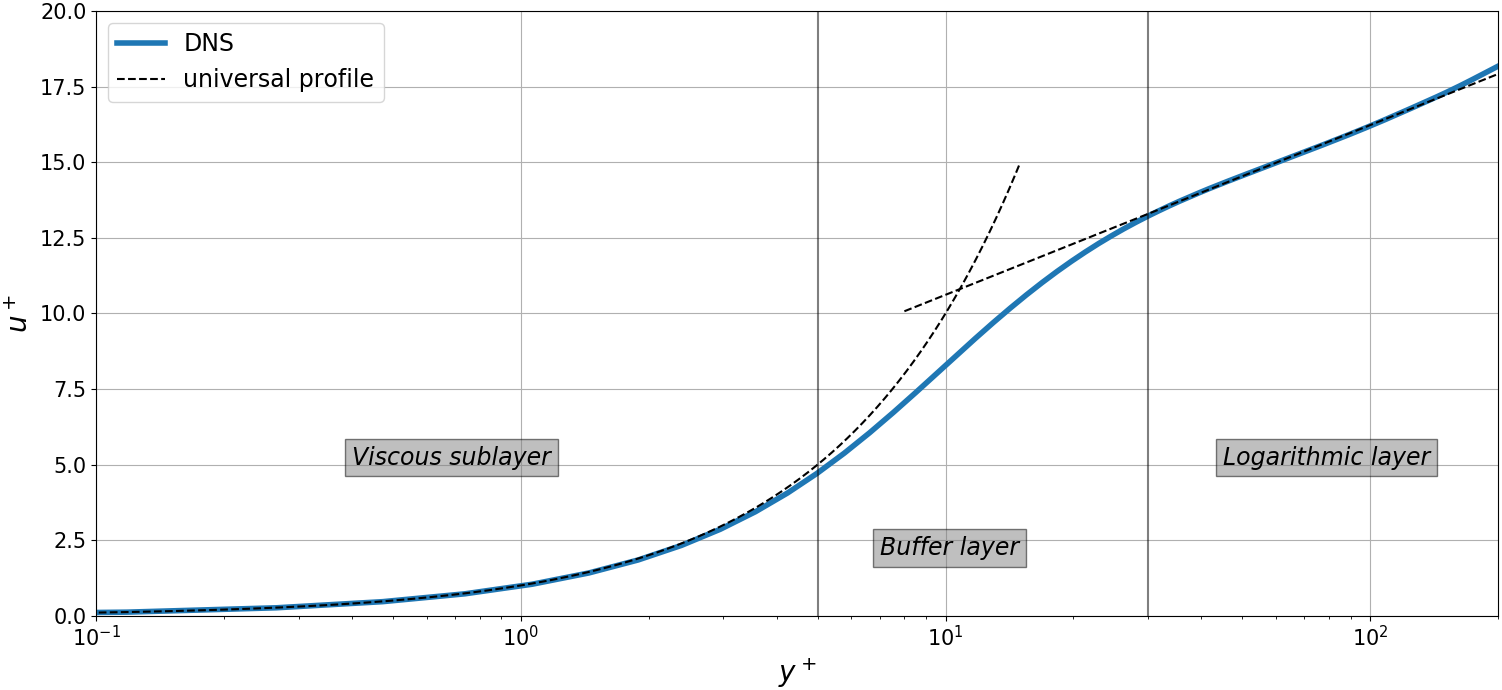}
    \caption{Comparison of DNS data ($Re_{\tau}=395$) with the universal velocity profile.}
    \label{fig:law_of_wall}
\end{figure}

\section{Turbulent boundary layer}
\subsection{Transition from laminar to turbulent}
The process of a laminar flow becoming turbulent is called laminar-turbulent transition. It is mainly used in the context of boundary layers, but applies to any fluid flow.  Generally, a laminar flow will transit to a turbulent flow if a certain limit of instabilities is exceeded and amplifying magnitudes cannot be damped anymore. The instability characteristics of a laminar flow mainly depend on its Reynolds number and on the intensity of mechanical excitation (disturbances in the flow, wall roughness, vibration,..). The interaction of all the influences determine the degree of instability. \\
The initial stage of the natural transition process is known as the receptivity phase. The receptivity gives a measure how the environmental disturbances are transformed to perturbations in the flow, deciding which form of transitional phase is taken. \\
If the initially generated disturbances are small enough, the next stage of the laminar-turbulent transition process is that of primary mode growth. The growth rate can be described by the linear stability theory assuming small amplitudes. In the case of boundary layers, the initial dominate instability mode will occur as a two-dimensional wave, called Tollmien-Schlichting wave, travelling in streamwise direction and its rotational axis crosswise to the flow. \\
The primary modes are for its part sensitive to disturbances, which leads to the second phase of transition so called secondary instabilities. As the linear modes grow and begin to slightly distort the mean flow, they start to exhibit nonlinearities and the linear stability theory no longer holds. The secondary instabilities often lead to coherent structures such as harpin vortices. This stage is rapidly followed by tertiary instabilities and the final breakdown into fully developed turbulent regime.
\begin{figure}[h]
    \centering
    \includegraphics[width=0.6\textwidth]{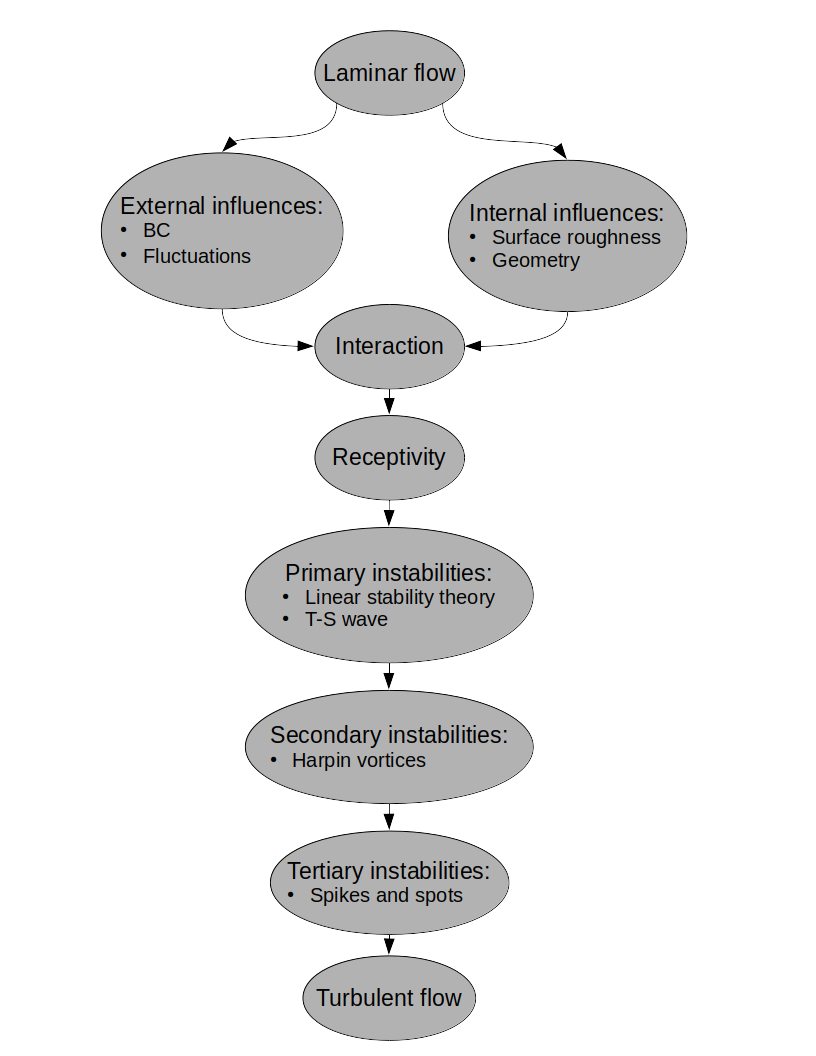}
    \caption{Illustration of laminar-turbulent transition process.}
    \label{fig:transition}
\end{figure}

\subsection{Near-wall structures and inverse energy cascade}
\label{sec:TKE}
Solid boundaries interact with fluid flows by retarding motion tangential to the surface via viscous shear and by blocking the motion of fluid normal to the interface. Typically, near-wall structures are streaks of relatively small velocity, which are diving in regions of higher velocity. These coherent structures commonly appear in turbulent flows and are named low-speed streaks. Such vortex structure can be seen in Figure~\ref{fig:coherent_structure}. The streaks are generated by the lifting of low-speed fluid near the wall induced from the streamwise vortices. Until they are sufficiently lifted ($y^+>20$), the streaks form to nearly hoof-shaped stretched arches. This arch vortex is oriented in such a way that low-speed near-wall fluid is moved away from the wall and fluid with higher kinetic energy is moved towards the wall. At the upper region, new turbulent shear layers form and spawn additional noisy vortices. Those coherent structures are bounded by the wall on one side and the outer flow on the other and already occur in the initial stage of the laminar-turbulent transition. In addition, near-wall structures are known to be sensitive to certain flow properties, such as pressure gradient and wall transpiration.
\begin{figure}[H]
     \centering
     \begin{subfigure}[t]{\textwidth}
         \centering
         \includegraphics[width=\textwidth]{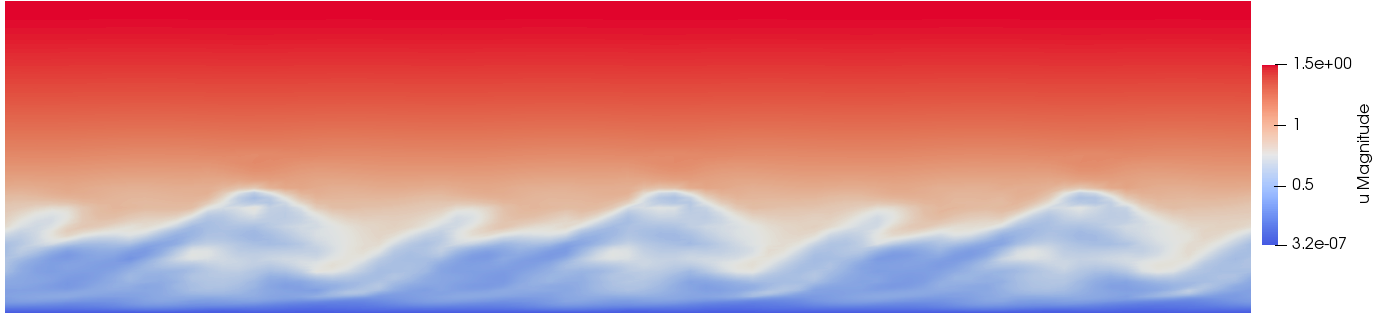}
         \caption{View in $x_1$,$x_2$-plane. Streamwise direction from left to right.}
         \label{fig:y equals x}
     \end{subfigure}

     \begin{subfigure}[b]{\textwidth}
         \centering
         \includegraphics[width=0.5\textwidth]{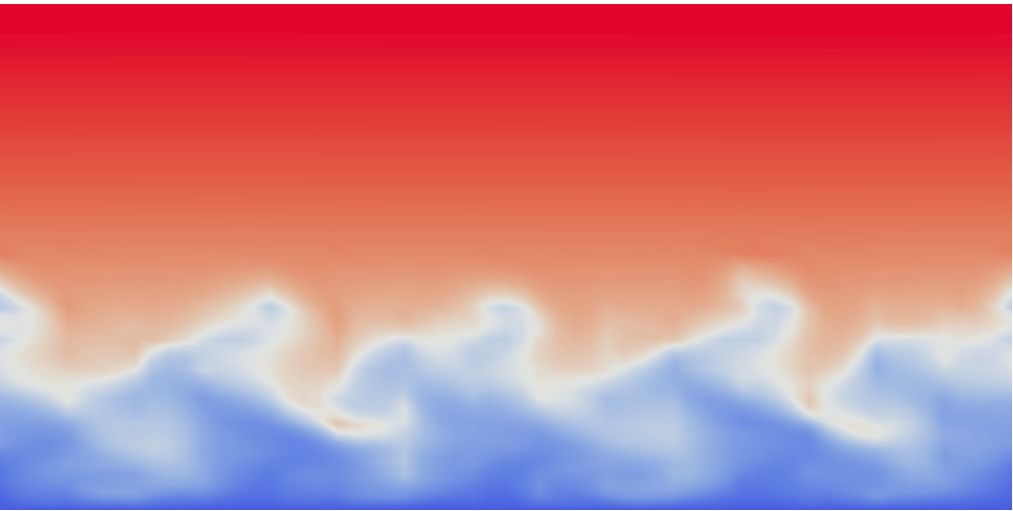}
         \caption{View in $x_2$,$x_3$-plane.}
         \label{fig:three sin x}
     \end{subfigure}
     \caption{Low-speed streaks.}
     \label{fig:coherent_structure}
\end{figure}
Despite of its prevalence, the shape of the turbulent energy spectrum shown previously is by far not universal. The transformation of bulk kinetic energy into turbulent energy can be mainly divided into two categories, both involving the existence of shear. The first are free-shear flows such as mixing layers or jets and the second are wall-bounded flows such as boundary layers or channel flow. In free-shear flows, the growth of instabilities is mainly an inviscid process. These instabilities generally start to develop into big unsteady quasi two-dimensional vortices and then degenerate into smaller eddies through the cascadic process described earlier. This implies that for equilibrium state, the small scales will generally tend to obey the law of Kolmogorov.\\
Wall-bounded flows do not allow the inviscid growth of instabilities. In the existence of solid boundaries, the primary instabilities develop through viscous processes instead. This means that turbulent energy is introduced into the system at small scales near the wall. Experiments have shown that the near-wall eddies are just as anisotropic and inhomogeneous as the large whirls in free-shear flows. This means that there must be some mechanism to transfer energy from small to large scales away from the wall, which is called the inverse energy cascade.\\
In the near-wall boundary layer, viscosity acts as a momentum sink to the core flow, in a similar way as the dissipative effect of the small scale end of the kinetic energy spectrum. The mean flow momentum and therefore mean kinetic energy is transfered to the surface layer by Reynolds stresses and there converted into turbulent kinetic energy and heat (through viscous dissipation). Most of the turbulent energy generated near the wall is lost to dissipation because of large velocity gradients in this region. A significant portion is however transported to the outer flow through turbulent diffusion before it dissipates. Since turbulent production in the outer flow is generally small, this makes the surface layer the main source of turbulent energy of the entire flow.\\[1ex]
In order to obtain a clearer picture of the balance of the turbulent kinetic energy, consider its transport equation. The turbulent kinetic energy $K$ is defined as
\begin{equation}
K=\frac{1}{2}\langle\underline{u}'\,\cdot\,\underline{u}'\rangle.
\end{equation}
Hence, the transport equation of $K$ may derived in the way that firstly substituting the Reynolds decomposition into the Navier-Stokes equations (Equation~\eqref{NSE_M}), then subtracting it by the RANS equations (Equation~\eqref{RANS_M}) and multiply with $\underline{u}'$.\\
The following equation is obtained
\begin{equation*}
\underline{u}'\,\cdot\,\bigg( \frac{\partial \underline{u}'}{\partial t} + \nabla\,\cdot\,\big(\langle\underline{u}\rangle\otimes\underline{u}'+\underline{u}'\otimes\langle\underline{u}\rangle+\underline{u}'\otimes\underline{u}'-\langle\underline{u}'\otimes\underline{u}'\rangle\big)\bigg) = \underline{u}'\,\cdot\,\bigg(-\frac{1}{\rho}\nabla p' + \nu\Delta\underline{u}'\bigg).
\end{equation*}
This equation may be further simplified with the relation $\underline{u}'\,\cdot\,\frac{\partial \underline{u}'}{\partial t}=\frac{1}{2}\frac{\partial (\underline{u}'\,\cdot\,\underline{u}')}{\partial t}$ and the incompressibility constraint to
\begin{align*}
\underline{u}'\,\cdot\,\bigg(\nabla\,\cdot\,\big(\langle\underline{u}\rangle\otimes\underline{u}'+\underline{u}'\otimes\langle\underline{u}\rangle\big)\bigg) &=
\underline{u}'\,\cdot\,\big(\nabla\,\cdot\,(\langle\underline{u}\rangle\otimes\underline{u}')\big)+\underline{u}'\,\cdot\,\big(\nabla\,\cdot\,(\underline{u}'\otimes\langle\underline{u}\rangle)\big) \\
&= \underline{u}'\,\cdot\,(\langle\underline{u}\rangle\,\cdot\,\nabla)\underline{u}'+\underline{u}'\,\cdot\,(\underline{u}'\,\cdot\,\nabla)\langle\underline{u}\rangle \\
&= \frac{1}{2}\langle\underline{u}\rangle\,\cdot\,\nabla(\underline{u}'\,\cdot\,\underline{u}')+(\underline{u}'\otimes\underline{u}'):\nabla\langle\underline{u}\rangle,
\end{align*}
while $:$ denotes the Frobenius inner product of two matrices.\\
This leads to the equation
\begin{align*}
& \frac{1}{2}\frac{\partial (\underline{u}'\,\cdot\,\underline{u}')}{\partial t}+\frac{1}{2}\langle\underline{u}\rangle\,\cdot\,\nabla(\underline{u}'\,\cdot\,\underline{u}')+(\underline{u}'\otimes\underline{u}'):\nabla\langle\underline{u}\rangle +\frac{1}{2}\nabla\,\cdot\,\big((\underline{u}'\,\cdot\,\underline{u}')\underline{u}'\big)-\underline{u}'\,\cdot\,\big(\nabla\,\cdot\,\langle\underline{u}'\otimes\underline{u}'\rangle\big) \\
&= -\frac{1}{\rho}\nabla\,\cdot\,(\underline{u}'p') + \nu\underline{u}'\,\cdot\,\Delta\underline{u}'.
\end{align*}
Reynolds averaging over the whole equation leads to
\begin{equation*}
\frac{\partial K}{\partial t}+\langle\underline{u}\rangle\,\cdot\,\nabla K=
-\langle\underline{u}'\otimes\underline{u}'\rangle:\nabla\langle\underline{u}\rangle-\frac{1}{2}\nabla\,\cdot\,\langle(\underline{u}'\,\cdot\,\underline{u}')\underline{u}'\rangle-\frac{1}{\rho}\nabla\,\cdot\,\langle\underline{u}'p'\rangle + \nu\langle\underline{u}'\,\cdot\,\Delta\underline{u}'\rangle.
\end{equation*}
The term $\langle\underline{u}'\,\cdot\,\Delta\underline{u}'\rangle$ may be further reduced to
\begin{align*}
\frac{1}{2}\langle\Delta(\underline{u}'\,\cdot\,\underline{u}')\rangle
&= \frac{1}{2}\big\langle\nabla\,\cdot\,\big(\nabla(\underline{u}'\,\cdot\,\underline{u}')\big)\big\rangle \\
&= \big\langle\nabla\,\cdot\,\big((\nabla\underline{u}')\underline{u}'\big)\rangle \\
&=\langle\underline{u}'\,\cdot\,\Delta\underline{u}'\rangle+\langle\nabla\underline{u}':\nabla\underline{u}'\rangle.
\end{align*}
Finally, the turbulent kinetic energy equation is
\begin{equation}
\frac{\partial K}{\partial t}+\langle\underline{u}\rangle\,\cdot\,\nabla K=
\Pi-\epsilon+\nu\Delta K + \nabla\,\cdot\,\underline{\Gamma}.
\label{TKE}
\end{equation}
The physically interpretation of each single term of Equation~\eqref{TKE} is:
\begin{itemize}
\item Material derivative of $K$:
\begin{equation*}
\frac{DK}{Dt}=\frac{\partial K}{\partial t}+\langle\underline{u}\rangle\,\cdot\,\nabla K.
\end{equation*}
\item Production of $K$ (generally $\Pi>0$):
\begin{equation*}
\Pi = -\langle\underline{u}'\otimes\underline{u}'\rangle:\nabla\langle\underline{u}\rangle.
\end{equation*}
\item Molecular dissipation ($\epsilon\geq0$):
\begin{equation*}
\epsilon=\nu\langle\nabla\underline{u}':\nabla\underline{u}'\rangle.
\end{equation*}
\item Viscous diffusion:
\begin{equation*}
\nu\Delta K.
\end{equation*}
\item Redistribution terms: Sum of pressure-diffusion and turbulent transport term
\begin{equation*}
\nabla\,\cdot\,\underline{\Gamma} =\nabla\,\cdot\,\big(-\frac{1}{\rho}\langle\underline{u}'p'\rangle-\frac{1}{2}\langle(\underline{u}'\,\cdot\,\underline{u}')\underline{u}'\rangle\big).
\end{equation*}
\end{itemize}
If the flow is parallel and in equilibrium (e.g. channel flow), the time-averaged streamwise derivatives and cross-stream velocities tend to zero. Assuming the contribution of viscous diffusion is small compared to the turbulent contribution, Equation~\eqref{TKE} reduces to
\begin{equation*}
\frac{\partial \Gamma_2}{\partial x_2}=\Pi-\epsilon,
\end{equation*}
so that the only significant mean flux is the wall-normal component $\Gamma_2$. Figure~\ref{fig:TKE_balance} shows the balance between $\Pi$ and $\epsilon$ produced by numerical data from \cite{DNSmoser} and $Re_{\tau}=395$.
\begin{figure}
     \centering
     \begin{subfigure}[b]{0.45\textwidth}
         \centering
         \includegraphics[width=\textwidth]{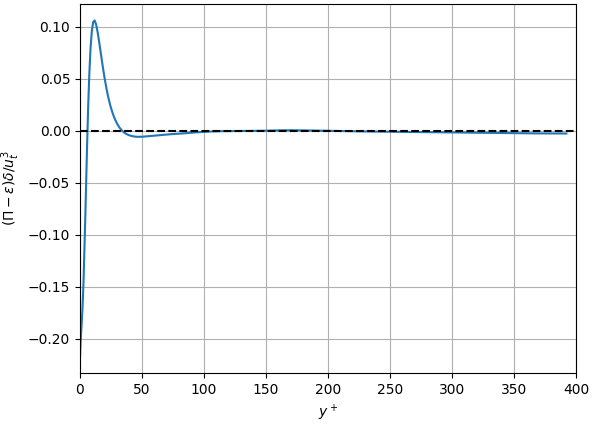}
         \caption{}
         \label{fig:balance}
     \end{subfigure}
     \hfill
     \begin{subfigure}[b]{0.45\textwidth}
         \centering
         \includegraphics[width=\textwidth]{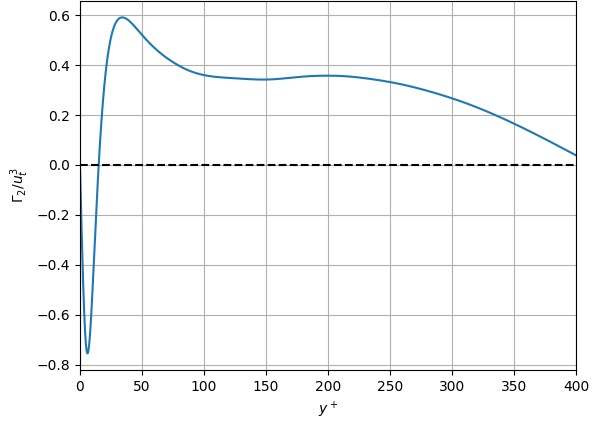}
         \caption{}
         \label{fig:flux}
     \end{subfigure}
     \caption{(a): Normalized energy balance $\Pi-\epsilon$. (b): Normalized energy flux $\Gamma_2$.}
     \label{fig:TKE_balance}
\end{figure}
It clearly shows that net turbulent production occurs only in the outer viscous and buffer region, while dissipation exceeds production in the viscous sublayer. This interpretation is reinforced by the energy flux, which is positive through most of the domain (except very near the wall), signifying a movement of energy away from the surface. Hardly any turbulence is produced in the central part of the channel. Hence, most of the turbulent energy is provided by the flux away from the walls.\\
Since most of the turbulent energy is contained in the largest scale eddies and the eddy size is limited by the distance from the wall, the near-wall energy containing eddies will be small while those centered in the core will be large. Therefore, the transfer of energy from the small to large scales is an example of an inverse energy cascade as opposed to the classical Kolmogorov cascade. Since the nature of the fluxes that are being transferred are different from that in the Kolmogorov theory, the spectral slope should be different too. In the work of \cite{eugene}, dimensional considerations have brought that the inverse energy spectrum has the following shape
\begin{equation}
\hat{E}(k)\sim k^{-1}.
\end{equation}
}

%% file: chapters/simulation.tex
\chapter{Simulation principles and modelling}
\label{chapter:simulation}
%%%%%%%%%%%%%%%%%%%%%%%%%%%%%%%%%%%%%%%%%%%%%%%%%%%%%%%%%%%%%%%%%%%%%%%%%%%%%%%%%%%%%%%%%%%%%%%%%%%%%%%%%%%%%%
%%%%%%%%%%%%%%%%%%%%%%%%%%%%%%%%%%%%%%%%%%%%%%%%%%%%%%%%%%%%%%%%%%%%%%%%%%%%%%%%%%%%%%%%%%%%%%%%%%%%%%%%%%%%%%

{
This chapter gives a brief overview on the several principles to numerically simulate turbulence. A CFD simulation generally consists of many steps, from modelling to post-processing. One decisive step is the translation from a mathematical model to an algebraic system of equations, so called discretization. The discretization methods used in this thesis are given in more detail in Chapter~\ref{chapter:discretization} and \ref{chapter:discretization2}. The most well-known principles of simulating incompressible turbulent flows will be discussed in the upcoming subsections.
\\[1ex]
Basically, three different approaches are considered:
\begin{itemize}
\item Direct numerical simulation (DNS): Resolving all relevant turbulent scales. A huge computational effort is necessary therefore.
\item RANS simulation: Modelling all scales. Hence, the computational cost is significantly low.
\item Large eddy simulation (LES): Resolve the large turbulent scales and modelling the small scales.
\end{itemize}
A comparison of the velocity signal of the different approaches is shown in Figure~\ref{fig:velocity_signal}.
\begin{figure}[ht]
    \centering
    \includegraphics[width=0.6\textwidth]{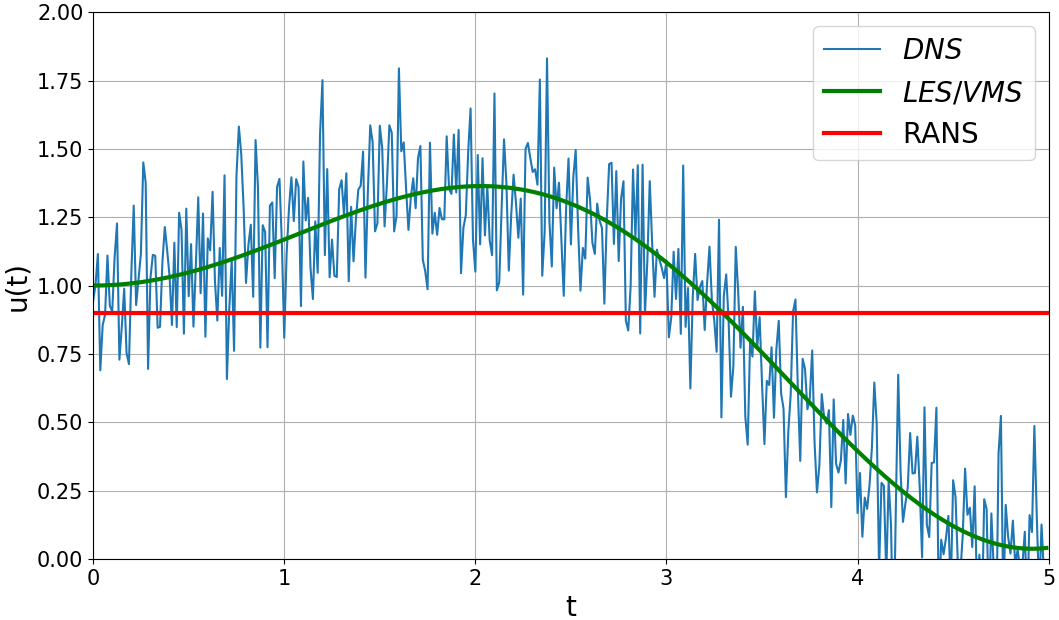}
    \caption{Comparison of velocity signal $u_1$ of different turbulence simulation principles over time $t$.}
    \label{fig:velocity_signal}
\end{figure}

\section{Direct numerical simulation}
The DNS of a turbulent fluid flow directly solves the Navier-Stokes equations of the form of Equation~\eqref{NSE_C} and \eqref{NSE_M} without any modelling concept or assumptions. To obtain a physically correct flow, all relevant scales from the Kolmogorov scale up to the integral scale have to be fully numerically resolved. For example, in atmospheric flows the scale range includes more than nine orders of magnitude. Hence, particular requirements of the efficiency of solving algorithms have to be made. In most of the engineering applications, no such temporal and spatial richness of detail is desired. Therefore, DNS is more or less used for validation and foundational research.
\\[1ex]
An estimation of the required resolution of computational domain may be given, assuming the mesh size of $h \approx \eta$. Then, the number of grid points in one dimension is given by the ratio of biggest to smallest length scale $N_{1D}\approx \frac{L}{\eta}$. Through dimensional analysis it follows
\begin{equation*}
N_{1D}\approx \frac{L}{\eta}=\frac{L}{(\frac{\nu^3}{\epsilon})^{1/4}}=\frac{L\big(\frac{U_b^3}{L}\big)^{1/4}}{\nu^{3/4}}=\bigg(\frac{U_bL}{\nu}\bigg)^{3/4}=Re_b^{3/4},
\end{equation*}
while the dissipation rate is $\epsilon = \frac{U_b^3}{L}$.\\
One similar ansatz is made for the time scale
\begin{equation*}
N_T\approx\frac{T}{T_\eta}=\frac{L}{U_b}\bigg(\frac{U_b^3}{L\nu}\bigg)^{1/2}=\sqrt{\frac{U_bL}{\nu}}=\sqrt{Re_b}.
\end{equation*}
Furthermore, the total computational cost may scale with
\begin{equation*}
N_TN_{1D}^3\approx Re_b^{11/4}.
\end{equation*}
Since it grows almost cubic with respect to the Reynolds number, high turbulent flow in complex geometries are computationally unfeasible in the near future. Herein, it has to be mentioned that it only gives a rough estimation, since the total complexity depends on a variety of things (solving algorithm, discretization,..). Additionally, in order to reduce the computational effort, for several cases a full resolution of the flow domain is often not of major importance even for a DNS.

\section{RANS simulation}
As the DNS of turbulent flows is still prohibitively expensive, models based on the RANS equations are very commonly employed in CFD nowadays. For engineering purposes, the detailed resolved scales are not desired and often only statistically values are of main interest. In RANS simulation, the whole range of turbulent scales is modeled by decomposing and averaging the governing equations of fluid motion, leading to the previous derived RANS equations of the from of Equation~\eqref{RANS_C} and \eqref{RANS_M}.\\
Hence, its main goal is the numerical solution of its equations. The aim of the statistically turbulence modelling is to solve the closure problem of Equation~\eqref{RANS_M}. In order to obtain a physically rightful model, it has to fulfill several physical and mathematical constraints (e.g. tensorial consistency) and development of such models is still a field of research.\\
It may be divided into two classes, one are the eddy-viscosity models and the other are Reynolds-stress models. In this thesis, only models of the first mentioned class will be discussed.

\subsection{Eddy viscosity hypothesis}
Boussinesq \cite{boussinesq} firstly introduced the concept of eddy viscosity. It is based on the analogy of turbulent to gas kinetic processes and he proposed a relation of the turbulence stresses to the mean flow. The transfer of momentum in between molecules results in friction. In macroscopic scales, the colliding of  "turbulence clusters" somehow emerges tensions in rather similar way as in the microscopic scale. The principle component of modelling is to determine a velocity scale, which is characteristic for the intensity of turbulent mixture and an appropriate typically length scale, where turbulence takes place.\\
Equation~\eqref{RANS_M} may be rewritten in the form
\begin{equation*}
\frac{\partial\langle\underline{u}\rangle}{\partial t} + (\langle\underline{u}\rangle\,\cdot\,\nabla)\langle\underline{u}\rangle
	= -\frac{1}{\rho}\nabla\langle p\rangle + \nabla\,\cdot\,\big(2\nu\langle\doubleunderline{S}\rangle-\langle\underline{u}'\otimes\underline{u}'\rangle\big),
\end{equation*}
neglecting the volume force term and introducing the mean strain rate tensor
\begin{equation*}
\langle\doubleunderline{S}\rangle=\frac{1}{2}(\nabla\langle\underline{u}\rangle+\nabla\langle\underline{u}\rangle^T).
\end{equation*}
According to the linear proportionality of the molecular diffusion term to the viscosity and mean strain rate tensor, the same proportionality is assumed for the Reynolds stresses. The ansatz is
\begin{equation}
\langle\underline{u}'\otimes\underline{u}'\rangle-\frac{2}{3}K\doubleunderline{I}=-2\nu_T\langle\doubleunderline{S}\rangle,
\label{EVH}
\end{equation}
where $\nu_T\in C(\Omega \times [0,T],\mathbb{R})$, $\nu_T>0$ and proportional to a typically velocity and length scale $\nu_T\sim l_TU_T$. In incompressible flows, the trace of the strain rate tensor is zero (divergence-free velocity) and therefore the viscous stresses are only described via the deviatoric part of the tensor. The left-hand side of Equation~\eqref{EVH} represents the deviatoric part of the RST (its trace is generally not zero), in order to obtain tensorial consistency.
\\[1ex]
Find $\nu_T=\nu_T(\underline{x},t)$  and Equation~\eqref{EVH} closes the RANS equations of form of Equation~\eqref{RANS_C} and \eqref{RANS_M}.
\begin{problem}[\textbf{Reynolds-averaged Navier-Stokes equations}]
Let $\nu,\rho\in \mathbb{R}\setminus \{ 0 \}$, $\nu_T\in C^1(\Omega \times [0,T],\mathbb{R})$, $\langle\underline{u}\rangle_D\in C^0(\partial\Omega_{D},\mathbb{R}^{3})$, $\underline{g}\in C^0(\partial\Omega_{N},\mathbb{R}^{3})$ and $\langle\underline{u}\rangle_0\in C^{2}(\Omega,\mathbb{R}^{3})$ find $\langle\underline{u}\rangle\in C^2(\Omega \times [0,T],\mathbb{R}^3)$ and $\hat{p}\in C^1(\Omega \times [0,T],\mathbb{R})$ such that
\begin{align*}
	& \nabla\,\cdot\,\langle\underline{u}\rangle=0 \numberthis & \textnormal{in}\:\Omega\times[0,T], & \\
	& \frac{\partial\langle\underline{u}\rangle}{\partial t} + (\langle\underline{u}\rangle\,\cdot\,\nabla)\langle\underline{u}\rangle
= -\frac{1}{\rho}\nabla \hat{p} +\nabla\,\cdot\,\big(2(\nu+\nu_T)\langle\doubleunderline{S}\rangle\big) \numberthis & \textnormal{in}\:\Omega\times[0,T],  & \\
	& \langle\underline{u}\rangle=\langle\underline{u}\rangle_0 & \textnormal{in}\:\Omega, t=0, & \\
	& \langle\underline{u}\rangle=\langle\underline{u}\rangle_D & \textnormal{in}\:\partial\Omega_{D}\times[0,T], & \\
	& \frac{\partial\langle\underline{u}\rangle}{\partial\underline{n}}=\underline{g} & \textnormal{in}\:\partial\Omega_{N}\times[0,T].
\end{align*}
\label{RANS}
\end{problem}
Since the pressure has no thermodynamic significance in incompressible flows, the isotropic part of the RST is incoporated by the modified pressure $\hat{p}=\langle p\rangle+\frac{2}{3}\rho K$.\\
The weakness of the Boussinesq assumption is that it is not valid in general. There is no evidence that the Reynolds stress tensor must be proportional to the strain rate tensor. In most of the simple flow cases, the linear proportionality gives a good approximation. For flow scenarios with strong curvature or strongly accelerated or decelerated flows, the Boussinesq assumption is simply not valid.\\
Contrary to the molecular viscosity, the turbulent viscosity is no material property and depends on the flow itself. In order to be able to determine $\nu_T$, further equations have to be provided. Nowadays, a large amount of different turbulence models exist, each with its advantages and drawbacks. In this thesis, the three most important two-equation turbulence models for wall-bounded flows are explained.

\subsection{The $K-\epsilon$ model}
Probably the most well-known turbulence model, the $K-\epsilon$ model and its variations were firstly introduced by Harlow and Nakayama \cite{harlow1968transport}. Through dimensional analysis, the following proportionalities for the characteristic velocity and length scale are given
\begin{equation*}
l_T \sim \frac{K^{3/2}}{\epsilon},
\end{equation*}
\begin{equation*}
U_T \sim \sqrt{K}.
\end{equation*}
Core assumption of this model is the following ansatz for the eddy viscosity
\begin{equation}
\nu_T=C_\mu\frac{K^2}{\epsilon},
\label{KE_nut}
\end{equation}
where $C_\mu$ is a constant.\\
To be able to compute the turbulent viscosity as in Equation~\eqref{KE_nut}, two additional transport equations for $K$ and $\epsilon$ need to be deducted. For the kinetic turbulent energy, Equation~\eqref{TKE} is used and its redistribution term is modelled as
\begin{equation*}
\underline{\Gamma}\approx\frac{\nu_T}{\sigma_K}\nabla K,
\end{equation*}
with $\sigma_K$ to be a constant. This new modeled term should act as a turbulent diffusion.
The production term $\Pi$ may be rewritten and closed with Equation~\eqref{EVH}
\begin{align*}
\Pi &= -\langle\underline{u}'\otimes\underline{u}'\rangle:\nabla\langle\underline{u}\rangle \\
&=-(\langle\underline{u}'\otimes\underline{u}'\rangle-\frac{2}{3}K\doubleunderline{I}):\langle\doubleunderline{S}\rangle \\
&=2\nu_T\langle\doubleunderline{S}\rangle:\langle\doubleunderline{S}\rangle.
\end{align*}
Secondly, an equation for $\epsilon$ has to be determined. Although a transport equation may be derived for $\epsilon$ as well, it contains even more unclosed terms and is subtantially more complicated than the transport equation for $K$. Therefore, analogous to the $K$ transport equation, a formula for $\epsilon$ is specified in the following
\begin{equation*}
\frac{\partial \epsilon}{\partial t}+\langle\underline{u}\rangle\,\cdot\,\nabla \epsilon=
C_{\epsilon 1}\frac{\Pi\epsilon}{K}-C_{\epsilon 2}\frac{\epsilon^2}{K}+\nabla\,\cdot\,\big((\nu+\frac{\nu_T}{\sigma_{\epsilon}})\nabla \epsilon\big),
\end{equation*}
where $C_{\epsilon 1}$, $C_{\epsilon 2}$ and $\sigma_{\epsilon}$ are model constants.
\\[1ex]
The coupled system of equations is summarized in the following problem.
\begin{problem}[\textbf{$K-\epsilon$ transport equations}]
Let $\sigma_K$, $\sigma_{\epsilon}$, $C_{\epsilon 1}$, $C_{\epsilon 2}$, $\nu$ $\in \mathbb{R}\setminus \{ 0 \}$, $\nu_T\in C^1(\Omega \times [0,T],\mathbb{R})$, $\langle\underline{u}\rangle\in C^1(\Omega \times [0,T],\mathbb{R}^3)$, $K_D\in C^0(\partial\Omega_{D},\mathbb{R})$, $g_k\in C^0(\partial\Omega_{N},\mathbb{R})$, $\epsilon_D\in C^0(\partial\Omega_{D},\mathbb{R})$, $g_{\epsilon}\in C^0(\partial\Omega_{N},\mathbb{R})$, $K_0\in C^{2}(\Omega,\mathbb{R})$ and $\epsilon_0\in C^{2}(\Omega,\mathbb{R})$ find $K\in C^2(\Omega \times [0,T],\mathbb{R})$ and $\epsilon\in C^2(\Omega \times [0,T],\mathbb{R})$ such that
\begin{align*}
& \frac{\partial K}{\partial t}+\langle\underline{u}\rangle\,\cdot\,\nabla K=
\Pi-\epsilon+\nabla\,\cdot\,\big((\nu+\frac{\nu_T}{\sigma_K})\nabla K\big) \label{TKE_modelled} \numberthis & \textnormal{in}\:\Omega\times[0,T], & \\
& \frac{\partial \epsilon}{\partial t}+\langle\underline{u}\rangle\,\cdot\,\nabla \epsilon=
C_{\epsilon 1}\frac{\Pi\epsilon}{K}-C_{\epsilon 2}\frac{\epsilon^2}{K}+\nabla\,\cdot\,\big((\nu+\frac{\nu_T}{\sigma_{\epsilon}})\nabla \epsilon\big) \label{epsilon} \numberthis& \textnormal{in}\:\Omega\times[0,T], & \\
& \Pi = \frac{1}{2}\nu_T(\nabla\langle\underline{u}\rangle+\nabla\langle\underline{u}\rangle^T):(\nabla\langle\underline{u}\rangle+\nabla\langle\underline{u}\rangle^T) & \textnormal{in}\:\Omega\times[0,T], & \\
& K=K_0 & \textnormal{in}\:\Omega, t=0, & \\
& \epsilon=\epsilon_0 & \textnormal{in}\:\Omega, t=0, & \\
& K=K_D & \textnormal{in}\:\partial\Omega_{D}\times[0,T], & \\
& \epsilon=\epsilon_D & \textnormal{in}\:\partial\Omega_{D}\times[0,T], & \\
& n\,\cdot\,\nabla K = g_k & \textnormal{in}\:\partial\Omega_{N}\times[0,T], & \\
& n\,\cdot\,\nabla \epsilon = g_{\epsilon} & \textnormal{in}\:\partial\Omega_{N}\times[0,T].
\end{align*}
\label{KEpsilon}
\end{problem}

\subsection{The $K-\omega$ model}
The $K-\epsilon$ model suffers from some major drawbacks, one of them is the lack of sensitivity to adverse pressure gradients as it is observed that under such conditions the model tends to overestimate the shear stress and by that delay flow separation.\\
The $K-\omega$ model as devised by Wilcox \cite{wilcox} overcomes this weakness of the $K-\epsilon$ model by the definition of the specific dissipation rate $\omega$
\begin{equation}
\omega = \frac{\epsilon}{C_{\mu}K}.
\label{omega_def}
\end{equation}
The eddy-viscosity reads
\begin{equation}
\nu_T=\frac{K}{\omega}.
\end{equation}
Again, the transport equation for $\omega$ is derived in a very similar way as for Equation~\eqref{epsilon}.
\begin{problem}[\textbf{$K-\omega$ transport equations}]
Let $\beta^*$, $\sigma_K$, $\sigma_{\omega}$, $C_{\omega 1}$, $C_{\omega 2}$, $\nu$ $\in \mathbb{R}\setminus \{ 0 \}$, $\nu_T\in C^1(\Omega \times [0,T],\mathbb{R})$, $\langle\underline{u}\rangle\in C^1(\Omega \times [0,T],\mathbb{R}^3)$, $K_D\in C^0(\partial\Omega_{D},\mathbb{R})$, $g_k\in C^0(\partial\Omega_{N},\mathbb{R})$, $\omega_D\in C^0(\partial\Omega_{D},\mathbb{R})$, $g_{\omega}\in C^0(\partial\Omega_{N},\mathbb{R})$, $K_0\in C^{2}(\Omega,\mathbb{R})$ and $\omega_0\in C^{2}(\Omega,\mathbb{R})$ find $K\in C^2(\Omega \times [0,T],\mathbb{R})$ and $\omega\in C^2(\Omega \times [0,T],\mathbb{R})$ such that
\begin{align*}
& \frac{\partial K}{\partial t}+\langle\underline{u}\rangle\,\cdot\,\nabla K=
\Pi-\beta^*K\omega+\nabla\,\cdot\,\big((\nu+\frac{\nu_T}{\sigma_K})\nabla K\big) \numberthis & \textnormal{in}\:\Omega\times[0,T], & \\
& \frac{\partial \omega}{\partial t}+\langle\underline{u}\rangle\,\cdot\,\nabla \omega=
C_{\omega 1}\frac{\omega}{K}\Pi-C_{\omega 2}\omega^2+\nabla\,\cdot\,\big((\nu+\frac{\nu_T}{\sigma_{\omega}})\nabla \omega\big) \label{omega} \numberthis & \textnormal{in}\:\Omega\times[0,T], & \\
& \Pi = \frac{1}{2}\nu_T(\nabla\langle\underline{u}\rangle+\nabla\langle\underline{u}\rangle^T):(\nabla\langle\underline{u}\rangle+\nabla\langle\underline{u}\rangle^T) & \textnormal{in}\:\Omega\times[0,T], & \\
& K=K_0 & \textnormal{in}\:\Omega, t=0, & \\
& \omega=\omega_0 & \textnormal{in}\:\Omega, t=0, & \\
& K=K_D & \textnormal{in}\:\partial\Omega_{D}\times[0,T], & \\
& \omega=\omega_D & \textnormal{in}\:\partial\Omega_{D}\times[0,T], & \\
& n\,\cdot\,\nabla K = g_k & \textnormal{in}\:\partial\Omega_{N}\times[0,T], & \\
& n\,\cdot\,\nabla \omega = g_{\omega} & \textnormal{in}\:\partial\Omega_{N}\times[0,T].
\end{align*}
\label{KOmega}
\end{problem}
This model has superior performance for wall-bounded flows with relatively low Reynolds numbers.

\subsection{The $K-\omega$ SST model}
The shear stress transport (SST) $K-\omega$ model, firstly revised by Menter \cite{menter}, combines the two previous models such that the $K-\omega$ model is used in the proximity of walls and switches to the $K-\epsilon$ in the free shear region. Menter recognized that the $\epsilon$ transport equation from Equation~\eqref{epsilon} may be transformed into a new $\omega$ transport equation via substituting $\omega$ from Equation~\eqref{omega_def}. This new transformed equation looks very similar to the one from Equation~\eqref{omega}, expect it includes an additional non-conservative cross-diffusion term
\begin{equation*}
\frac{2C_{\omega 3}}{\omega}\nabla K : \nabla \omega,
\end{equation*}
introducing a new constant $C_{\omega 3}$.\\
The inclusion of this term will potentially make it able to smoothly switch between the models via a blending functions (highly nonlinear functions of the several parameter, including the wall distance). The additional functions are given by
\begin{align}
& F_1=\tanh(\zeta_1), & \\
& \zeta_1 = \bigg(\min\bigg[\max\big(\frac{\sqrt{K}}{\beta^*\omega d},\frac{500\nu}{\omega d^2}\big),\frac{4C_{\omega 2}K}{CDd^2}\bigg]\bigg)^4, & \\
& CD=\max\big(\frac{2C_{\omega 3}}{\omega}\nabla K : \nabla \omega,10^{-20}\big), & \\
& F_2 = \tanh(\zeta_2), & \\
& \zeta_2 = \bigg(\max\big(2\frac{\sqrt{K}}{\beta^*\omega d},\frac{500\nu}{\omega d^2}\big)\bigg)^2.
\end{align}
The variable $d$ describes the distance to the nearest wall. If $F_1 = 1$ the cross-diffusion term disappears and the $K-\omega$ model is recovered, otherwise if $F_1 = 0$ the $K-\epsilon$ model is reobtained.\\
The model parameters (e.g. $\sigma_K$, $\sigma_{\omega}$) are blended via
\begin{equation}
\chi=F_1\chi_1 + (1-F1)\chi_2,
\label{blendingfunction}
\end{equation}
where $\chi_1$ represents the $K-\omega$ model constants and $\chi_2$ the ones from the $K-\epsilon$ model.\\
The eddy viscosity is computed from
\begin{equation}
\nu_T=\frac{a_1 K}{max(a_1\omega,SF_2)},
\end{equation}
while $a_1$ is a model constant and $S$ is the magnitude of the strain rate tensor
\begin{equation}
S=\sqrt{2\langle\doubleunderline{S}\rangle:\langle\doubleunderline{S}\rangle}.
\end{equation}
\begin{problem}[\textbf{$K-\omega$ SST transport equations}]
Let $\beta^*$, $\sigma_K$, $\sigma_{\omega}$, $C_{\omega 1}$, $C_{\omega 2}$, $C_{\omega 3}$, $\nu$ $\in \mathbb{R}\setminus \{ 0 \}$, $\nu_T\in C^1(\Omega \times [0,T],\mathbb{R})$, $\langle\underline{u}\rangle\in C^1(\Omega \times [0,T],\mathbb{R}^3)$, $K_D\in C^0(\partial\Omega_{D},\mathbb{R})$, $g_k\in C^0(\partial\Omega_{N},\mathbb{R})$, $\omega_D\in C^0(\partial\Omega_{D},\mathbb{R})$, $g_{\omega}\in C^0(\partial\Omega_{N},\mathbb{R})$, $K_0\in C^{2}(\Omega,\mathbb{R})$ and $\omega_0\in C^{2}(\Omega,\mathbb{R})$ find $K\in C^2(\Omega \times [0,T],\mathbb{R})$ and $\omega\in C^2(\Omega \times [0,T],\mathbb{R})$ such that
\begin{align*}
& \frac{\partial K}{\partial t}+\langle\underline{u}\rangle\,\cdot\,\nabla K=
\Pi-\beta^*K\omega+\nabla\,\cdot\,\big((\nu+\frac{\nu_T}{\sigma_K})\nabla K\big) \numberthis & \textnormal{in}\:\Omega\times[0,T], & \\
& \frac{\partial \omega}{\partial t}+\langle\underline{u}\rangle\,\cdot\,\nabla \omega=\frac{C_{\omega 1}}{\nu_T}\Pi-C_{\omega 2}\omega^2& \\ & +\nabla\,\cdot\,\big((\nu+\frac{\nu_T}{\sigma_{\omega}})\nabla \omega\big)+\frac{2(1-F_1) C_{\omega 3}}{\omega}\nabla K : \nabla \omega \numberthis & \textnormal{in}\:\Omega\times[0,T], & \\
& \Pi = \frac{1}{2}\nu_T(\nabla\langle\underline{u}\rangle+\nabla\langle\underline{u}\rangle^T):(\nabla\langle\underline{u}\rangle+\nabla\langle\underline{u}\rangle^T) & \textnormal{in}\:\Omega\times[0,T], & \\
& K=K_0 & \textnormal{in}\:\Omega, t=0, & \\
& \omega=\omega_0 & \textnormal{in}\:\Omega, t=0, & \\
& K=K_D & \textnormal{in}\:\partial\Omega_{D}\times[0,T], & \\
& \omega=\omega_D & \textnormal{in}\:\partial\Omega_{D}\times[0,T], & \\
& n\,\cdot\,\nabla K = g_k & \textnormal{in}\:\partial\Omega_{N}\times[0,T], & \\
& n\,\cdot\,\nabla \omega = g_{\omega} & \textnormal{in}\:\partial\Omega_{N}\times[0,T].
\end{align*}
\label{SSTKOmega}
\end{problem}
The SST model exhibits the closest agreement with the test cases and works very well in wall-bounded shear flows with adverse pressure gradients. Although, its computational cost is slightly higher than the most linear eddy-viscosity models (due to the nonlinearities in its evaluation), the effort is reasonable in comparison with a LES.

\subsection{The scalar convection diffusion problem}
These two-equation eddy-viscosity models mainly describe two coupled scalar convection diffusion problems with (nonlinear) source and sink terms. The scalar quantity $\phi\in C(\Omega \times [0,T],\mathbb{R})$ gets transported along a given stream $\langle\underline{u}\rangle$ with the property $\nabla\,\cdot\,\langle\underline{u}\rangle=0$. The corresponding convective term is then simply $(\langle\underline{u}\rangle\,\cdot\,\nabla \phi)$. Convection is superimposed by diffusion which describes the exchange of the transported quantity on a molecular and turbulent level. The corresponding diffusive flux term is $\nabla\,\cdot\,\big((\nu+\nu_T)\nabla \phi\big)$. The term $f$ describes the sources and sinks and may consist of nonlinear terms.\\
Generally, the scalar convection diffusion problem reads:
\begin{problem}[\textbf{The scalar convection diffusion equation}]
Let $\nu$ $\in \mathbb{R}\setminus \{ 0 \}$, $\nu_T\in C^1(\Omega \times [0,T],\mathbb{R})$, $\langle\underline{u}\rangle\in C^1(\Omega \times [0,T],\mathbb{R}^3)$, $\phi_D\in C^0(\partial\Omega_{D},\mathbb{R})$, $g_{\phi}\in C^0(\partial\Omega_{N},\mathbb{R})$ and $\phi_0\in C^{2}(\Omega,\mathbb{R})$ find $\phi\in C^2(\Omega \times [0,T],\mathbb{R})$ such that
\begin{align*}
& \frac{\partial \phi}{\partial t}+\langle\underline{u}\rangle\,\cdot\,\nabla \phi=\nabla\,\cdot\,\big((\nu+\nu_T)\nabla \phi\big)+f \numberthis & \textnormal{in}\:\Omega\times[0,T], & \\
&\nabla\,\cdot\,\langle\underline{u}\rangle=0 & \textnormal{in}\:\Omega\times[0,T],& \\
& \phi=\phi_0 & \textnormal{in}\:\Omega, t=0, & \\
& \phi=\phi_D & \textnormal{in}\:\partial\Omega_{D}\times[0,T], & \\
& n\,\cdot\,\nabla \phi = g_{\phi} & \textnormal{in}\:\partial\Omega_{N}\times[0,T]. &
\end{align*}
\label{scalarconvectiondiffusion}
\end{problem}

\subsection{Near-wall treatments}
In the proximity of walls, most of the eddy-viscosity models suffer in performance, since the core assumptions of isotropic turbulence and high Reynolds number flow do not hold in this area. Especially the $K-\epsilon$ model is not valid close to solid walls. This situation gives rise to a plethora of near-wall conditions. Generally, two approaches are distinguished, the low Reynolds number (LRN) treatment and high Reynolds number (HRN) treatment.\\
Models using the LRN version integrate every equation up to the wall using the appropriate Dirichlet or Neumann boundary conditions for the physical quantities at the solid wall. Therefore, the first computational cell has to be in $y^+\sim 1$, resulting in fine resolved meshes close to walls. Additionally, some models use damping functions of the model parameters to guarantee asymptotic consistency with the turbulent boundary layer behavior.\\
The HRN approach uses wall functions, which are applied in the nearest cell at the wall instead of integration. These functions often rely on approximated log-law velocity profiles. No direct boundary conditions are therefore necessary, since the closest cell to the wall is computed according to the wall functions. This method enhances the computational effort, however it is not suitable for more complex scenarios. \\
In this thesis, we stick to the LRN approach for all RANS simulations.

\section{Large eddy simulation}
The LES is a technique intermediate between the DNS and the solution of RANS simulation. In LES, the contribution of the largest, kinetic energy-carrying eddies is computed exactly, while only the effect of the smaller scale whirls is modeled. Since the smaller structures tend to be more homogeneous and universal and less affected by the global conditions than the larger scaled eddies, it seems to be more attractive to model the small scale part. Similar to DNS, LES provides a three-dimensional, time-dependent solution of the Navier-Stokes equations.\\
To seperate the large scales of motion from the smaller ones, some kind of averaging has to be done. In contrast to the ensemble average in RANS, in LES the averaging operator is generally a spatial and temporal low-pass filter. Formally, each flow variable is decomposed in the large and small scale part
\begin{align}
	& \underline{u}(\underline{x},t)=\overline{\underline{u}}(\underline{x},t)+\underline{u}''(\underline{x},t), & p(\underline{x},t)=\overline{p}(\underline{x},t)+p''(\underline{x},t). &
\end{align}
The overbar donates the resolved, large components (grid scales (GS)) and the double prime the unresolved part (sub-grid scales (SGS)).\\
To be able to extract the low frequency components of the quantity, the filtering operation is defined as
\begin{equation}
\overline{\phi}(\underline{x},t)=\int_{\Omega}G(\underline{x}-\underline{x}',t-t';\Delta_{\delta})\phi(\underline{x}',t')d\underline{x}'dt',
\label{filteroperator}
\end{equation}
where $G$ is the filter convolution kernel, associated with the cutoff length scale $\Delta_{\delta}$ also called filter width. An important note is that the LES filtering operation does not satisfy the properties of a Reynolds operator as described in Definition~\ref{def3}. A LES filter must satisfy the following set of properties.
\begin{theorem}[\textbf{LES filter operator}]
Let $\phi,\psi \in C(\Omega\times[0,T],\mathbb{R})$, $c\in\mathbb{R}$ and $G$ the LES filter operator as defined in Definition~\ref{filteroperator}, then the following properties have to be satisfied:
\begin{itemize}
\item Linearity:
\begin{equation*}
\overline{\phi+\psi}=\overline{\phi}+\overline{\psi}
\end{equation*}
\item Commutation with derivatives:
\begin{align*}
& \frac{\overline{\partial\phi}}{\partial\underline{x}}=\frac{\partial\overline{\phi}}{\partial\underline{x}} & \frac{\overline{\partial\phi}}{\partial t}=\frac{\partial\overline{\phi}}{\partial t} &
\end{align*}
\item Constants:
\begin{equation*}
\overline{c} = c
\end{equation*}
which implies that,
\begin{equation*}
\int_{\Omega}G(\underline{x}';\Delta_{\delta})d\underline{x}'=1
\end{equation*}
\end{itemize}
Generally, this filter operator does not satisfy the following properties:
\begin{multicols}{2}
\begin{itemize}
\item $\overline{\overline{\phi}}\neq\overline{\phi}$
\item $\overline{\phi ''}\neq 0$
\end{itemize}
\end{multicols}
Meaning that it is not a Reynolds operator.
\end{theorem}
\begin{figure}
     \centering
     \begin{subfigure}[b]{0.7\textwidth}
         \centering
         \includegraphics[width=\textwidth]{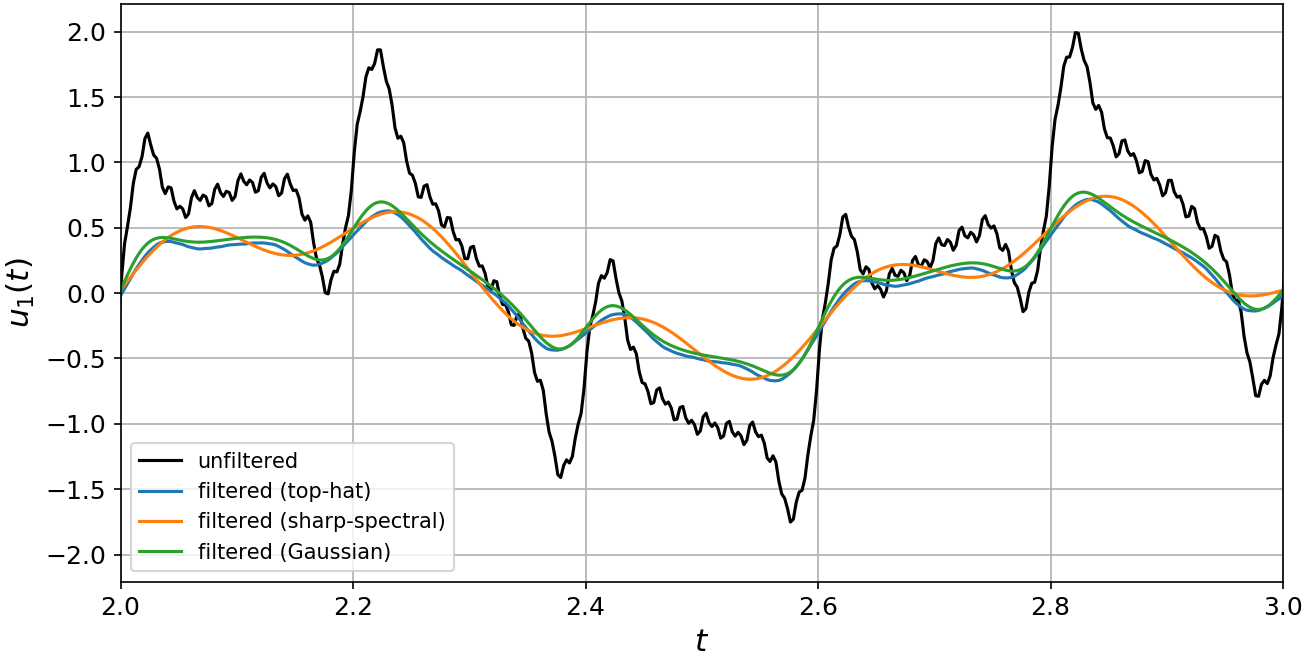}
         \caption{}
         \label{fig:filtered_real}
     \end{subfigure}
     \begin{subfigure}[b]{0.7\textwidth}
         \centering
         \includegraphics[width=\textwidth]{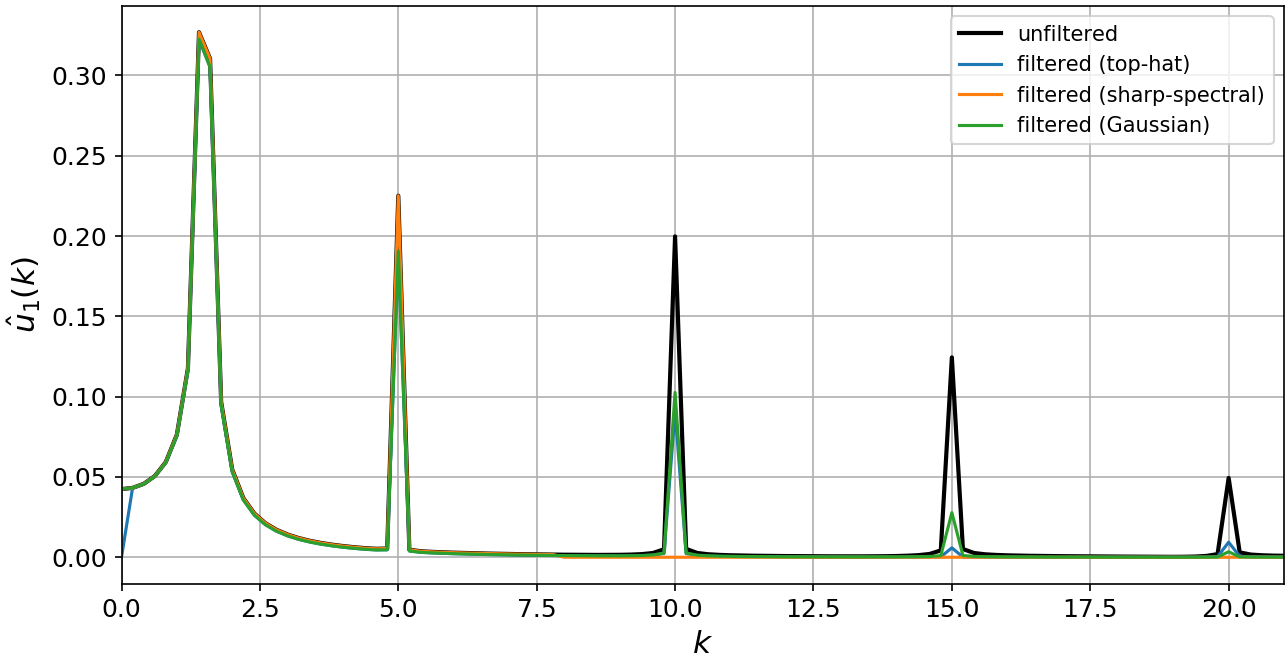}
         \caption{}
         \label{fig:filtered_spec}
     \end{subfigure}
     \caption{Comparison unfiltered and filtered (top-hat, sharp spectral and Gaussian filter kernel) velocity signal with filter width $\Delta_{\delta}=\frac{4}{10}$. (a): Time domain. (b): Frequency domain.}
     \label{fig:filtered}
\end{figure}
The Fourier transform of the filter operator is defined as
\begin{equation}
\hat{G}(\underline{k};\Delta_{\delta})=\int_{-\infty}^{\infty} G(\underline{x};\Delta_{\delta})e^{-j\underline{k}\cdot\underline{x}}\,d\underline{x}.
\end{equation}
In frequency domain, the filtering operation is simply obtained by 
\begin{equation}
\hat{\overline{\phi}}(\underline{k})=\hat{G}(\underline{k};\Delta_{\delta})\hat{\phi}(\underline{k}).
\end{equation}
The most common filter kernels that have been applied to LES are the Gaussian filter, sharp spectral filter or top-hat filter. The one-dimensional filter kernels are defined as:
\begin{table}[H]
	\centering
	\begin{tabular}{|c|c|c|}
		\hline
		Filter & Function & Fourier transform \\ \hline
		Gaussian & $G_G(x;\Delta_{\delta})=\sqrt{\frac{6}{\pi\Delta_{\delta}^2}}e^{-6x^2/\Delta_{\delta}^2}$ & $\hat{G}_G(k;\Delta_{\delta})=e^{-k^2\Delta_{\delta}^2/24}$ \\
		Sharp spectral & $G_S(x;\Delta_{\delta})=\frac{\sin\frac{x\pi}{\Delta_{\delta}}}{x\pi}$ & $\hat{G}_S(k;\Delta_{\delta})=\begin{cases}
    1,& k \leq \frac{\pi}{\Delta_{\delta}}\\
    0, & \mathrm{else}
\end{cases}$   \\
		Top-hat &  $G_T(x;\Delta_{\delta})=\begin{cases}
    \frac{1}{\Delta_{\delta}},& -\frac{\Delta_{\delta}}{2}\leq x \leq \frac{\Delta_{\delta}}{2}\\
    0, & \mathrm{else}
\end{cases}$ & $\hat{G}_T(k;\Delta_{\delta})=\frac{\sin\frac{k\Delta_{\delta}}{2}}{\frac{k\Delta_{\delta}}{2}}$   \\
		\hline
	\end{tabular}
\end{table}
If this filtering operator is applied to the Navier-Stokes equations of the form of Equation~\eqref{NSE_C} and \eqref{NSE_M} (again neglecting the volume force term), we obtain the spatial-filtered Navier-Stokes equations (SFNS)
\begin{align}
& \nabla\,\cdot\,\overline{u}=0, & \\
& \frac{\partial\overline{\underline{u}}}{\partial t} + \nabla\,\cdot\,(\overline{\underline{u}\otimes\underline{u}})
	= -\frac{1}{\rho}\nabla \overline{p} + \nabla\,\cdot\,\nu(\nabla\overline{\underline{u}}+\nabla\overline{\underline{u}}^T). \label{filtered_M}&
\end{align}
The effect of the spatial filtering can be also seen in a damped energy spectrum in Figure~\ref{fig:filtered_spectrum}.
\begin{figure}[ht]
    \centering
    \includegraphics[width=0.65\textwidth]{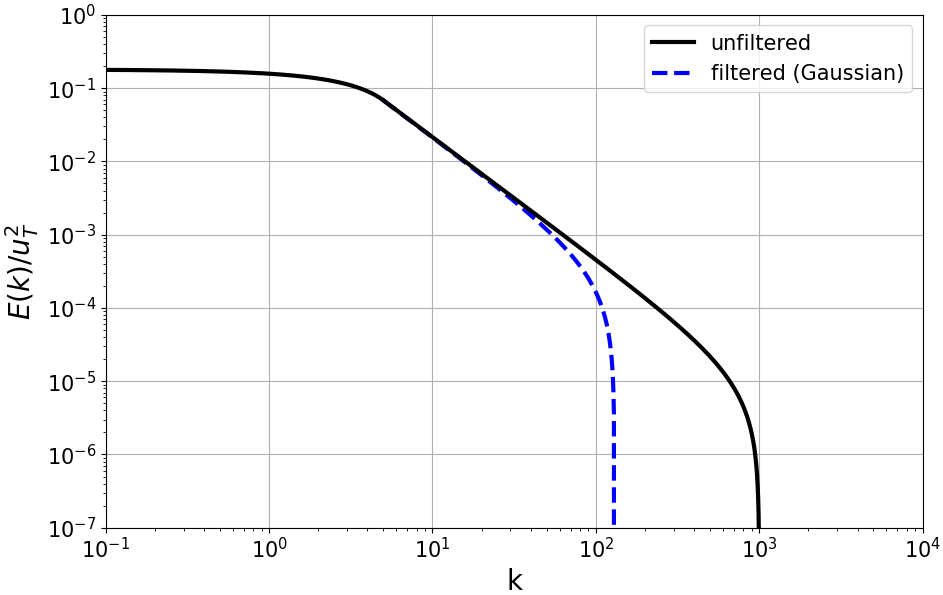}
    \caption{Comparison unfiltered and filtered energy spectrum.}
    \label{fig:filtered_spectrum}
\end{figure}
\\[1ex]
Although the definition of the quantities differs from that in the RANS equation, the closure problem is conceptually very similar. Since $\overline{\underline{u}\otimes\underline{u}}\neq\overline{\underline{u}}\otimes\overline{\underline{u}}$, a model approximation has to be taken into account. The common way to address this issue of closure is to introduce the so called SGS stress tensor,
\begin{equation}
\doubleunderline{T} =\overline{\underline{u}\otimes\underline{u}}-\overline{\underline{u}}\otimes\overline{\underline{u}}.
\end{equation}
The symmetric tensor $\doubleunderline{T}$ has to have the property that $|\doubleunderline{T}|\to0$ as $\Delta_{\delta} \to 0$, so that in the limit of mesh spacing the DNS solution is recovered. The SGS stress tensor is functionally very similar to the RST, but the physics of the problem is somewhat different.\\
Inserting $\doubleunderline{T}$ in Equation~\eqref{filtered_M}, it follows
\begin{equation}
\frac{\partial\overline{\underline{u}}}{\partial t} + (\overline{\underline{u}}\,\cdot\,\nabla)\overline{\underline{u}}
	= -\frac{1}{\rho}\nabla \overline{p} + \nabla\,\cdot\,\nu(\nabla\overline{\underline{u}}+\nabla\overline{\underline{u}}^T)-\nabla\,\cdot\,\doubleunderline{T}. \label{filtered_M_T}
\end{equation}
The filtered kinetic energy $\overline{E}$ can be decomposed into
\begin{align*}
\overline{E}&=\frac{1}{2}\overline{\underline{u}\,\cdot\,\underline{u}}\\
&=\frac{1}{2}(\overline{\underline{u}}\,\cdot\,\overline{\underline{u}}+\overline{\underline{u}\,\cdot\,\underline{u}}-\overline{\underline{u}}\,\cdot\,\overline{\underline{u}})\\
&=E_{GS}+\frac{1}{2}tr(\doubleunderline{T}),
\end{align*}
while $E_{GS}$ is the kinetic energy of the resolved filtered scales and $\frac{1}{2}tr(\doubleunderline{T})$ the SGS energy. The conservation equation for $E_{GS}$ can be obtained by multiplying the filtered momentum transport equation (Equation~\eqref{filtered_M_T}) by the filtered velocity $\overline{\underline{u}}$ to yield
\begin{equation*}
\overline{\underline{u}}\,\cdot\,\bigg(\frac{\partial\overline{\underline{u}}}{\partial t} + (\overline{\underline{u}}\,\cdot\,\nabla)\overline{\underline{u}}\bigg)
	= \overline{\underline{u}}\,\cdot\,\bigg(-\frac{1}{\rho}\nabla \overline{p} + \nu\Delta\overline{\underline{u}}-\nabla\,\cdot\,\doubleunderline{T}\bigg).
\end{equation*}
The derivation is quite similar to the one in Chapter~\eqref{sec:TKE}. Further rearrangements lead to
\begin{align*}
\overline{\underline{u}}\,\cdot\,(\nu\Delta\overline{\underline{u}})&=\frac{1}{2}\nu\Delta(\overline{\underline{u}}\,\cdot\,\overline{\underline{u}})-\nu\nabla\overline{\underline{u}}:\nabla\overline{\underline{u}}\\
&=\nu\Delta E_{GS}-\epsilon_{GS},
\end{align*}
and
\begin{align*}
\overline{\underline{u}}\,\cdot\,(\nabla\,\cdot\,\doubleunderline{T})&=\nabla\,\cdot\,(\doubleunderline{T}\,\overline{\underline{u}})-\doubleunderline{T}:\nabla\overline{\underline{u}}.
\end{align*}
The final transport equation is then
\begin{equation}
\frac{\partial E_{GS}}{\partial t} + (\overline{\underline{u}}\,\cdot\,\nabla)E_{GS} = \nabla\,\cdot\,\big(-\frac{1}{\rho}\overline{p}\,\overline{\underline{u}}-\doubleunderline{T}\,\overline{\underline{u}}\big)+\nu\Delta E_{GS}-\epsilon_{GS}+\doubleunderline{T}:\nabla\overline{\underline{u}}.
\label{EGS}
\end{equation}
The pyhisical interpretation of the first term on the right-handside of Equation~\eqref{EGS} is the redistribution term and the second one the viscous diffusion. The term $\epsilon_{GS}>0$ is the viscous dissipation and always results in a reduction of the GS energy. The very last term of Equation~\eqref{EGS} represents the SGS dissipation
\begin{equation*}
\epsilon_{SGS}=\doubleunderline{T}:\nabla\overline{\underline{u}}.
\end{equation*}
The SGS dissipation may be positive or negative, meaning if it is negative, energy dissipates from the resolved scales to the sub-grid scales, which is called forwardscatter. However, if it is positive, energy transfers from the sub-grid scales to the resolved ones, so called backscatter.\\
A much smaller part of the turbulent energy spectrum is covered by the SGS energy than the filtered kinetic energy, meaning that accuracy of the SGS model may be less crucial than in RANS.\\
In the following, a method to successfully model the SGS stress tensor is shown.

\subsection{Sub-grid scale modelling}
The introduced model in this subsection have many parallels with the RANS counterparts. Nevertheless, the fact that a much smaller part of the turbulent energy spectrum has to be modeled, contributes to a smaller error potential and simple models may produce good results.\\
In LES, the dissipative scales are generally not resolved, therefore the main role of the SGS model is to extract energy from the resolved scales. This can be accomplished with an eddy-viscosity model similar to the RANS model. To this end we assume
\begin{equation}
\doubleunderline{T}-\frac{1}{3}tr(\doubleunderline{T})I=-\nu_{SGS}(\nabla\overline{\underline{u}}+\nabla\overline{\underline{u}}^T)=-2\nu_{SGS}\doubleunderline{\overline{S}},
\end{equation}
where $\doubleunderline{\overline{S}}$ is the filtered strain rate tensor
\begin{equation}
\doubleunderline{\overline{S}}=\frac{1}{2}(\nabla\overline{\underline{u}}+\nabla\overline{\underline{u}}^T).
\end{equation}
The isotropic part of the SGS stress tensor is incoporated by the filtered pressure. The SFNS problem then reads in the following:
\begin{problem}[\textbf{Spatial-filtered Navier-Stokes equations}]
Let $\nu$, $\rho$ $\in \mathbb{R}\setminus \{ 0 \}$, $\nu_{SGS}\in C^1(\Omega \times [0,T],\mathbb{R})$, $\overline{\underline{u}}_D\in C^0(\partial\Omega_{D},\mathbb{R}^{3})$, $\underline{g}\in C^0(\partial\Omega_{N},\mathbb{R}^{3})$ and $\overline{\underline{u}}_0\in C^{2}(\Omega,\mathbb{R}^{3})$ find $\overline{\underline{u}}\in C^2(\Omega \times [0,T],\mathbb{R}^3)$ and $\hat{p}\in C^1(\Omega \times [0,T],\mathbb{R})$ such that
\begin{align*}
	& \nabla\,\cdot\,\overline{\underline{u}}=0 \numberthis & \textnormal{in}\:\Omega\times[0,T], & \\
	& \frac{\partial\overline{\underline{u}}}{\partial t} + (\overline{\underline{u}}\,\cdot\,\nabla)\overline{\underline{u}}
	= -\frac{1}{\rho}\nabla \hat{p} +\nabla\,\cdot\,\big((\nu+\nu_{SGS})(\nabla\overline{\underline{u}}+\nabla\overline{\underline{u}}^T)\big)  \numberthis & \textnormal{in}\:\Omega\times[0,T],  & \\
	& \overline{\underline{u}}=\overline{\underline{u}}_0 & \textnormal{in}\:\Omega, t=0, & \\
	& \overline{\underline{u}}=\overline{\underline{u}}_D & \textnormal{in}\:\partial\Omega_{D}\times[0,T], & \\
	& \frac{\partial\overline{\underline{u}}}{\partial\underline{n}}=\underline{g} & \textnormal{in}\:\partial\Omega_{N}\times[0,T].
\end{align*}
\label{SFNS}
\end{problem}
Two fundamental topics in LES have to be mentioned.\\
Firstly, there are mainly three different approaches, implicit, implicitly filtered and explicitly filtered LES. The common method is implicit LES, where the system of equations are never acted upon a filtering. The filtering is provided implicitly by two causes, the computational grid and the discretization used. Traditionally, the filtering is done by the grid itself and the inherent numerical diffusion acts implicitly as sub-grid scale model. It is also known as quasi or coarse DNS. In implicitly filtered LES, the filtered Navier-Stokes equations given in Problem~\ref{SFNS} with an appropriate SGS model are solved numerically. Formally, the governing system of equations do not differ from the unsteady RANS equations. Nonetheless, an explicit filter can still be applied in order to derive variables used in the SGS model. In explicitly filtered LES, the application of an numerical filter on the equations is performed at each time step. Therefore, it is  possible to control the shape and type of the filter. In this thesis, we investigate the first two procedures.\\
Secondly, the derivation of the SFNS equations takes the advantage that the filter operator commutes with differentiation, which holds in absence of boundaries. Briefly, in presence of boundaries, it does not commute and leads to the so called commutation error. In scenarios with periodic boundary conditions and nearly homogeneous turbulence, this error is negligible. The numerical analysis of the commutation error in LES is given in \cite{LESmathematics}.
\\[1ex]
A rich variety of SGS models has been developed. The first proposed eddy-viscosity model is the Smagorinsky model \cite{LESJohn}. Via dimensional analysis, it follows for the dissipation rate that
\begin{equation*}
\epsilon\sim\frac{U_I^3}{L_I}.
\end{equation*}
The same relation also holds for the filter width $\Delta_{\delta}$ and the corresponding characteristic velocity of the unresolved scales $U_{\delta}$
\begin{equation*}
\epsilon\sim\frac{U_{\delta}^3}{\Delta_{\delta}}.
\end{equation*}
It follows that
\begin{equation*}
\nu_{SGS}\sim U_{\delta}\Delta_{\delta}\sim U_IL_I^{-1/3}\Delta_{\delta}^{4/3}.
\end{equation*}
The mixing length assumption is then
\begin{equation*}
U_I\sim L_I\overline{S},
\end{equation*}
where $\overline{S}$ is the magnitude of the filtered strain rate tensor.\\
One gets by replacing the mixing length assumption into the viscosity relation
\begin{equation*}
\nu_{SGS}=CL_I^{2/3}\Delta_{\delta}^{4/3}\overline{S}.
\end{equation*}
The integral length scale is approximated by $L_I\sim \Delta_{\delta}$.\\
Then, the artificial viscosity is defined as
\begin{equation}
\nu_{SGS}=(C_S\Delta_{\delta})^2\overline{S},
\label{Smagorinsky}
\end{equation}
where $C_S$ is the Smagorinsky coefficient. The filter width is thereby computed via the geometrical mean
\begin{equation}
\Delta_{\delta}=\sqrt[3]{\Delta_{x1}\Delta_{x2}\Delta_{x3}},
\end{equation}
where $\Delta_{xi}\,,\,i=1,2,3$ is the mesh cell size of the corresponding spatial dimension.\\
This model assumes that the unresolved scales dissipate entirely and instantaneously all the energy received from the larger scales, therefore it prevents backscatter. Despite that the Smagorinsky model has some drawbacks, it is widely used and dissipates energy at probably the right average rate.
Generally, it has to be mentioned that the SGS models have to satisfy some properties (e.g. realizability, reversibility).\\
In the existence of boundary layers, the Smagorinsky model predicts large amounts of dissipation. This often prevents the formation of eddies and coherent structures and may eliminate any turbulence. To avoid excessive dissipation, a damping function is used to reduce the Smagorinsky constant $C_S\rightarrow 0$ as the boundary is approached. It is called Van Driest scaling \cite{pope_2000} and reads,
\begin{equation}
C_S=C_S(y^+)=C_S(1-e^{-y^+/30}).
\end{equation}
It improves the performance of the model in situations with simple geometries, where more or less the boundary layer theory holds.\\
Throughout all LES computations performed in this thesis, the Smagorinsky model with Van Driest damping was used.

\section{Variational multiscale}
Variational Multiscale (VMS) approach is a comparatively new method to simulate incompressible turbulent flow. The fundamental idea is based on scale separation similar to LES, but referring to the variational framework of the underlying equations. VMS concepts for LES were primarily introduced by Hughes \cite{VMShughes}. Instead of using the filtered governing equations, the weak form of the equations and variational projection into subspaces is the basic concept behind VMS. Due to the variational formulation, the finite element method framework may be preferable, although it is also suitable for other discretization techniques. Nowadays, many different classes of VMS methods exist, in this thesis the focus will be on the projection-based VMS method by \cite{VMSjohn}.
\\[1ex]
The first step is the variational formulation of the Navier-Stokes equations of form of Equation~\eqref{NSE_C} and \eqref{NSE_M}. For the case, we only consider homogeneous Dirichlet and homogeneous Neumann boundary conditions in the following. For this let $\partial\Omega=\partial\Omega_D\cup\partial\Omega_N$ with $|\partial\Omega_N| > 0$. As we have a system of partial differential equations for the velocity and pressure, we define two spaces for our solutions and test functions
\begin{align}
& V=H_{0,\partial\Omega_D}^1(\Omega,\mathbb{R}^3)=\{\underline{v}\in H^1(\Omega,\mathbb{R}^3):\underline{v}=\underline{0}\;\textrm{on}\;\partial\Omega_D\}. \label{H1}& \\ & Q=L^2(\Omega,\mathbb{R}). \label{L2} &
\end{align}
We multiply Equation~\eqref{NSE_M} with test function $\underline{v}\in V$ and Equation \eqref{NSE_C} with $q\in Q$ respectively, integrate over the whole domain $\Omega$ and integrate by parts, to get
\begin{align*}
& \int_{\Omega}\frac{\partial\underline{u}}{\partial t}\,\cdot\,\underline{v}\;d\underline{x}+\int_{\Omega}2\nu\doubleunderline{S}(\underline{u}):\doubleunderline{S}(\underline{v})\;d\underline{x}-\int_{\Omega}(\nabla\,\cdot\,\underline{v})p\;d\underline{x} + \int_{\Omega}(\underline{u}\,\cdot\,\nabla)\underline{u}\,\cdot\,\underline{v}\;d\underline{x} = \\ & \int_{\Omega}\underline{f}\,\cdot\,\underline{v}\;d\underline{x} & \forall \underline{v}\in V,& \\
& -\int_{\Omega}(\nabla\,\cdot\,\underline{u})q\;d\underline{x} & \forall q\in Q,&
\end{align*}
$\doubleunderline{S}(\underline{u})$ is the strain rate tensor
\begin{equation*}
\doubleunderline{S}(\underline{u})=\frac{1}{2}(\nabla\underline{u}+\nabla\underline{u}^T).
\end{equation*}
We consequently define the bilinear forms $a(\cdot,\cdot):V\times V\to\mathbb{R}$, $b(\cdot,\cdot):V\times Q\to\mathbb{R}$, the trilinear form $c(\cdot,\cdot,\cdot):V\times V\times V\to\mathbb{R}$ and linear form $f(\cdot):V\to\mathbb{R}$:
\begin{align}
& a(\underline{u},\underline{v})=\int_{\Omega}2\nu\doubleunderline{S}(\underline{u}):\doubleunderline{S}(\underline{v})\;d\underline{x}, & \\
& b(\underline{u},q)=-\int_{\Omega}(\nabla\,\cdot\,\underline{u})q\;d\underline{x}, & \\
& c(\underline{u},\underline{u},\underline{v})=\int_{\Omega}(\underline{u}\,\cdot\,\nabla)\underline{u}\,\cdot\,\underline{v}\;d\underline{x}, & \\
& f(\underline{v})=\int_{\Omega}\underline{f}\,\cdot\,\underline{v}\;d\underline{x}. &
\end{align}
Generally, the $L^2$ inner product is defined as
\begin{equation}
(\phi,\psi)=\int_{\Omega}\phi\cdot\psi\;d\underline{x} \;\;\;\; \forall \phi,\psi\in L^2.
\end{equation}
\\[1ex]
The variational problem of the Navier-Stokes equations then reads:
\begin{problem}[\textbf{Weak formulation of the Navier-Stokes equations}]
Find $\underline{u}\in V$ and $p\in Q$ satisfying
\begin{align}
&(\frac{\partial\underline{u}}{\partial t},\underline{v})+a(\underline{u},\underline{v})+b(\underline{v},p)+c(\underline{u},\underline{u},\underline{v})=f(\underline{v}) & & \forall \underline{v}\in V, & \\
& b(\underline{u},q)=0 & & \forall q\in Q. &
\end{align}
\label{weakformNS}
\end{problem}
The equations from Problem~\ref{weakformNS} may also be written in short form as,
\begin{align*}
& A((\underline{u},p),(\underline{v},q))& \\ & =(\frac{\partial\underline{u}}{\partial t},\underline{v})+a(\underline{u},\underline{v})+b(\underline{v},p)+c(\underline{u},\underline{u},\underline{v})+b(\underline{u},q) \label{bigWF} \numberthis & &\forall (\underline{v},q)\in V\times Q.&
\end{align*}
Systems of this form are called saddle-point problem and its analysis is seen in the next chapter.\\
For VMS methods, the corresponding trial and test spaces are decomposed into three parts, large scales, small scales and unresolved scales
\begin{align*}
&V=V_L\oplus V_S\oplus V_U,& & Q=Q_L\oplus Q_S\oplus Q_U. &
\end{align*}
For the solution and test functions, we obtain
\begin{align*}
&\underline{u}=\underline{u}_L+\underline{u}_S+\underline{u}_U,& &p=p_L+p_S+p_U,& \\
&\underline{v}=\underline{v}_L+\underline{v}_S+\underline{v}_U,& &q=q_L+q_S+q_U.&
\end{align*}
Inserting the scale seperation into Equation~\eqref{bigWF}, it may be written as a system of three variational equations
\begin{align}
& A((\underline{u}_L,p_L),(\underline{v}_L,q_L))+A((\underline{u}_S,p_S),(\underline{v}_L,q_L))+A((\underline{u}_U,p_U),(\underline{v}_L,q_L))=f(\underline{v}_L), & \\
& A((\underline{u}_L,p_L),(\underline{v}_S,q_S))+A((\underline{u}_S,p_S),(\underline{v}_S,q_S))+A((\underline{u}_U,p_U),(\underline{v}_S,q_S))=f(\underline{v}_S), & \\
& A((\underline{u}_L,p_L),(\underline{v}_U,q_U))+A((\underline{u}_S,p_S),(\underline{v}_U,q_U))+A((\underline{u}_U,p_U),(\underline{v}_U,q_U))=f(\underline{v}_U). \label{unresolved}&
\end{align}
As it is not intended to explicitly solve the unresolved scales, Equation~\eqref{unresolved} is neglected.\\
Another assumption is that the unresolved scales do not influence the large scales directly, therefore
\begin{equation*}
A((\underline{u}_U,p_U),(\underline{v}_L,q_L))=0.
\end{equation*}
The influence of the unresolved scales onto the resolved has to be modeled. A widely used way is to use the eddy-viscosity model as previously mentioned in Equation~\eqref{Smagorinsky}
\begin{equation*}
A((\underline{u}_U,p_U),(\underline{v}_S,q_S))\approx (2\nu_U\doubleunderline{S}(\underline{u}_S),\doubleunderline{S}(\underline{v}_S)).
\end{equation*}
As a result of neglecting the unresolved scales, we get the new system of equations.
\begin{problem}[\textbf{Three-scale VMS formulation of the Navier-Stokes equations}]
Find $\underline{u}_L+\underline{u}_S\in V_L\oplus V_S$ and $p_L+p_S\in Q_L\oplus Q_S$ such that
\begin{align}
& A((\underline{u}_L,p_L),(\underline{v}_L,q_L))+A((\underline{u}_S,p_S),(\underline{v}_L,q_L))=f(\underline{v}_L), & \\
& A((\underline{u}_L,p_L),(\underline{v}_S,q_S))+A((\underline{u}_S,p_S),(\underline{v}_S,q_S))+(2\nu_U\doubleunderline{S}(\underline{u}_S),\doubleunderline{S}(\underline{v}_S))=f(\underline{v}_S), &
\end{align}
\label{VMSthreescale}
for all $\underline{v}_L+\underline{v}_S\in V_L\oplus V_S$ and $q_L+q_S\in Q_L\oplus Q_S$.
\end{problem}
A crucial point of VMS methods is the definition of the appropriate spaces for the large scales and small scales . The strategy used in this thesis is a coarse space projection-based method.
\\[1ex]
Consider a standard pair of conforming finite element spaces $V^h\times Q^h \subset V\times Q$ for all scales of velocity and pressure which fulfills several conditions for saddle-point problems (discussed in detail in the next chapter). In addition, let $L$ be a finite element space of symmetric $d\times d$ tensor-valued functions
\begin{equation}
L^h\subset L=\{\doubleunderline{l}\in L^2(\Omega,\mathbb{R}^{d\times d}):\doubleunderline{l}=\doubleunderline{l}^T\}.
\end{equation}
Let $V_L^h\in H^1(\Omega,\mathbb{R}^3)$ be the discrete space for the large scales such that the condition $L^h=\{\doubleunderline{S}(\underline{v}^h_L):\underline{v}^h_L\in V^h_L\}\subseteq\{\doubleunderline{S}(\underline{v}^h):\underline{v}^h\in V^h\}$ holds. There are two possibilities of choosing the coarse finite element space $V_L^h$. On the one hand, if for $V^h$ a higher order finite element space is chosen, one may take a lower order finite element space for $V_L^h$ on the same grid. This approach is the one-level projection-based VMS method. On the other hand, in the case of the same order for the resolved and large scales spaces, $V_L^h$ may be defined on a coarser grid, which is called the two-level projection-based method. Due to reasons of simplicity, we stick to the first mentioned approach in this thesis.\\
However, the discrete space $V_L^h$ does not incoporate boundary conditions, thus generally speaking $V_L^h$ is no subset of $V^h$.
\\[1ex]
Define the projection operator $\Pi_V:V^h\to V_L^h$ such that,
\begin{align}
& \big(\doubleunderline{S}(\underline{u}^h-\Pi_V\underline{u}^h),\doubleunderline{S}(\underline{v^h_L})\big)=0 & &\forall \underline{v}^h_L\in V_L^h.&
\label{Pi_V}
\end{align}
In addition, let $\Pi_L:L\to L^h$ be the $L^2$-projection from $L$ to $L^h$ respectively,
\begin{align}
& \big(\doubleunderline{S}(\underline{u}^h)-\Pi_L\doubleunderline{S}(\underline{u}^h),\doubleunderline{l}^h\big)=0 & &\forall \doubleunderline{l}^h\in L^h. &
\label{Pi_L}
\end{align}
One important aspect is that the strain rate tensor of the large scales defined in Equation~\eqref{Pi_V} equals the large scales of the strain rate tensor defined in Equation~\eqref{Pi_L}. It follows that the definition by projection of the large scales and differentiation commutes. As mentioned previously, this does not hold in classic LES generally.
\begin{lemma}
Let $\underline{v}^h\in V^h$ and $L^h=\{\doubleunderline{S}(\underline{v}^h_L):\underline{v}^h_L\in V^h_L\}$ then it holds
\begin{equation}
\Pi_L\doubleunderline{S}(\underline{v}^h)=\doubleunderline{S}(\Pi_V\underline{v}^h).
\end{equation}
\end{lemma}
A simple proof is given in \cite{VMSjohn}.
\\[1ex]
Taking the system of equations of Problem~\ref{VMSthreescale}, reunite the decomposition $V^h=V_L^h\oplus V_S^h$ and $Q^h=Q_L^h\oplus Q_S^h$ with the small scale part defined by the projection $V_S^h=(I-\Pi_V)V^h$, we obtain
\begin{align*}
& A((\underline{u}^h,p^h),(\underline{v}^h,q^h))+(2\nu_U\doubleunderline{S}(\underline{u}_S^h),\doubleunderline{S}(\underline{v}_S^h))=f(\underline{v}^h) & &\forall (\underline{v},q)\in V^h\times Q^h.&
\end{align*}
The modeled term may be rewritten as
\begin{align*}
\big(2\nu_U\doubleunderline{S}(\underline{u}_S^h),\doubleunderline{S}(\underline{v}_S^h)\big) & = \big(2\nu_U\doubleunderline{S}\big((I-\Pi_V)\underline{u}^h\big),\doubleunderline{S}\big((I-\Pi_V)\underline{v}^h\big)\big) \\
& = \big(2\nu_U(I-\Pi_L)\doubleunderline{S}(\underline{u}^h),(I-\Pi_L)\doubleunderline{S}(\underline{v}^h)\big) \\
& = \big(2\nu_U\doubleunderline{S}(\underline{u}^h),\doubleunderline{S}(\underline{v}^h)\big)-\big(2\nu_U\Pi_L\doubleunderline{S}(\underline{u}^h),\doubleunderline{S}(\underline{v}^h)\big).
\end{align*}
Let $\doubleunderline{g}^h\in L^h$ such that $\doubleunderline{g}^h=\Pi_L\doubleunderline{S}(\underline{u}^h)$, we finally obtain the projection-based VMS method.
\begin{problem}[\textbf{Projection-based VMS formulation of the Navier-Stokes equations}]
Let $(\underline{u}^h,p^h,\doubleunderline{g}^h)\in V^h\times Q^h\times L^h$, such that
\begin{align}
& (\frac{\partial\underline{u}^h}{\partial t},\underline{v}^h)+\big(2(\nu+\nu_U)\doubleunderline{S}(\underline{u}^h),\doubleunderline{S}(\underline{v}^h)\big)& \\ & +c(\underline{u}^h,\underline{u}^h,\underline{v}^h)+b(\underline{v}^h,p^h)-\big(2\nu_U\doubleunderline{g}^h,\doubleunderline{S}(\underline{v}^h)\big)=f(\underline{v}^h) & & \forall \underline{v}^h\in V^h,& \\
& b(\underline{u}^h,q^h)=0 & &\forall q\in Q^h, & \\
& \big(\doubleunderline{g}^h-\doubleunderline{S}(\underline{u}^h),\doubleunderline{l}^h\big) & & \forall \doubleunderline{l}^h\in L^h. &
\end{align}
\label{VMS}
\end{problem}
Refering to the numerical analysis of Problem~\ref{VMS} in \cite{VMSahmed}. The principal way of performing the analysis is the same as for the Galerkin discretization of the Navier-Stokes equations.\\
In order to obtain an efficient implementation for solving Problem~\ref{VMS}, the space $L^h$ has to be a discontinuous finite element space with a $L^2$-orthogonal basis. This ensures that the mass matrix is a diagonal matrix and its inverse may be computed easily. Using a discontinuous space for $L^h$ makes also sense from the point of view that the functions of $L^h$ are $L^2$-projections of strain rate tensor of finite element functions, which are usually discontinuous functions too.\\
All in all, the combination of choosing the space $L^h$ and the eddy-viscosity model results in the projection-based VMS method. Nevertheless, this VMS method is less sensitive to choice of the eddy-viscosity model than in traditional LES. This is due to the fact that the modeled scales influence much less scales directly in VMS than in LES. On the one hand, if $L^h=\{0\}$ the eddy-viscosity model influences all scales, recovering the classic LES model. On the other hand, if $L^h=\{\doubleunderline{S}(\underline{v}^h):\underline{v}^h\in V^h\}$ the modeled viscosity is switched off and the Navier-Stokes equations are reobtained. Therefore, a low-order space of $L^h$ means that the turbulence model has a larger influence and for a higher order space of $L^h$ the model has less influence.\\
The local turbulence intensity may be estimated with the size of the local small resolved scales,
\begin{equation}
\eta_T=\parallel \doubleunderline{g}^h-\doubleunderline{S}(\underline{u}^h)\parallel_{L^2(T)},
\end{equation}
where $\mathcal{T}$ is a triangulation of a domain $\Omega$ and $T\in\mathcal{T}$. If the size of the small scales is large, many unresolved scales can be expected and vice versa. A method choosing the adaptive projection space may be found in \cite{VMSadaptive}. In this thesis, we stick to the conventional non-adaptive version of the three-scale projection-based VMS method.
}

%% file: chapters/discretization.tex
\chapter{Continuous Galerkin method for the Navier-Stokes equations}
\label{chapter:discretization}
%%%%%%%%%%%%%%%%%%%%%%%%%%%%%%%%%%%%%%%%%%%%%%%%%%%%%%%%%%%%%%%%%%%%%%%%%%%%%%%%%%%%%%%%%%%%%%%%%%%%%%%%%%%%%%
%%%%%%%%%%%%%%%%%%%%%%%%%%%%%%%%%%%%%%%%%%%%%%%%%%%%%%%%%%%%%%%%%%%%%%%%%%%%%%%%%%%%%%%%%%%%%%%%%%%%%%%%%%%%%%
{
In this chapter, we present a continuous Galerkin discretization technique for the unsteady Navier-Stokes equations. Firstly, we consider the discretization of the steady Stokes problem and discuss the characteristic conditions to obtain a stable method. Further on, a time discretization of the governing equations of motion is introduced.\\
For ease of simplicity and practical reasons, inhomogeneous Dirichlet and homogeneous Neumann boundary conditions are assumed in all derivations.

\section{Weak formulation}
At first, we consider the weak formulation of Stokes problem and the unsteady Navier-Stokes problem (as already derived in the previous chapter in Problem~\ref{weakformNS}). The appropriate spaces for the velocity and pressure are of the form of Equation~\eqref{H1} and \eqref{L2}. Additionally, the Dirichlet boundary condition is incoporated in the solution space $V_D$ as defined
\begin{equation*}
V_D=\{\underline{v}\in H^1(\Omega,\mathbb{R}^3):\underline{v}=\underline{v}_D\;\textrm{on}\;\partial\Omega_D\}.
\end{equation*}
We obtain the weak formulation of the steady Stokes equations by neglecting the time derivative and convective term of the Navier-Stokes equations.
\begin{problem}[\textbf{Weak formulation of the steady Stokes equations}]
Find $\underline{u}\in V_D$ and $p\in Q$ satisfying
\begin{align}
&a(\underline{u},\underline{v})+b(\underline{v},p)=f(\underline{v}) & & \forall \underline{v}\in V, & \\
& b(\underline{u},q)=0 & & \forall q\in Q. &
\end{align}
\label{weakformS}
\end{problem}
Systems of the form seen in Problem~\ref{weakformS} are called saddle-point problems, since the solution $\underline{u},p\in V\times Q$ is also a minimizer of
\begin{equation*}
J(\underline{u})=a(\underline{u},\underline{u})+f(\underline{u})\to min,
\end{equation*}
subject to the constraint
\begin{equation*}
b(\underline{u},q)=0 \;\;\; \forall q\in Q.
\end{equation*}
As for optimization problems with constraints, we apply a Lagrangian
\begin{equation*}
L(\underline{u},q)=J(\underline{u})+b(\underline{u},q).
\end{equation*}
Then we have that each solution $\underline{u},p\in V\times Q$ is a saddle-point of the Lagrangian and it holds
\begin{equation*}
L(\underline{u},q)\leq L(\underline{u},p)\leq L(\underline{v},p) \;\;\; \forall (\underline{v},q)\in V\times Q.
\end{equation*}
The pressure $p\in Q$ may be interpret as Lagrangian multiplier associated with the incompressibility constraint $\nabla\,\cdot\,\underline{u}=0$.
\\[1ex]
In order to prove the existence and uniqueness for elliptic partial differential equations, we could show the continuity and coercivity of the bilinear form and the lemma of Lax-Milgram will guarantee the unique solvability. Unfortunately, this holds not for saddle-point problems and an additional condition has to be fullfilled. Brezzi's theorem for mixed methods is used therefore.\\
A general mixed variational form involves the two bilinear forms $a(\cdot,\cdot):V\times V\to \mathbb{R}$ and $b(\cdot,\cdot):V\times Q\to \mathbb{R}$ and two linear forms $f(\cdot):V\to\mathbb{R}$ and $g(\cdot):Q\to\mathbb{R}$. Thus, with these forms we define the following mixed problem
\begin{align}
&a(\underline{u},\underline{v})+b(\underline{v},p)=f(\underline{v}) & & \forall \underline{v}\in V, \label{mixed_system_a}& \\
& b(\underline{u},q)=g(q) & & \forall q\in Q. &
\label{mixed_system_b}
\end{align}
We define the space $V_0$ of the kernel of the bilinear form $b(\cdot,\cdot)$
\begin{equation}
V_0=\{\underline{v}\in V:b(\underline{v},q)=0 \;\; \forall q\in Q\}.
\end{equation}
\begin{definition}[\textbf{Brezzi’s Theorem}]
Let $a(\cdot,\cdot):V\times V\to \mathbb{R}$ and $b(\cdot,\cdot):V\times Q\to \mathbb{R}$ be two given bilinear forms, that fulfill the conditions:\\
\begin{enumerate}[label=(\roman*)]
\item The bilinear forms are continuous,
\begin{align}
&a(\underline{u},\underline{v})\leq \alpha_1\parallel\underline{u}\parallel_V\parallel\underline{v}\parallel_V \;\;\; \forall\underline{u},\underline{v}\in V,&
\\
&b(\underline{u},q)\leq \beta_1\parallel\underline{u}\parallel_V\parallel q\parallel_Q \;\;\; \forall\underline{u}\in V,q\in Q.&
\end{align}
\item $a(\cdot,\cdot)$ is coercive on the kernel, i.e. there exists an $\alpha_2 >0$ so that,
\begin{equation}
a(\underline{u},\underline{u})\geq \alpha_2\parallel\underline{u}\parallel_V^2 \;\;\; \forall\underline{u}\in V_0.
\end{equation}
\item The Ladyshenskaja-Babuška-Brezzi (LBB) condition of the constraint $b(\cdot,\cdot)$ is fullfilled, i.e. there exists an $\beta_2 >0$ such that,
\begin{equation}
\sup_{\underline{v}\in V}\frac{b(\underline{v},q)}{\parallel\underline{v}\parallel_V}\geq \beta_2\parallel q\parallel_Q \;\;\; \forall q\in Q.
\end{equation}
\end{enumerate}
Then the mixed method from Equation \eqref{mixed_system_a} and \eqref{mixed_system_b} is uniquely solvable and the solution fullfills the stability estimate,
\begin{equation}
\parallel\underline{v}\parallel_V+\parallel p\parallel_Q \leq C(\parallel f\parallel_{V^*}+\parallel g\parallel_{Q^*}).
\end{equation}
\label{Brezzi}
\end{definition}
Here denotes $V^*$ and $Q^*$ the corresponding dual space to $V$ and $Q$. For more details we refer to \cite{MixedFEM}. The proof of Theorem~\ref{Brezzi} is given in \cite{NumPDE} and the analysis of the Stokes problem are shown in \cite{masterthesis_lederer}.
\\[1ex]
The weak formulation of the Navier-Stokes equations is already given in Problem~\ref{weakformNS}.

\section{Approximation of the weak formulation}
In the following, we apply the continuous Galerkin discretization for the mixed problem of form of Equation~\eqref{mixed_system_a} and \eqref{mixed_system_b}. We define the finite-dimensional subspaces $V_h\subset V$, $V_{hD}=\{\underline{v}\in V_h:\underline{v}=\underline{v}_D \: \mathrm{on} \: \partial\Omega_D\}$ and $Q_h\subset Q$. The $h$ refers to a discrete space. The discrete form of the variational problem is:
\begin{problem}[\textbf{Discrete formulation of the steady Stokes equations}]
Find $\underline{u}_h\in V_{hD}$ and $p_h\in Q_h$ satisfying
\begin{align}
&a(\underline{u}_h,\underline{v}_h)+b(\underline{v}_h,p_h)=f(\underline{v}_h) & & \forall \underline{v}_h\in V_h, \label{finite_mixed_system_a}& \\
& b(\underline{u}_h,q_h)=g(q_h) & & \forall q_h\in Q_h. &
\label{finite_mixed_system_b}
\end{align}
\label{discreteS}
\end{problem}
Discrete stability of Problem~\ref{discreteS} is not inherited from the continuous problem. The continuity of the bilinear forms follows from the infinite dimensional case since $V_h\subset V$. The discrete kernel ellipticity is
\begin{equation}
a(\underline{u}_h,\underline{u}_h)\geq \alpha_{2h}\parallel\underline{u}_h\parallel_{V}^2 \;\;\; \forall\underline{u}_h\in V_{h0},
\end{equation}
where $V_{h0}$ is defined as
\begin{equation}
V_{h0}=\{\underline{v}_h\in V_h:b(\underline{v}_h,q_h)=0 \;\; \forall q_h\in Q_h\}.
\end{equation}
The discrete LBB-condition reads
\begin{equation}
\sup_{\underline{v}_h\in V_h}\frac{b(\underline{v}_h,q_h)}{\parallel\underline{v}_h\parallel_{V}}\geq \beta_{2h}\parallel q_h\parallel_{Q} \;\;\; \forall q_h\in Q_h.
\end{equation}
It follows for the Stokes problem the condition
\begin{equation}
\sup_{\underline{v}_h\in V_h}\frac{\int_{\Omega}(\nabla\,\cdot\,\underline{v}_h)q_h\:d\underline{x}}{\parallel\underline{v}_h\parallel_{V}}\geq \beta_{2h}\parallel q_h\parallel_{Q} \;\;\; \forall q_h\in Q_h,
\end{equation}
which is the constraint that arises for the definitions of the finite spaces. Since the LBB-condition for Stokes equations holds in the continuous level, for
a fixed pressure space the velocity space can be enlarged to get discrete LBB-condition. The enlargement can be done by increasing the polynomial order or
refining the mesh. Therefore, the same polynomial degree of order for the velocity and pressure space may lead to an unstable discretization.\\
The discrete formulation of the unsteady Navier-Stokes equations is given below.
\begin{problem}[\textbf{Discrete formulation of the Navier-Stokes equations}]
Find $\underline{u}_h\in V_{hD}$ and $p_h\in Q_h$ such that
\begin{align}
&(\frac{\partial\underline{u}_h}{\partial t},\underline{v}_h)+a(\underline{u}_h,\underline{v}_h)+b(\underline{v}_h,p_h)+c(\underline{u}_h,\underline{u}_h,\underline{v}_h)=f(\underline{v}_h), &  \\
& b(\underline{u}_h,q_h)=0, &
\end{align}
is satisfied for all $\underline{v}_h\in V_{h}$ and $q_h\in Q_h$.
\label{TaylorHood_NS}
\end{problem}

\subsection{Exact divergence-free}
The discretized weak formulation of Problem~\ref{discreteS} can only be solved if the velocity $\underline{u}_h$ fulfills the incompressibility constraint,
\begin{equation*}
\int_{\Omega}(\nabla\,\cdot\,\underline{u}_h)q_h\:d\underline{x}=0 \;\;\; \forall	q_h\in Q_h,
\end{equation*}
which is called a discrete divergence-free property. Nevertheless, this does not generally hold in a strong formalism.\\
Assume we have the property that the space of the divergence of the test functions of the velocity space is a subspace of the pressure space, thus
\begin{equation}
\nabla\,\cdot\,V_h\subset Q_h.
\label{divfree}
\end{equation}
Then a discrete divergence-free velocity is also exactly divergence-free, namely
\begin{equation*}
\int_{\Omega}(\nabla\,\cdot\,\underline{u}_h)q_h\:d\underline{x}=0 \;\;\; \forall	q_h\in Q_h \;\;\;\; \Rightarrow \;\;\;\; \nabla\,\cdot\,\underline{u}_h=0.
\end{equation*}
This property has many advantages, especially $\underline{u}_h$ leads to a better approximation of the velocity and proper physical behavior.\\
From \cite{lehrenfeld} and \cite{masterthesis_lederer}, the impact of exactly divergence-free velocity approximation may be seen in the kinetic energy loss. Consider the momentum equation of the Navier-Stokes equations from Equation~\eqref{NSE_M} with constant density $\rho=1$ and no volume force $\underline{f}$. Due to the friction of the fluid, the kinetic energy should decrease over time. The rate of change of energy with respect to time is then
\begin{equation*}
\frac{dE}{dt}=\int_{\Omega}\frac{1}{2}\frac{\partial(\underline{u}\,\cdot\,\underline{u})}{\partial t}\:d\underline{x}=\int_{\Omega}\underline{u}\,\cdot\,\frac{\partial \underline{u}}{\partial t}\:d\underline{x}=\int_{\Omega}\big(-\nu\nabla\underline{u}:\nabla\underline{u}-\underline{u}\,\cdot\,((\underline{u}\,\cdot\,\nabla)\underline{u})-(\nabla\,\cdot\,\underline{u})p\big)\;d\underline{x}.
\end{equation*}
The convective term can be rewritten as
\begin{equation*}
\int_{\Omega}\underline{u}\,\cdot\,((\underline{u}\,\cdot\,\nabla)\underline{u})\:d\underline{x}=\frac{1}{2}\int_{\Omega}(\underline{u}\,\cdot\,\nabla)(\underline{u}\,\cdot\,\underline{u})\:d\underline{x}=-\frac{1}{2}\int_{\Omega}(\underline{u}\,\cdot\,\underline{u})(\nabla\,\cdot\,\underline{u})\:d\underline{x}.
\end{equation*}
Using the incompressibility constraint, it follows
\begin{equation*}
\frac{dE}{dt}=\int_{\Omega}-\nu\nabla\underline{u}:\nabla\underline{u}\:d\underline{x}\leq 0.
\end{equation*}
This does not automatically hold in the approximate sense
\begin{equation*}
\frac{dE_h}{dt}=\int_{\Omega}-\nu\nabla\underline{u}_h:\nabla\underline{u}_h+\frac{1}{2}(\underline{u}_h\,\cdot\,\underline{u}_h)(\nabla\,\cdot\,\underline{u}_h)\:d\underline{x}.
\end{equation*}
since $(\underline{u}_h\,\cdot\,\underline{u}_h)\notin Q_h$ and $\underline{u}_h$ is only discrete divergence-free.\\
If Equation~\eqref{divfree} is fullfilled, then we obtain the physically correct behavior
\begin{equation*}
\frac{dE_h}{dt}=\int_{\Omega}-\nu\nabla\underline{u}_h:\nabla\underline{u}_h\:d\underline{x}\leq 0.
\end{equation*}
Especially in turbulent flows, where conditions with relatively small viscosity (low molecular diffusion) and very rapid mixing of the transport quantities (high turbulent diffusion), the property of exact divergence-free velocity is essential.

\section{Finite elements}
\label{section:TH}
We have seen in the previous section that the used finite spaces have to fulfill the discrete LBB-condition in order to achieve an unique and stable solution.\\
We assume a triangulation $\mathcal{T}$ which is regular and  consists of elements $T\in\mathcal{T}$ with the corresponding set of vertices $\mathcal{V}$ and edges/faces $\mathcal{E}/\mathcal{F}$.
\\[1ex]
We introduce the most common finite element pairing for the Stokes problem, the Taylor-Hood element \cite{TaylorHood}. This discretization consists of standard $H^1$-conforming elements for the velocity and pressure space, while the polynomial order of the pressure space is one degree lower (at least linear polynomials for pressure space). The notation $P_p$-$P_{p-1}$ is used for the finite element pairing, in this case for the Taylor-Hood element.
\\[1ex]
Let $\mathbb{P}^p(\mathcal{T})$ be the space of all element-wise polynomials on $\mathcal{T}$ up to degree $p$.
The finite element spaces are then chosen as
\begin{align}
&V_h=[\mathbb{P}^p(\mathcal{T})]^3\cap [C^0(\Omega)]^3, \\
&Q_h=\mathbb{P}^{p-1}(\mathcal{T})\cap C^0(\Omega).
\end{align}
As usual, the test functions for the velocity live in the space
\begin{equation}
V_{h0}=\{\underline{v}_h\in V_h:\underline{v}_h=0 \; \mathrm{on} \; \partial\Omega_D\}. 
\end{equation} 
Due to the continuity of the velocity and pressure, the variational formulation of the Navier-Stokes (Problem~\ref{weakformNS}) does not have to be changed for the discretization with Taylor-Hood elements.\\
The Taylor-Hood element satisfies the discrete LBB-condition for the Stokes Problem~\ref{finite_mixed_system_a} and \ref{finite_mixed_system_b}, its proof may be seen in \cite{TaylorHoodproof}. For $P_2$-$P_{1}$, it can be shown that these elements have a quadratic convergence rate if the solution is smooth enough
\begin{equation}
\parallel\underline{u}-\underline{u}_h\parallel_{H^1(\Omega)}+\parallel p-p_h \parallel_{L^2(\Omega)}\leq h^2|\underline{u}|_{H^3(\Omega)}+h^2|p|_{H^2(\Omega)}.
\end{equation}
The big drawback of this choice of elements is that it only peserves discrete divergence-free property since $\nabla\,\cdot\,V_h\not\subset Q_h$.

\section{Time discretization}
After the spatial discretization of the unsteady Navier-Stokes equations (Problem~\ref{TaylorHood_NS}), there are still two aspects to consider to obtain a complete discretization. Firstly, an appropriate discretization of the time derivative has to be discussed. Secondly, an approach to solve the nonlinear convective term of the momentum equation. Therefore, an implicit-explicit (IMEX) splitting scheme will be discussed (refer to \cite{IMEX}). For the sake of simplicity, the volume force term $\underline{f}$ will be neglected here.
\\[1ex]
The approximated velocity $\underline{u}_h(\underline{x},t)$ and pressure $p(\underline{x},t)$ is given by
\begin{align*}
&\underline{u}_h(\underline{x},t)=\sum^{N_V}_{i=1}\underline{u}_i(t)\underline{\phi}_i(\underline{x}),&
&p_h(\underline{x},t)=\sum^{N_Q}_{i=1}p_i(t)\psi_i(\underline{x}),&
\end{align*}
while $\{\underline{\phi}_i(\underline{x})\}^{N_V}_{i=1}$ and $\{\psi_i(\underline{x})\}^{N_Q}_{i=1}$ are the basis functions of the finite element spaces $V_h$ and $Q_h$. By that we define the matrices
\begin{align*}
&M\in\mathbb{R}^{N_V\times N_V} \;\;\; M_{ij}=\int_{\Omega}\underline{\phi}_i(\underline{x})\,\cdot\,\underline{\phi}_j(\underline{x})\:d\underline{x} \;\;\; \forall i,j=1\:...\:N_V,& \\
&A\in\mathbb{R}^{N_V\times N_V} \;\;\; A_{ij}=\int_{\Omega}\nu\nabla\underline{\phi}_i(\underline{x})\,:\,\nabla\underline{\phi}_j(\underline{x})\:d\underline{x} \;\;\; \forall i,j=1\:...\:N_V,& \\
&B\in\mathbb{R}^{N_V\times N_Q} \;\;\; B_{ij}=-\int_{\Omega}\big(\nabla\,\cdot\,\underline{\phi}_i(\underline{x})\big)\psi_j(\underline{x})\:d\underline{x} \;\;\; \forall i=1\:...\:N_V\; ;\;j=1\:...\:N_Q,& \\
&C(\underline{u}_h)\in\mathbb{R}^{N_V} \;\;\; C_{i}=\int_{\Omega}\big((\underline{u}_h\,\cdot\,\nabla)\underline{u}_h\big)\,\cdot\,\underline{\phi}_i(\underline{x})\:d\underline{x} \;\;\; \forall i=1\:...\:N_V.&
\end{align*}
For appropriate initial conditions, the problem is given as:
\begin{problem}[\textbf{Matrix form of spatial discretized Navier-Stokes equations}]
Find $\underline{u}_i\in \mathbb{R}^{N_V} \;\; \forall i=1\:...\:N_V$ and $p_i\in \mathbb{R}^{N_Q} \;\; \forall i=1\:...\:N_Q$ satisfying
\begin{align*}
&M\frac{\partial \underline{u}_i(t)}{\partial t}+A\underline{u}_i(t)+Bp_i(t)+C\big(\underline{u}_i(t)\big)\underline{u}_i(t)=\underline{0} & & in\;[0,T],& \\
&B^T\underline{u}_i(t)=0& & in\;[0,T],&\\
&\underline{u}_i(t=0)=\underline{u}_{i,0}.&
\end{align*}
\label{IMEX_NS}
\end{problem}
For the time discretization, we use the first order IMEX scheme. The main idea is to handle the convective term explicitly and use it as a kind of force term, while the diffusion and the incompressibility constraint are processed implicitly.\\
Generally, explicit methods are computationally cheap and can incoporate nonlinearities without solving a nonlinear system. However, they are conditionally stable and may get unstable if the time step is not restricted.\\
Implicit methods have the advantage that they are generally unconditionally stable, but are very expensive since at each time step a system of equations has to be solved.\\
Such decomposition methods are called IMEX splitting schemes. We define the time step $\Delta_t \geq 0$ and apply the first order method on Problem~\ref{IMEX_NS}, we get
\begin{align*}
&M\frac{\underline{u}_i(t+\Delta_t)-\underline{u}_i(t)}{\Delta_t}+A\underline{u}_i(t+\Delta_t)+Bp_i(t+\Delta_t)=-C\big(\underline{u}_i(t)\big)\underline{u}_i(t), & \\
&B^T\underline{u}_i(t+\Delta_t)=0,&
\end{align*}
which can be rewritten in matrix form
\begin{equation*}
\left(\begin{matrix}M+\Delta_t A&\Delta_t B \\ \Delta_t B^T&0\end{matrix}\right)\left(\begin{matrix}\underline{u}_i(t+\Delta_t)\\p_i(t+\Delta_t)\end{matrix}\right)=\Delta_t\left(\begin{matrix}-C\big(\underline{u}_i(t)\big)\underline{u}_i(t)+\frac{1}{\tau}M\underline{u}_i(t)\\0\end{matrix}\right).
\end{equation*}
In order to obtain the residual form, the equations are extended
\begin{equation*}
\left(\begin{matrix}M+\Delta_t A&\Delta_t B \\ \Delta_t B^T&0\end{matrix}\right)\left(\begin{matrix}\underline{u}_i(t+\Delta_t)-\underline{u}_i(t)\\p_i(t+\Delta_t)-p_i(t)\end{matrix}\right)=\Delta_t\left(\begin{matrix}-C\big(\underline{u}_i(t)\big)\underline{u}_i(t)-A\underline{u}_i(t)-Bp_i(t)\\-B^T\underline{u}_i(t)\end{matrix}\right).
\end{equation*}
We define
\begin{equation*}
M^*=\left(\begin{matrix}M+\Delta_t A&\Delta_t B \\ \Delta_t B^T&0\end{matrix}\right),
\end{equation*}
and
\begin{equation*}
D=\left(\begin{matrix}-C\big(\underline{u}_i(t)\big)-A &-B \\ -B^T & 0\end{matrix}\right).
\end{equation*}
We obtain the final system of equations
\begin{equation}
\left(\begin{matrix}\underline{u}_i(t+\Delta_t)\\p_i(t+\Delta_t)\end{matrix}\right)=(\doubleunderline{I}+\Delta_t M^{*-1}D)\left(\begin{matrix}\underline{u}_i(t)\\p_i(t)\end{matrix}\right).
\end{equation}
After each time step, only the convective part has to be updated with the new velocity, if there is no change in the time step size.\\ Unfortunately, for RANS and LES/VMS, the eddy-viscosity gets recalculated in each step, forcing an update of $M^*$, $D$ and therefore $M^{*-1}$.

\section{Discretization of the turbulence models}
\label{Discretization_turb}
The approximation of the RANS/SFNS equations (Problem~\ref{RANS} and \ref{SFNS}) is very similar to the discretization of the Navier-Stokes equations of the previous sections. The only difference is the diffusion term. In turbulence modelling, the divergence of the Reynolds stress tensor (sub-grid scale tensor in LES/VMS) is approximated as a turbulent diffusion by the Boussinesq hypothesis. This new term gets incoporated by the molecular diffusion, resuming in a condensed diffusion term. The total viscosity consists of molecular- and eddy-viscosity
\begin{equation*}
\nu_{total}=\nu+\nu_T.
\end{equation*}
In the Navier-Stokes equations, $\nu = const$ and therefore the diffusion term could be simplified to
\begin{equation*}
\nabla\,\cdot\,\big(\nu(\nabla\underline{u}+\nabla\underline{u}^T)\big)=\nu\Delta\underline{u},
\end{equation*}
with the incompressibility constraint $\nabla\,\cdot\,\underline{u}=0$.
\\[1ex]
In the case of the RANS/SFNS equations, this simplification does not hold since $\nu_{total}$ is no constant anymore
\begin{equation*}
\nabla\,\cdot\,\big(\nu_{total}(\nabla\underline{u}+\nabla\underline{u}^T)\big)\neq\nu_{total}\Delta\underline{u}.
\end{equation*}
The bilinear form $a(\underline{u}_h,\underline{v}_h)$ then changes to
\begin{equation*}
a_{turb}(\underline{u}_h,\underline{v}_h)=\int_{\Omega}2(\nu+\nu_T)\doubleunderline{S}(\underline{u}_h):\doubleunderline{S}(\underline{v}_h)\;d\underline{x},
\end{equation*}
with $\doubleunderline{S}(\underline{u}_h)=\frac{1}{2}(\nabla\underline{u}_h+\nabla\underline{u}_h^T)$.
The theory for the incompressible Navier-Stokes equations with non-constant viscosity is given in \cite{kaiser}.
\\[1ex]
The discretization of the two-equation eddy-viscosity models (Problem~\ref{KEpsilon}, \ref{KOmega} and \ref{SSTKOmega}) using the continuous Galerkin method with $H^1$-conforming elements is relatively straight forward. For ease of presentation we only consider the approximation of the $K-\epsilon$ model, the derivation of the other models is analogously.
\\[1ex]
Firstly, as usual we start with the weak formulation. Define the function space for the turbulent kinetic energy and for the dissipation rate
\begin{align*}
&R_D=H^1(\Omega,\mathbb{R})=\{r\in H^1(\Omega,\mathbb{R}):r=r_D\;\textrm{on}\;\partial\Omega_D\},& \\
&R=H_0^1(\Omega,\mathbb{R})=\{r\in H^1(\Omega,\mathbb{R}):r=0\;\textrm{on}\;\partial\Omega_D\}.&
\end{align*}
As for the Navier-Stokes equations, we assume Dirichlet and homogeneous Neumann boundary conditions $\partial\Omega=\partial\Omega_D\cup\partial\Omega_N$, $g_K|_{\partial\Omega_N},g_{\epsilon}|_{\partial\Omega_N}=0$ on a bounded domain $\Omega\subset\mathbb{R}^3$.\\
We multiply Equation~\eqref{TKE_modelled} with test functions $k\in R$ and Equation~\eqref{epsilon} with test functions $e\in R$, integrate over the whole domain and integrate by parts. Then we obtain:
\begin{problem}[\textbf{Weak formulation of the $K-\epsilon$ equations}]
Find $K\in R_D$ and $\epsilon\in R_D$ such that
\begin{align*}
&\int_{\Omega}\frac{\partial K}{\partial t}k\:d\underline{x}+\int_{\Omega}(\langle\underline{u}\rangle\,\cdot\,\nabla K)k\:d\underline{x}=
\int_{\Omega}(\Pi-\epsilon)k\:d\underline{x}-\int_{\Omega}(\nu+\frac{\nu_T}{\sigma_K})\nabla K\,\cdot\,\nabla k\:d\underline{x},
& \\
&\int_{\Omega}\frac{\partial \epsilon}{\partial t}e\:d\underline{x}+\int_{\Omega}(\langle\underline{u}\rangle\,\cdot\,\nabla \epsilon)e\:d\underline{x}=
\int_{\Omega}(C_{\epsilon 1}\frac{\Pi\epsilon}{K}-C_{\epsilon 2}\frac{\epsilon^2}{K})e\:d\underline{x}-\int_{\Omega}(\nu+\frac{\nu_T}{\sigma_{\epsilon}})\nabla \epsilon\,\cdot\,\nabla e\:d\underline{x},&
\end{align*}
is satisfied for all $k\in R,$ and $e\in R$.
\end{problem}
This problem has generally the form of two scalar convection diffusion problems, where the source and sink terms depend on each other.\\
The weak variational formulation of the steady convection diffusion problem of form of Problem~\ref{scalarconvectiondiffusion} is seen below.
\begin{problem}[\textbf{Weak formulation of the scalar convection diffusion problem}]
Find $\phi\in R_D$ such that
\begin{equation}
\int_{\Omega}(\langle\underline{u}\rangle\,\cdot\,\nabla \phi)\psi\:d\underline{x}+\int_{\Omega}(\nu+\nu_T)\nabla \phi\,\cdot\,\nabla \psi\:d\underline{x}=
\int_{\Omega}f\psi\:d\underline{x},
\label{WeakSCD}
\end{equation}
is satisfied for all $\psi\in R$.
\end{problem}
In order to prove existence and uniqueness of a solution of the weak Problem in Equation~\eqref{WeakSCD}, we have to check if the bilinear form is continuous and coercive. Let $a_{CD}(\cdot,\cdot):R\times R\to \mathbb{R}$ be a bilinear form defined as,
\begin{equation*}
a_{CD}(\phi,\psi)=\int_{\Omega}(\langle\underline{u}\rangle\,\cdot\,\nabla \phi)\psi\:d\underline{x}+\int_{\Omega}(\nu+\nu_T)\nabla \phi\,\cdot\,\nabla \psi\:d\underline{x}.
\end{equation*}
If $\langle\underline{u}\rangle=\underline{0}$ then we obtain a symmetric bilinear form.\\
Let $\langle\underline{u}\rangle\in L^{\infty}(\Omega)$ then one gets with the Cauchy–Schwarz inequality, Hölder’s inequality, and the Poincaré–Friedrichs inequality (refer to \cite{brenner_scott}),
\begin{align*}
a_{CD}(\phi,\psi)
& \leq(\nu+\nu_T)\parallel\nabla\phi\parallel_{L^2(\Omega)}\parallel\nabla \psi\parallel_{L^2(\Omega)} + \parallel\langle\underline{u}\rangle\parallel_{L^{\infty}(\Omega)}\parallel\nabla\phi\parallel_{L^2(\Omega)}\parallel \psi\parallel_{L^2(\Omega)}&\\
& \leq(\nu+\nu_T)\parallel\nabla\phi\parallel_{L^2(\Omega)}\parallel\nabla \psi\parallel_{L^2(\Omega)} + \alpha_{PF}\parallel\langle\underline{u}\rangle\parallel_{L^{\infty}(\Omega)}\parallel\nabla\phi\parallel_{L^2(\Omega)}\parallel\nabla \psi\parallel_{L^2(\Omega)}&\\
& = \alpha_*\parallel\nabla\phi\parallel_{L^2(\Omega)}\parallel\nabla \psi\parallel_{L^2(\Omega)}&\\
& = \alpha_*\:|\phi|_{H^1(\Omega)}| \psi|_{H^1(\Omega)},&
\end{align*}
for all $\phi,\psi\in R$ with $\alpha_*=(\nu+\nu_T+\alpha_{PF}\parallel\langle\underline{u}\rangle\parallel_{L^{\infty}(\Omega)})$. Hence, the bilinear form is bounded.\\
The convective part may be rewritten as,
\begin{equation*}
\int_{\Omega}(\langle\underline{u}\rangle\,\cdot\,\nabla \psi)\psi\:d\underline{x}=
-\frac{1}{2}\int_{\Omega}(\nabla\,\cdot\,\langle\underline{u}\rangle)\psi^2\:d\underline{x}.
\end{equation*}
Inserting this relation into the bilinear form
\begin{equation*}
a_{CD}(\psi,\psi)=\int_{\Omega}(\nu+\nu_T)\nabla \psi\,\cdot\,\nabla \psi -\frac{1}{2}(\nabla\,\cdot\,\langle\underline{u}\rangle)\psi^2\:d\underline{x},
\end{equation*}
then for all $\psi\in R$ it follows if $(\nabla\,\cdot\,\langle\underline{u}\rangle)\geq 0$
\begin{equation*}
a_{CD}(\psi,\psi)\geq \int_{\Omega}\big((\nu+\nu_T)\nabla \psi\,\cdot\,\nabla \psi\:d\underline{x}=\alpha_2\parallel \psi\parallel_R^2.
\end{equation*}
Thus, $a_{CD}(\psi,\psi)$ is coercive. It must be mentioned that for convection-dominated flux, as it is usually the case for turbulent flows, coercivity loss may occur. Although the exact problem is well-posed, instabilities are possible.\\
The lemma of Lax-Milgram then states that for each bounded functional $f\in R^*$ there exists an unique solution $\phi\in R$.
\\[1ex]
For the discretization, the standard continuous Galerkin method is used, which just replaces the space $R$ by $R_h\subset R$ in the variational formulation. The space $R_{hD}$ incoporates the inhomogeneous Dirichlet boundary condition.
\begin{problem}[\textbf{Discrete formulation of the scalar convection diffusion problem}]
Find $\phi_h\in R_{hD}$, such that
\begin{equation*}
a_{CD}(\phi_h,\psi_h)=\int_{\Omega}f\psi_h\:d\underline{x},
\end{equation*}
is satisfied for all $\psi_h \in R_h$.
\end{problem}
The existence of an unique solution of the discrete problem follows directly
from the theorem of Lax–Milgram, since $R_h$ is a closed subspace of the Hilbert space $R$ and the properties of the bilinear form carry over from $R$ to $R_h$. The use of standard $H^1$-conforming finite elements is sufficient.
\\[1ex]
As for the Navier-Stokes equations, a first order IMEX splitting scheme is used for the two-equation eddy-viscosity models. The diffusive term is handled implicitly and the convective term as well as the source and sink terms are handled explicitly.\\
For each time step, the turbulent quantities of the eddy-viscosity models are updated firstly with respect to the values of the previous time step. After that, the RANS/SFNS equations are calculated with the new updated eddy-viscosity.
}

%% file: chapters/discretization2.tex
\chapter{Hybrid discontinuous Galerkin method for the Navier-Stokes equations}
\label{chapter:discretization2}
%%%%%%%%%%%%%%%%%%%%%%%%%%%%%%%%%%%%%%%%%%%%%%%%%%%%%%%%%%%%%%%%%%%%%%%%%%%%%%%%%%%%%%%%%%%%%%%%%%%%%%%%%%%%%%
%%%%%%%%%%%%%%%%%%%%%%%%%%%%%%%%%%%%%%%%%%%%%%%%%%%%%%%%%%%%%%%%%%%%%%%%%%%%%%%%%%%%%%%%%%%%%%%%%%%%%%%%%%%%%%
{
In the previous chapter, we discussed the concept of the standard continuous Galerkin finite element method. It uses an approximation of the weak formulation of the PDE, which is achieved by replacing the infinite dimensional space in which the variational formulation is posed by a finite dimensional subspace. This finite element space normally uses piecewise polynomials, which are continuous across element interfaces. As we already have observed in the previous chapter, its disadvantage is the conservation property and that no stabilization can be used for convection dominated flows.\\
Discontinuous Galerkin method overcomes this problem by using a discretization, which is continuous on each element but discontinuous across elements. While more degrees of freedom are required, it offers generally more flexibility. Its drawback is the high computational effort, because of larger system of equations with less sparsity due to a lot more couplings of unknowns.\\
In this chapter we introduce a relatively new discretization method that was introduced by Joachim Schöberl and Christoph Lehrenfeld in \cite{lehrenfeld} and \cite{LEHRENFELD2016339}, which manages to solve those drawbacks. It uses a hybridized version of the divergence-conforming DG method from \cite{CK}, called an $H(\mathrm{div})$-conforming hybrid discontinuous Galerkin (HDG) finite element method.

\section{Introduction}
Firstly, we want to use a discontinuous finite element approximation of the weak formulation of the Navier-Stokes equations. Therefore, we have to look at the space of element-piecewise $H^1(T)$ functions, which form the broken Sobolev space
\begin{equation*}
H^1(\mathcal{T})=\{v\in L^2(\Omega), v\in H^1(T)\:\forall T\in\mathcal{T}\}.
\end{equation*}
Due to the reason that functions are no longer continuous over $\Omega$, applying integration by parts is no longer valid. Therefore, we are allowed to integrate by parts on each element $T\in \mathcal{T}$. As the functions do not belong to $H^1(\Omega)$ anymore, we use the interior penalty method introduced in \cite{interiorPenalty} to weakly enforce continuity, which would lead to a DG formulation.\\
However, in this method we use a semi-discontinuous approach called $H(\mathrm{div})$-conforming discretization. The basic idea is to decompose the velocity into continuous normal velocity components and discontinuous tangential velocity components over edges of finite elements
\begin{equation*}
H^1(\Omega,\mathbb{R}^3)= \{\underline{u}\in H^1(T,\mathbb{R}^3):\llbracket\underline{u}\,\cdot\,\underline{n}\rrbracket=0 \; \text{on} \; E\in\mathcal{F}\}\cap\{\underline{u}\in H^1(T,\mathbb{R}^3):\llbracket\underline{u}\times\underline{n}\rrbracket=\underline{0} \; \text{on} \; E\in\mathcal{F}\},
\end{equation*}
where $E$ is an element of the triangulation skeleton $\mathcal{F}$ and $\llbracket\cdot\rrbracket$ is the jump on a common edge $E$ of two neighbouring elements $T_1$ and $T_2$
\begin{equation*}
\llbracket\underline{u}\,\cdot\,\underline{n}\rrbracket_{E_{12}}=(\underline{u}|_{T_1}-\underline{u}|_{T_2})\,\cdot\,\underline{n}_{E_{12}},
\end{equation*}
and
\begin{equation*}
\llbracket\underline{u}\times\underline{n}\rrbracket_{E_{12}}=(\underline{u}|_{T_1}-\underline{u}|_{T_2})\times\underline{n}_{E_{12}}.
\end{equation*}
The sobolev space for $H(\mathrm{div})$-conforming functions is defined as
\begin{equation}
H(\mathrm{div},\Omega)=\{\underline{v}\in L^2(\Omega,\mathbb{R}^3):\nabla\,\cdot\,\underline{v}\in L^2(\Omega,\mathbb{R})\}.
\end{equation}
The compatibility condition for $H(\mathrm{div})$-conformity is the continuity of the normal component over element edges. We introduce the finite dimensional space
\begin{equation}
W_h=\{\underline{v}^T_h\in [\mathbb{P}^p(\mathcal{T})]^3:\llbracket\underline{v}^T_h\,\cdot\,\underline{n}\rrbracket=0 \; \text{on} \; E\in\mathcal{F}\},
\end{equation}
for that holds $W_h \subset H(\mathrm{div},\Omega)$. Note that here the superscript $T$ should indicate the element $T\in\mathcal{T}$ and should not be confused with transposing.\\
For the pressure space, a discontinuous finite element space is used
\begin{equation}
Q_h=\{q_h\in \mathbb{P}^{p-1}(\mathcal{T})\}\subset L^2(\Omega).
\end{equation}
It directly follows that $\nabla\,\cdot\,W_h=Q_h$ and thus the exact divergence-free property is fulfilled.\\
Tangential continuity is not included in the velocity space so we have to account for it in another way. This is done in a hybrid DG formulation. Of course, a standard DG formulation for weakly enforcing continuity of the tangential component of the velocity is possible. Nevertheless, all degrees of freedom of neighbouring elements would couple directly. In order to reduce the coupling of the system matrix, we introduce an additional finite tangential facet space on the skeleton $\mathcal{F}$
\begin{equation}
F_h=\{\underline{v}^F_h\in [\mathbb{P}^p(\mathcal{F})]^3:\underline{v}^F_h\,\cdot\,\underline{n}=0 \; \text{on} \; E\in\mathcal{F}\}.
\end{equation}
The HDG scheme has a computational advantage. Although this comes with additional degrees of freedoms, the coupling is significantly reduced since element DOF do not couple with each other and the same for facet DOF. By static condensation, the resulting linear system of equations only accounts the facet DOF. The size of the corresponding schur complement is then fairly reduced.
\begin{figure}[ht]
    \centering
    \includegraphics[width=0.5\textwidth]{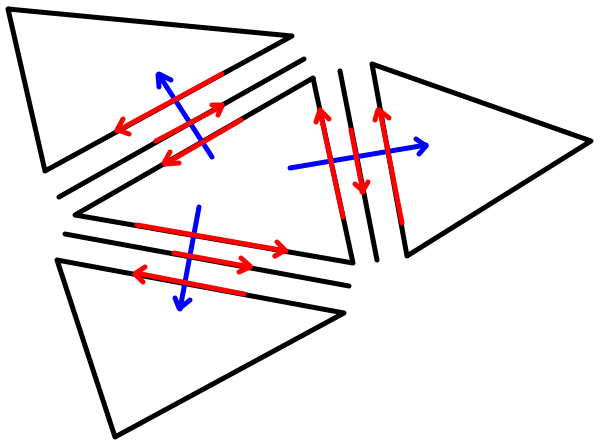}
    \caption{Normal and tangential continuity of $H(\mathrm{div})$-conforming hybrid DG method.}
    \label{fig:hdivHDG}
\end{figure}

\section{Derivation of $H(\mathrm{div})$-conforming HDG method for the Navier-Stokes
problem}
As for the previous derivations, the boundary of the domain $\Omega$ is divided into Dirichlet and homogeneous Neumann boundaries. The new finite compound space for the velocity is $V_h=W_h\times F_h$. The following notation $\underline{u}_h=(\underline{u}_h^T,\underline{u}_h^F)\in V_h$ for the solution and $\underline{v}_h=(\underline{v}_h^T,\underline{v}_h^F)\in V_h$ for the test functions is used. Dirichlet boundary conditions are posed on the facet functions only. Thus, the discrete space is given as
\begin{align}
& V_{h}=\{\underline{u}_h\in V_h:\underline{u}_h^{T,n}=0,\:\underline{u}_h^F=0 \; \text{on} \; \partial\Omega_D\}, \\
& V_{hD}=\{\underline{u}_h\in V_h:\underline{u}_h^{T,n}=\underline{u}_{h,D}^{T,n},\:\underline{u}_h^F=(\underline{u}_{h,D}^F)^t \; \text{on} \; \partial\Omega_D\}.
\end{align}
The jump of the tangential component on the element to the facet is defined as
\begin{equation*}
\llbracket\underline{u}_h^t\rrbracket=(\underline{u}_h^{T,t}-\underline{u}_h^F),
\end{equation*}
and
\begin{equation*}
\llbracket\underline{v}_h^t\rrbracket=(\underline{v}_h^{T,t}-\underline{v}_h^F).
\end{equation*}
The superscript $t$ and $n$ indicates the tangential and normal velocity component respectively
\begin{equation*}
\underline{u}^n_h=(\underline{u}_h\;\cdot\;\underline{n})\underline{n},
\end{equation*}
and
\begin{equation*}
\underline{u}^t_h = \underline{u}_h-\underline{u}^n_h.
\end{equation*}
The continuity of the normal component is automatically fulfilled by the definition of $W_h$.
\\[1ex]
For the viscous part, we define the bilinear form $a_{HDG}(\cdot,\cdot):V_h\times V_h \to \mathbb{R}$ as in \cite{LEHRENFELD2016339}. First we integrate by parts on each element
\begin{equation*}
-\sum_{T\in\mathcal{T}}\int_T\nu\Delta\underline{u}_h^T\,\cdot\,\underline{v}_h^T\:d\underline{x}=\sum_{T\in\mathcal{T}}\bigg(\int_T\nu\nabla\underline{u}_h^T:\nabla\underline{v}_h^T\:d\underline{x}-\int_{\partial T}\nu(\nabla\underline{u}_h^T\,\underline{n})\,\cdot\,\underline{v}_h^T\:d\underline{s}\bigg),
\end{equation*}
while $(\nabla\underline{u}_h^T\,\underline{n})$ denotes the matrix vector product of the velocity vector gradient and element boundary normal vector.\\
We add a consistency term with the facet variables $\underline{\hat{v}}_h=\underline{v}_h^{T,n}+\underline{v}_h^F$
\begin{equation*}
\sum_{T\in\mathcal{T}}\int_{\partial T}\nu(\nabla\underline{u}_h^T\,\underline{n})\,\cdot\,\underline{\hat{v}}_h\:d\underline{s}=\int_{\partial\Omega_N}\nu(\nabla\underline{u}_h^T\,\underline{n})\,\cdot\,\underline{\hat{v}}_h\:d\underline{s},
\end{equation*}
which is in our case zero (homogeneous Neumann boundary conditions) to the previous equation. Then we obtain
\begin{equation*}
\sum_{T\in\mathcal{T}}\bigg(\int_T\nu\nabla\underline{u}_h^T:\nabla\underline{v}_h^T\:d\underline{x}-\int_{\partial T}\nu(\nabla\underline{u}_h^T\,\underline{n})\,\cdot\,\llbracket\underline{v}_h^t\rrbracket\:d\underline{s}\bigg),
\end{equation*}
since $\underline{v}_h^T-\underline{\hat{v}}_h=\underline{v}_h^{T,t}-\underline{v}_h^F$.\\
We add two additional terms for symmetry and stability, which are all zero for the exact solution. The final bilinear form then reads
\begin{align*}
a_{HDG}(\underline{u}_h,\underline{v}_h)= & \sum_{T\in\mathcal{T}}\bigg(\int_T\nu\nabla\underline{u}_h^T:\nabla\underline{v}_h^T\:d\underline{x} & \\ & -\int_{\partial T}\nu(\nabla\underline{u}_h^T\,\underline{n})\,\cdot\,\llbracket\underline{v}_h^t\rrbracket\:d\underline{s}-\int_{\partial T}\nu(\nabla\underline{v}_h^T\,\underline{n})\,\cdot\,\llbracket\underline{u}_h^t\rrbracket\:d\underline{s}& \\ & +\int_{\partial T}\nu\frac{\alpha p^2}{h}\llbracket\underline{u}_h^t\rrbracket\,\cdot\,\llbracket\underline{v}_h^t\rrbracket\:d\underline{s}\bigg), &
\end{align*}
where $\alpha$ is the stabilization parameter.
\\[1ex]
For the pressure we again integrate by parts on each element and define the bilinear form $b_{HDG}(\cdot,\cdot):V_h\times Q_h \to \mathbb{R}$ such as
\begin{equation*}
b_{HDG}(\underline{u}_h,q_h)=-\sum_{T\in\mathcal{T}}\int_T(\nabla\,\cdot\,\underline{u}_h^T)q_h\:d\underline{x}.
\end{equation*}
The final discrete $H(\mathrm{div})$-conforming HDG formulation of the Stokes problem is:
\begin{problem}[\textbf{Discrete HDG formulation of the steady Stokes equations}]
Find $\underline{u}_h\in V_{hD}$ and $p_h \in Q_h$ such that
\begin{align*}
& a_{HDG}(\underline{u}_h,\underline{v}_h)+b_{HDG}(\underline{v}_h,p_h)=f(\underline{v}_h)& &\forall \underline{v}_h\in V_{h}.& \\
&b_{HDG}(\underline{u}_h,q_h)=0& &\forall q_h\in Q_h.&
\end{align*}
\end{problem}
To show well-posedness of the method, we will make use of Brezzi’s theorem for saddle-point problems again. By using this theorem, the necessary conditions are the coercivity of the bilinear form $a_{HDG}$ and the LBB-condition of $b_{HDG}$. For the continuous problem we already showed that these conditions hold.\\
In \cite{lehrenfeld} it is shown that for a sufficiently large stabilization parameter the coercivity condition holds independently of $h$. Using the norm
\begin{equation*}
\parallel\underline{v}_h\parallel^2_{HDG}=\sum_{T\in\mathcal{T}}\big(\parallel\nabla\underline{v}_h\parallel^2_{L^2(T)}+\frac{p^2}{h}\parallel\llbracket\underline{v}_h^t\rrbracket\parallel^2_{L^2(\partial T)}\big),
\end{equation*}
the discrete LBB-condition is
\begin{equation*}
\sup_{\underline{v}_h\in V_h}\frac{b_{HDG}(\underline{v}_h,q_h)}{\parallel\underline{v}_h\parallel_{HDG}}\geq \beta\parallel q_h\parallel_{L^2} \;\;\; \forall q_h\in Q_h,
\end{equation*}
and the paramater $\beta$ is independent of $h$.\\
In \cite{masterthesis_lederer}, a p-version of the discrete LBB-condition is given.
\\[1ex]
The discretization of the convective term of the Navier-Stokes equations uses an upwind stabilization technique. Due to the normal continuity of the velocity, it is only required to have to treat the tangential part in the upwind fashion. The upwind function $\underline{u}_h^{up}$ is defined as
\begin{equation*}
    \underline{u}_h^{up}=\underline{u}_h^{T,n}+
\begin{cases}
    \underline{u}_h^F,& \underline{w}\,\cdot\,\underline{n} < 0\\
    \underline{u}_h^{T,t}, & \underline{w}\,\cdot\,\underline{n} \geq 0.
\end{cases}
\end{equation*}
where $\underline{w}\,\cdot\,\underline{n}$ denotes the normal component of the wind of the convection. This means that the value which comes from the direction where the wind originates is chosen. On outflow edges of the element boundary, we choose the tangential element value, but on inflow edges the facet value is taken.\\
The DG upwind scheme is derived by partial integration on each element and choosing the upwind value for the element boundary integral. The bilinear (trilinear) form $c_{HDG}(\underline{w},\cdot,\cdot):V_h\times V_h \to \mathbb{R}$ as
\begin{equation*}
c_{HDG}(\underline{w},\underline{u}_h,\underline{v}_h)=\sum_{T\in\mathcal{T}}\bigg(-\int_T(\underline{w}\,\otimes\,\underline{u}_h^T):\nabla\underline{v}_h^T\:d\underline{x}+\int_{\partial T}(\underline{w}\,\cdot\,\underline{n})\underline{u}_h^{up}\,\cdot\,\underline{v}_h^T\:d\underline{s}\bigg).
\end{equation*}
\begin{figure}
     \centering
     \begin{subfigure}[b]{0.40\textwidth}
         \centering
         \includegraphics[width=\textwidth]{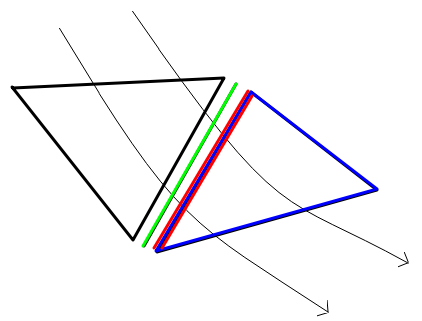}
         \caption{}
         \label{fig:upwind_inflow}
     \end{subfigure}
     \hfill
     \begin{subfigure}[b]{0.40\textwidth}
         \centering
         \includegraphics[width=\textwidth]{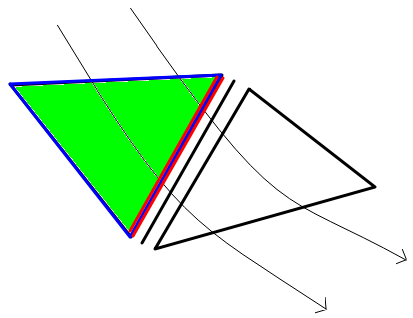}
         \caption{}
         \label{fig:upwind_outflow}
     \end{subfigure}
     \caption{The HDG upwind scheme. The curved arrows are the wind and the blue lined triangle is the reference element. (a): At the inflow edge (red) the facet variable is taken. (b): At the outflow edge (red) the tangential component of the element variable is taken.}
     \label{fig:upwind}
\end{figure}
Obviously, the unknowns of different elements do not couple at all, because facet unknowns just couple with one neighbouring element, the downwind element. At the outflow of the element boundary, an additional constraint is applied to overcome this problem. This constraint glues the facet values on the trace of the upwind element (in a weak sense). The additive constraint reads
\begin{equation*}
\sum_{T\in\mathcal{T}}\int_{\partial T_{out}}(\underline{w}\,\cdot\,\underline{n})(\underline{u}_h^F-\underline{u}_h^{T,t})\,\cdot\,\underline{v}_h^F\:d\underline{s},
\end{equation*}
while $\partial T_{out}$ denotes the outflow edges.\\
\begin{figure}[ht]
    \centering
    \includegraphics[width=0.4\textwidth]{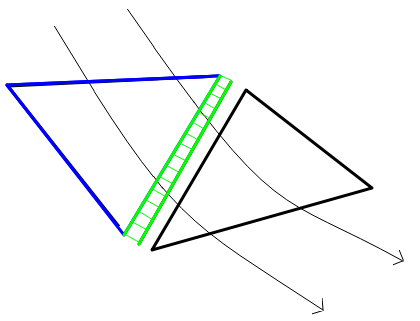}
    \caption{The facet is glued to the outflow boundary.}
    \label{fig:upwind_glue}
\end{figure}
Thus, the final HDG bilinear form for the convective part is
\begin{align*}
c_{HDG}(\underline{w},\underline{u}_h,\underline{v}_h)= & \sum_{T\in\mathcal{T}}\bigg(-\int_T(\underline{w}\,\otimes\,\underline{u}_h^T):\nabla\underline{v}_h^T\:d\underline{x} & \\
& +\int_{\partial T}(\underline{w}\,\cdot\,\underline{n})\underline{u}_h^{up}\,\cdot\,\underline{v}_h^T\:d\underline{s} & \\ &
+\int_{\partial T_{out}}(\underline{w}\,\cdot\,\underline{n})(\underline{u}_h^F-\underline{u}_h^{T,t})\,\cdot\,\underline{v}_h^F\:d\underline{s}\bigg). &
\end{align*}
Finally we obtain the spatial discretization of the steady Navier-Stokes equations by setting the wind as the velocity itself.
\begin{problem}[\textbf{Discrete HDG formulation of the Navier-Stokes equations}]
Find $\underline{u}_h\in V_{hD}$ and $p_h \in Q_h$ such that,
\begin{align*}
& a_{HDG}(\underline{u}_h,\underline{v}_h)+b_{HDG}(\underline{v}_h,p_h)+c_{HDG}(\underline{u}_h,\underline{u}_h,\underline{v}_h)=f(\underline{v}_h)& &\forall \underline{v}_h\in V_{h},& \\
&b_{HDG}(\underline{u}_h,q_h)=0& &\forall q_h\in Q_h.&
\end{align*}
\end{problem}
The time discretization for the $H(\mathrm{div})$-conforming HDG method uses the same IMEX splitting technique as already described in the previous chapter.

\section{Low order $H(\mathrm{div})$-conforming finite elements}
The most common examples for $H(\mathrm{div})$-conforming finite elements are the Raviart-Thomas $RT$ elements from \cite{RT} and the Brezzi-Douglas-Marini $BDM$ elements, see \cite{BDM}. As we have seen before that $H(\mathrm{div},\Omega)$ vector fields have a continuous normal component and a discontinuous tangential component. For ease of use and simplicity, we assume $\Omega\subset\mathbb{R}^2$ in this section. For the construction of higher order $H(\mathrm{div})$-conforming elements we refer to \cite{zaglmayer}.
\\[1ex]
The $RT_p$ element of order $p\geq 0$ is defined with the finite space
\begin{equation*}
RT_p(\mathcal{T})=[\mathbb{P}^p(\mathcal{T})]^2+\underline{x}\hat{\mathbb{P}}^p(\mathcal{T}),
\end{equation*}
with the number of degrees of freedom per element $\mathrm{dim}\big(RT_p(T)\big)=(p+1)(p+3)$ and $\hat{\mathbb{P}}^p(\mathcal{T})$ denotes homogeneous polynomials.\\
The lowest order $RT_0$ element is defined by one degree of freedom per edge $E_i$ of the element (low order edge-based DOF), such that
\begin{equation*}
N_E^0:\underline{\psi}\to\int_{E_i}\underline{\psi}\,\cdot\,\underline{n}\:d\underline{s}.
\end{equation*}
By the definition of the barycentric coordinates, the shape functions associated with the edge is
\begin{equation*}
\underline{\psi_E^0}(\lambda_1,\lambda_2)=\lambda_2\nabla\times\left(\begin{matrix}\lambda_1\\\lambda_1\end{matrix}\right)-\lambda_1\nabla\times\left(\begin{matrix}\lambda_2\\\lambda_2\end{matrix}\right).
\end{equation*}
The global $RT_0$ finite element space is then
\begin{equation*}
RT_0(\mathcal{T})=\text{span}\{\underline{\psi_E^0}:\forall E\in \mathcal{F}\} \;\; \subset H(\mathrm{div},\Omega).
\end{equation*}
\\[1ex]
The $BDM_p$ element of order $p\geq 1$ is defined with the finite space
\begin{equation*}
BDM_p(\mathcal{T})=\big\{\underline{v}\in[\mathbb{P}^p(\mathcal{T})]^2:\llbracket\underline{v}\,\cdot\,\underline{n}\rrbracket=0 \; \text{on} \; E\in\mathcal{F}\}.
\end{equation*}
with the number of degrees of freedom per element $\mathrm{dim}\big(BDM_p(T)\big)=(p+1)(p+2)$.\\
The linear $BDM_1$ is defined by one additional degree of freedom per edge $E_i$ of the element (high order edge-based DOF), namely
\begin{equation*}
N_E:\underline{\psi}\to\int_{E_i}\underline{\psi}\,\cdot\,\underline{n}v\:d\underline{s} \;\; \forall v\in \mathbb{P}^1(E_i).
\end{equation*}
The corresponding shape functions are
\begin{equation*}
\underline{\psi}_E(\lambda_1,\lambda_2)=\nabla\times\left(\begin{matrix}\lambda_1\lambda_2\\\lambda_1\lambda_2\end{matrix}\right).
\end{equation*}
The global $BDM_1$ finite element space is then
\begin{equation*}
BDM_1(\mathcal{T})=\text{span}\{\underline{\psi_E^0},\underline{\psi_E}:\forall E\in \mathcal{F}\} \;\; \subset H(\mathrm{div},\Omega).
\end{equation*}
Indeed, there holds
\begin{equation*}
\nabla\,\cdot\,\underline{\psi}_E=0 \;,\; N_E(\underline{\psi}_E)=0 \;\;\; \forall E\in \partial T.
\end{equation*}
\begin{figure}[ht]
    \centering
    \includegraphics[width=0.35\textwidth]{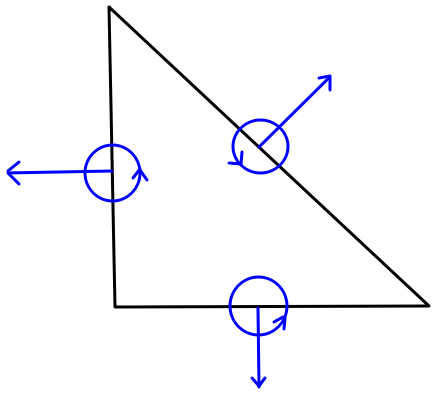}
    \caption{Degrees of freedom for the first order BDM element.}
    \label{fig:BDM}
\end{figure}
\\[1ex]
Both elements fulfill the exact divergence-free condition
\begin{equation*}
\nabla\,\cdot\,RT_0(\mathcal{T})=\nabla\,\cdot\,BDM_1(\mathcal{T})=\mathbb{P}^0(\mathcal{T}).
\end{equation*}
But the approximation is better for $BDM_1$ as
\begin{equation*}
[\mathbb{P}^0(\mathcal{T})]^2\subset RT_0(\mathcal{T})\subset BDM_1(\mathcal{T})=[\mathbb{P}^1(\mathcal{T})]^2.
\end{equation*}
However, the $RT_0$ elements need less degrees of freedom to achieve the divergence-free condition.\\
For both finite elements, higher order versions exist. The construction of the higher order shape functions mimics the exact sequence property of the spaces $H^1$, $H(\mathrm{curl})$, $H(\mathrm{div})$ and $L^2$ called de Rham Complex (see \cite{zaglmayer}). As we have seen above, the $BDM_1$ element uses low and high order edge-based DOF. For $p>1$, high order cell-based divergence-free and non divergence-free DOF are used.\\ One very interesting remark about the physical interpretation of the space separation was mentioned in \cite{lehrenfeld}. As already known, in turbulent flow eddies of different sizes occur. In a discretized domain, the largest eddies may cover one or more vertices, while smaller whirls are located on an edge. The smallest ones lie within one element. To resolve the largest eddies, the lowest order $RT_0$ element is sufficient. Higher order elements properly represent the eddies located at one edge or within an element.

\section{HDG discretization of the turbulence models}
The same issue as already explained in the Section~\ref{Discretization_turb} applies as well for the HDG scheme. Instead of using the Laplacian of the velocity $\nu\Delta\underline{u}$, the strain rate tensor is applied $\nabla\,\cdot\,\big((\nu+\nu_T)(\nabla\underline{u}+\nabla\underline{u}^T)\big)$. Therefore, only the bilinear form $a_{HDG}(\underline{u}_h,\underline{v}_h)$ changes to
\begin{align*}
a_{HDG,turb}(\underline{u}_h,\underline{v}_h)= & \sum_{T\in\mathcal{T}}\bigg(\int_T2(\nu+\nu_T)\doubleunderline{S}(\underline{u}_h^T):\doubleunderline{S}(\underline{v}_h^T)\:d\underline{x} & \\ & -\int_{\partial T}2(\nu+\nu_T^F)(\doubleunderline{S}(\underline{u}_h^T)\,\underline{n})\,\cdot\,\llbracket\underline{v}_h^t\rrbracket\:d\underline{s}-\int_{\partial T}2(\nu+\nu_T^F)(\doubleunderline{S}(\underline{v}_h^T)\,\underline{n})\,\cdot\,\llbracket\underline{u}_h^t\rrbracket\:d\underline{s}& \\ & +\int_{\partial T}2(\nu+\nu_T^F)\frac{\alpha p^2}{h}\llbracket\underline{u}_h^t\rrbracket\,\cdot\,\llbracket\underline{v}_h^t\rrbracket\:d\underline{s}\bigg), &
\end{align*}
with $\doubleunderline{S}(\underline{u}_h)=\frac{1}{2}(\nabla\underline{u}_h+\nabla\underline{u}_h^T)$. The variable $\nu_T^F$ corresponds to the eddy-viscosity calculated by the facet variable of the transported turbulent quantities. 
\\[1ex]
The HDG version of the two-equation turbulence models for RANS simulation are very analogously to the HDG version of the scalar convection diffusion equation. The derivation of this scheme is very similar to Stokes and Navier-Stokes problem and is fairly detailed explained in \cite{lehrenfeld}.\\
We define the compounded finite space for the transported quantity
\begin{equation*}
R_h=\{(\phi,\phi^F):\phi\in\mathbb{P}^p(\mathcal{T}), \phi^F\in \mathbb{P}^p(\mathcal{F})\},
\end{equation*}
and the two bilinear forms $a_{CD,HDG}:R_h\times R_h \to \mathbb{R}$ and $c_{CD,HDG}:R_h\times R_h \to \mathbb{R}$
\begin{align*}
a_{CD,HDG}(\phi_h,\psi_h)= & \sum_{T\in\mathcal{T}}\bigg(\int_T(\nu+\nu_T)\nabla\phi_h\,\cdot\,\nabla \psi_h\:d\underline{x} & \\ & -\int_{\partial T}(\nu+\nu_T^F)(\nabla\phi_h\,\cdot\,\underline{n})\llbracket \psi_h\rrbracket\:d\underline{s}-\int_{\partial T}(\nu+\nu_T^F)(\nabla \psi_h\,\cdot\,\underline{n})\llbracket\phi_h\rrbracket\:d\underline{s} & \\ & +\int_{\partial T}(\nu+\nu_T^F)\frac{\alpha p^2}{h}\llbracket\phi_h\rrbracket\llbracket \psi_h\rrbracket\:d\underline{s}\bigg), &
\end{align*}
with $\llbracket \phi_h\rrbracket=\phi_h-\phi_h^F$ and $\llbracket \psi_h\rrbracket=\psi_h-\psi_h^F$ and 
\begin{align*}
c_{CD,HDG}(\phi_h,\psi_h)= & \sum_{T\in\mathcal{T}}\bigg(-\int_T\langle\underline{u}\rangle\phi_h\nabla \psi_h\:d\underline{x} & \\ & + \int_{\partial T}(\langle\underline{u}\rangle\,\cdot\,\underline{n})\phi_h^{up}\psi_h\:d\underline{s} & \\ & +\int_{\partial T_{out}}(\langle\underline{u}\rangle\,\cdot\,\underline{n})\llbracket\phi_h\rrbracket \psi_h^F\:d\underline{s}\bigg), &
\end{align*}
as the upwind value $\phi_h^{up}$ is defined as
\begin{equation*}
    \phi_h^{up}=
\begin{cases}
    \phi_h^F,& \langle\underline{u}\rangle\,\cdot\,\underline{n} < 0\\
    \phi_h, & \langle\underline{u}\rangle\,\cdot\,\underline{n} \geq 0.
\end{cases}
\end{equation*}
Adding diffusion and convection together, we obtain the final equation:
\begin{problem}[\textbf{Discrete HDG formulation of the steady scalar convection diffusion equations}]
Find $\phi_h\in R_{hD}$ such that
\begin{equation*}
a_{CD,HDG}(\phi_h,r_h)+c_{CD,HDG}(\phi_h,r_h)=f(r_h) \;\;\; \forall r_h\in R_{h}
\end{equation*}
\label{HDG_SCD}
\end{problem}
The two-equation models are adapted from Problem~\ref{HDG_SCD} and the time derivative and corresponding source and sink terms are appended.
}

%% file: chapters/results.tex
\chapter{Numerical test case}
\label{chapter:results}
%%%%%%%%%%%%%%%%%%%%%%%%%%%%%%%%%%%%%%%%%%%%%%%%%%%%%%%%%%%%%%%%%%%%%%%%%%%%%%%%%%%%%%%%%%%%%%%%%%%%%%%%%%%%%%
%%%%%%%%%%%%%%%%%%%%%%%%%%%%%%%%%%%%%%%%%%%%%%%%%%%%%%%%%%%%%%%%%%%%%%%%%%%%%%%%%%%%%%%%%%%%%%%%%%%%%%%%%%%%%%

{
This chapter is dedicated to the setup and results of the plane channel flow test case. This case demonstrates the accuracy and validity of the different modelling principles and discretization techniques for wall-bounded turbulent flows. Furthermore, its relatively less costly computational effort is also beneficial.\\
One main focus of this thesis is on the comparison of the standard discretization method and the HDG approach previously described in Chapter~\ref{chapter:discretization} and \ref{chapter:discretization2}.

\section{Basic setup}
Firstly, the simulation of fully developed turbulent flow in a plane channel at friction Reynolds number $Re_{\tau}=395$ is considered. The case consists of two infinite parallel plates bordering an equilibrium flow. In order to approximate this configuration, a finite sub-domain is taken and periodic boundaries are applied in streamwise $x_1$ and spanwise $x_3$ directions. In normal direction to the walls $x_2$, no-slip Dirichlet boundary conditions are chosen for the velocity. The overall dimensions of the computational domain are $4\delta\times 2\delta\times 2\delta$, allowing sufficiently large structures to envelope. For all computations, a channel half width of $\delta=\frac{1}{2}$ is taken and the domain was discretized in form of uniform hexahedral meshes.\\
All results given in this thesis are also compared to the well-resolved DNS  dataset performed by Moser \cite{DNSmoser} with the same friction Reynolds number. The results of this DNS benchmark may be considered as an exact solution of the Navier-Stokes equations for the purpose of this thesis.\\
All channel flow simulations have the following properties:
\begin{table}[!ht]
	\centering
	\begin{tabular}{|c|l|l|}
		\hline
		Description & Quantity & Value   \\ \hline
		friction Reynolds number            & $Re_{\tau}$   & $395$    \\
		bulk velocity             & $U_b$ & $1$   \\
		bulk Reynolds number            & $Re_b$   & $13350$ \\
		viscosity & $\nu$       & $7.5\times 10^{-5}$    \\
		friction velocity & $u_{\tau}$       & $0.05925$    \\
		\hline
	\end{tabular}
	\caption{Flow properties}
	\label{table:properties}
\end{table}
\\[1ex]
To ensure consistency, the streamwise bulk velocity $U_b$ through the channel is adjusted to be equal to the value in Table~\ref{table:properties} by varying an imposed streamwise pressure gradient. The time-averaged value of this quantity is equivalent to the mean wall shear stress.
\\[1ex]
As initial conditions for the velocity, a sightly pertubated laminar parabolic velocity profile is taken (except for RANS). Normally, the laminar flow takes many flow-through times before small disturbances produced by numerical errors trigger the transition to the turbulent state. Random pertubations seem very ineffective, the pertubated flow field does not obey divergence-free velocity and has no structure. To accelerate the laminar-turbulent transition, a method by \cite{eugene} creates initial wavelike structures in the near-wall region. These structures have the statistical characteristics of the near-wall streaks described in Chapter~\ref{sec:TKE} and their interaction with the superposed laminar channel flow cause linear instabilities.\\
The transition to turbulence occurs relatively rapidly and is shown for different times and planes in Figure~\ref{transition_figure}. In the first few time units, the streaks form to a very regular pattern. After about three flow-through times, the flow is becoming chaotic and the near-wall structures strongly effect the mean flow.  The laminar flow regime completely breaks down and vortical structures and eddies are becoming widespread in the flow domain. At $t\approx 25$, the flow is completely turbulent.
\\[1ex]
Once a statistically steady-state is reached, time averaging over 200 time units is performed, followed by spatial streamwise and spanwise averaging over the entire channel. In the scope of this chapter, the definition of the Reynolds stress tensor has slightly changed. The components of the new defined RST are the covariance of each resolved velocity fluctuation component defined by time and spatial averaging and not by the ensemble average as given in Chapter~\ref{section:RANS}. Furthermore, the first and second order statistical moments are calculated. Except for RANS simulations, there is no averaging needed obviously.\\
The results are presented via the non-dimensional mean streamwise velocity and the normalized turbulent intensities (second order statistics) and turbulent kinetic energy. For RANS and LES/VMS, the normalized eddy-viscosity and the modeled shear component of the RST and SGS tensor are given. Additionally, the total energy spectrum in streamwise and spanwise direction at the particularly chosen $y^+$-positions is shown. The three different positions are $y^+=5$ (viscous sublayer), $y^+=40$ (buffer layer) and $y^+=100$ (logarithmic layer). The energy spectrum is calculated from the discrete Fourier transformation of the velocity fluctuations in streamwise and spanwise direction.\\
The following sections are divided into the four different simulation principles of turbulent flow.

\begin{figure}[h]
    \centering
    \includegraphics[width=0.6\textwidth]{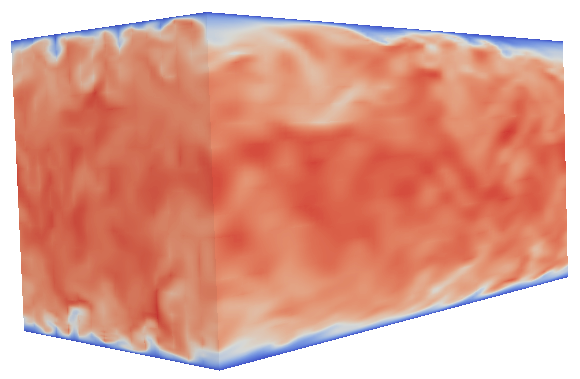}
    \caption{Fully turbulent channel flow.}
    \label{fig:DNS_3d}
\end{figure}

\begin{figure}
     \centering

     \begin{subfigure}[b]{\textwidth}
         \centering
         \includegraphics[width=0.3\textwidth]{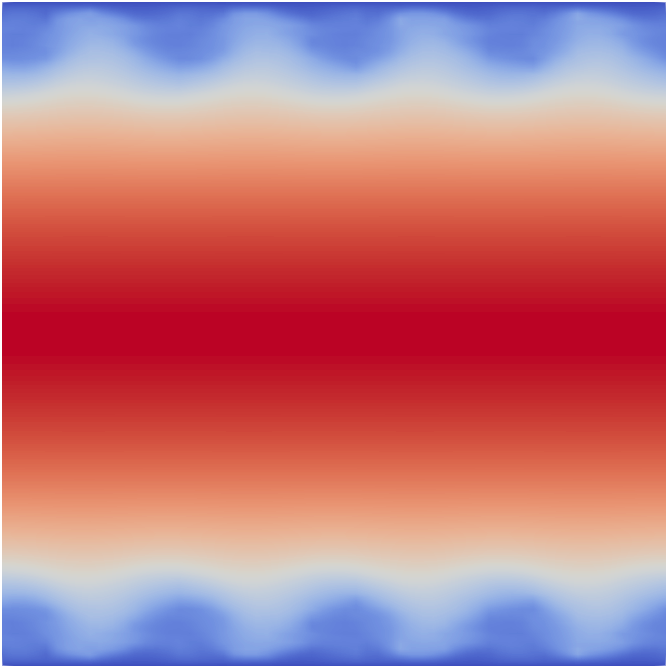}
         \hfill
         \includegraphics[width=0.6\textwidth]{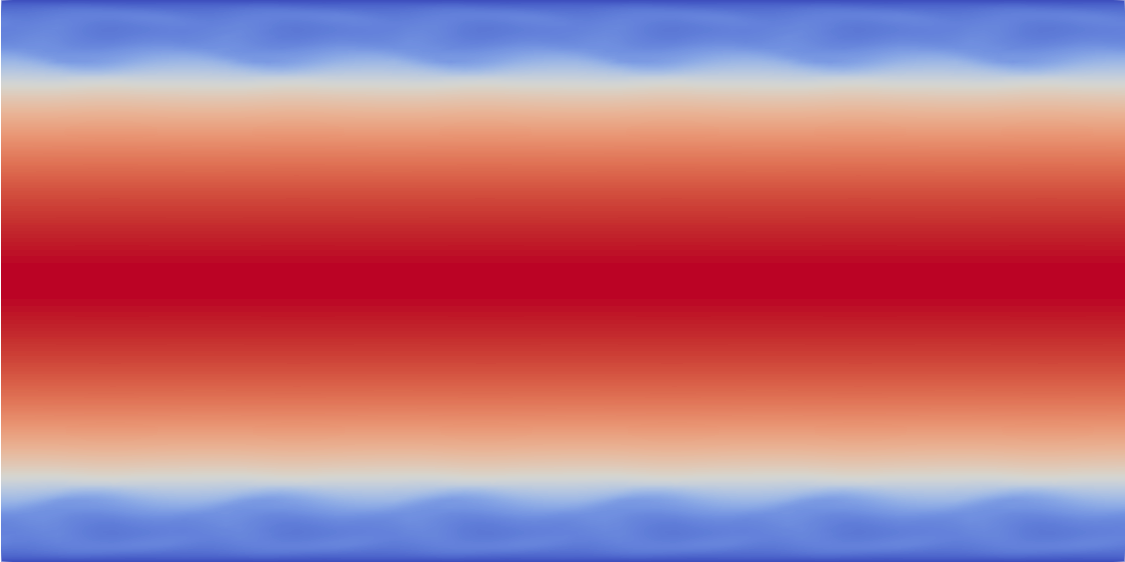}
         \caption{$t=3$}
     \end{subfigure}

     \begin{subfigure}[b]{\textwidth}
         \centering
         \includegraphics[width=0.3\textwidth]{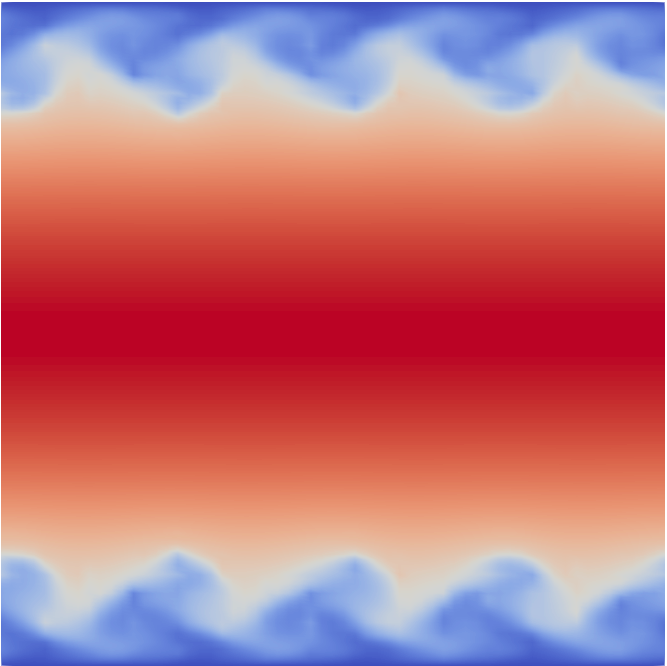}
         \hfill
         \includegraphics[width=0.6\textwidth]{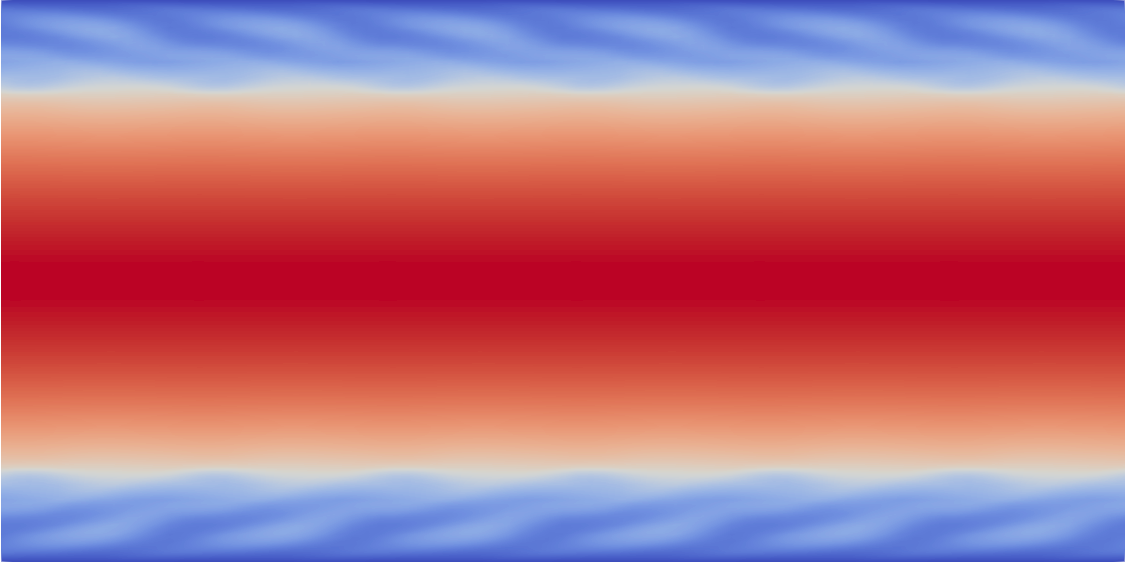}
         \caption{$t=4$}
     \end{subfigure}

     \begin{subfigure}[b]{\textwidth}
         \centering
         \includegraphics[width=0.3\textwidth]{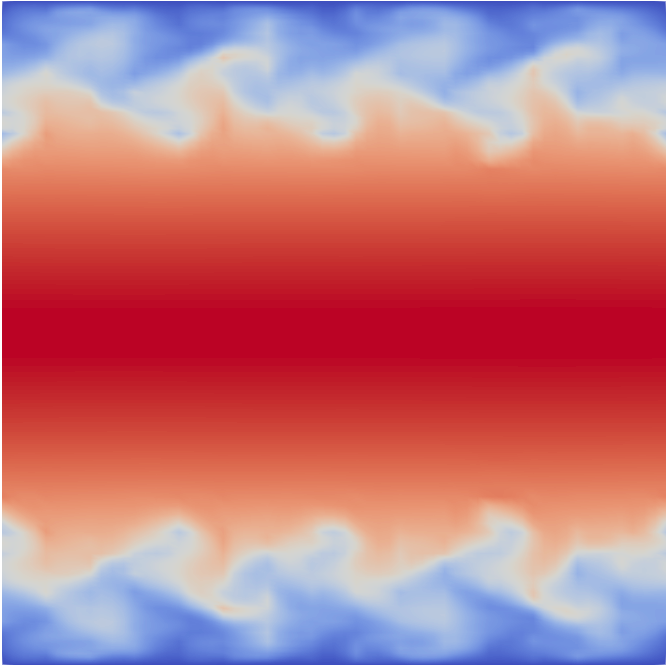}
         \hfill
         \includegraphics[width=0.6\textwidth]{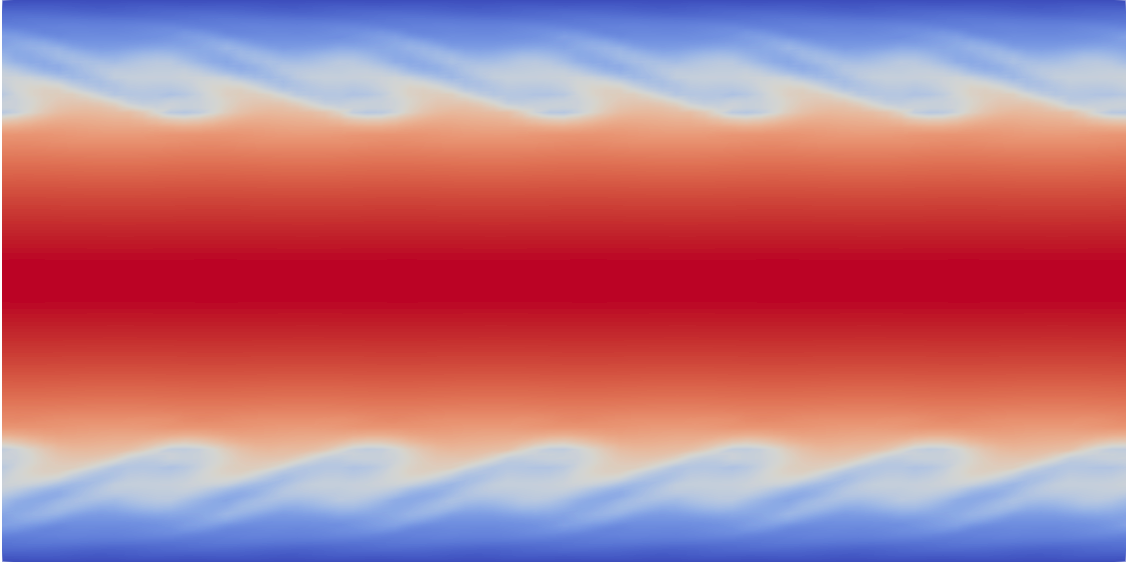}
         \caption{$t=5$}
     \end{subfigure}

     \begin{subfigure}[b]{\textwidth}
         \centering
         \includegraphics[width=0.3\textwidth]{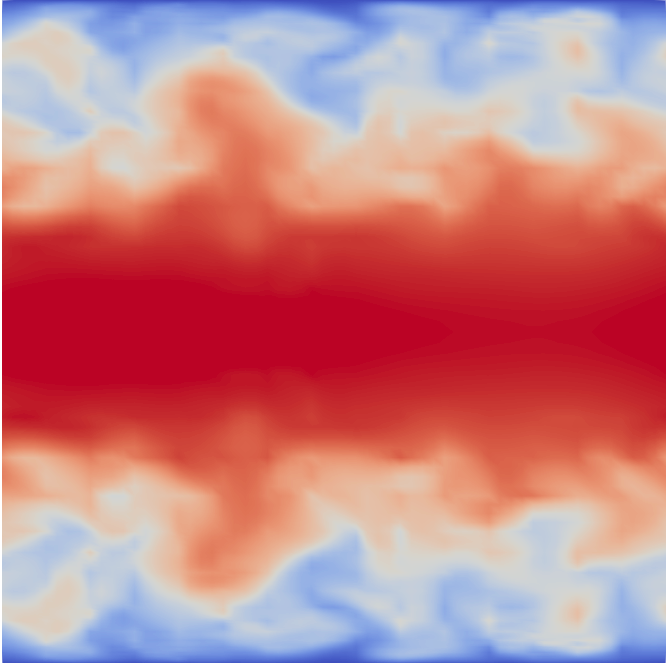}
         \hfill
         \includegraphics[width=0.6\textwidth]{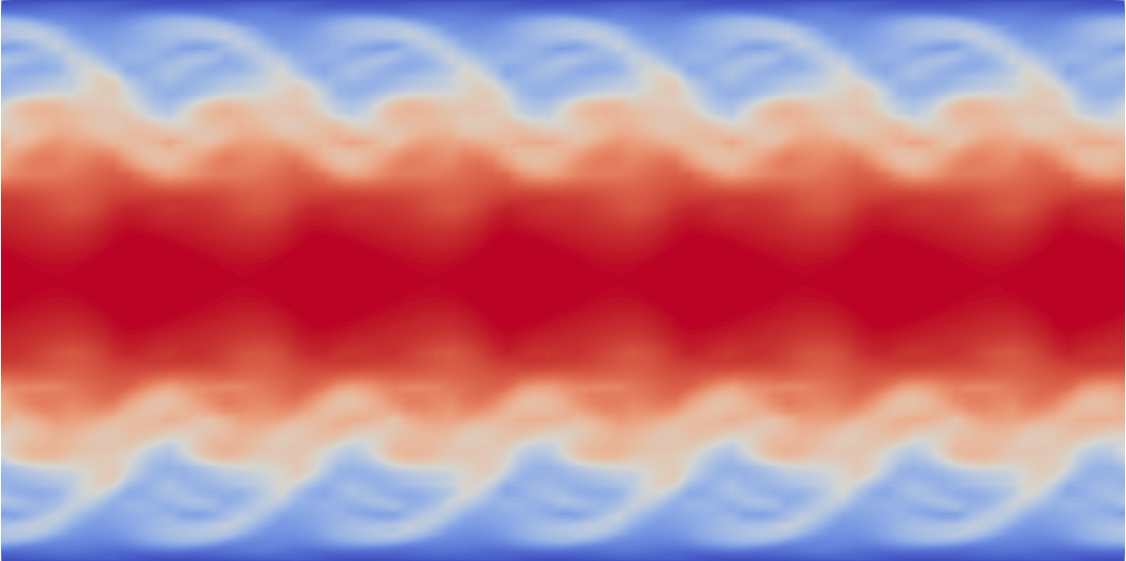}
         \caption{$t=8$}
     \end{subfigure}

\end{figure}

\begin{figure}\ContinuedFloat
     \centering

     \begin{subfigure}[b]{\textwidth}
         \centering
         \includegraphics[width=0.3\textwidth]{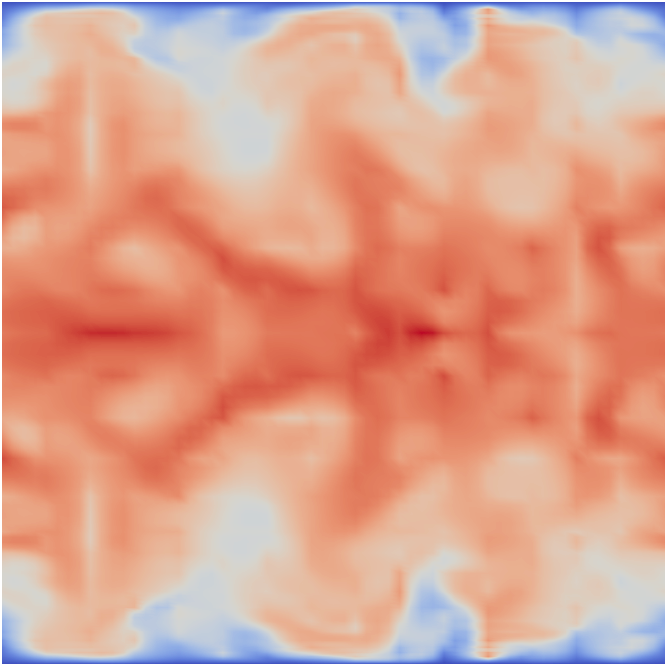}
         \hfill
         \includegraphics[width=0.6\textwidth]{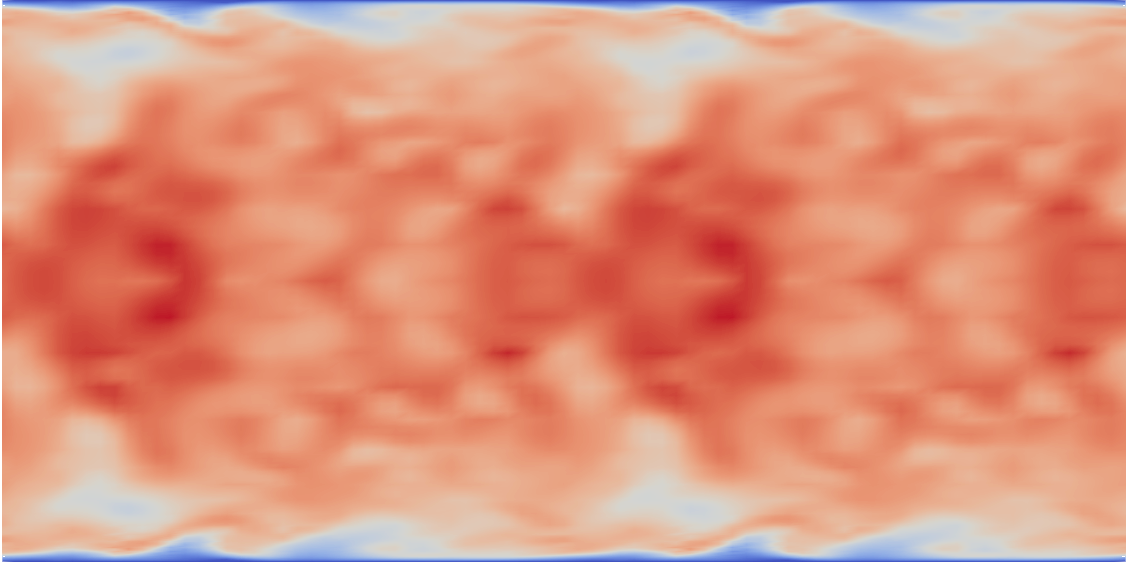}
         \caption{$t=14$}
     \end{subfigure}

     \begin{subfigure}[b]{\textwidth}
         \centering
         \includegraphics[width=0.3\textwidth]{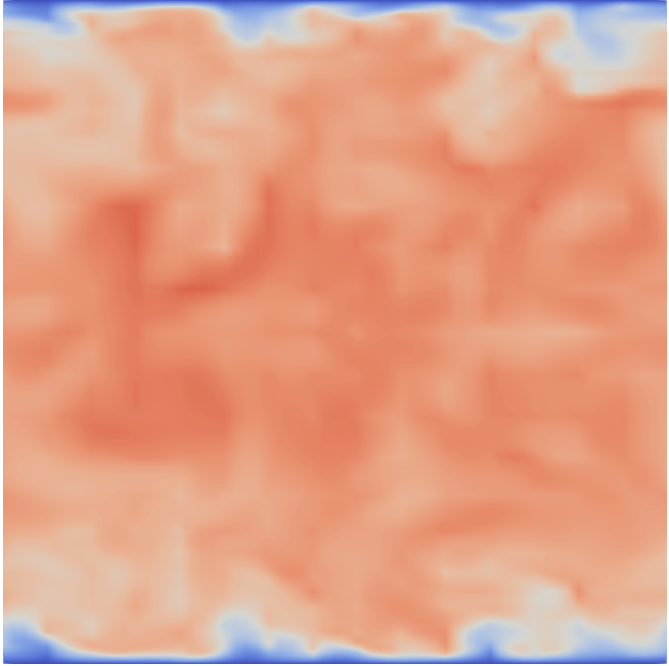}
         \hfill
         \includegraphics[width=0.6\textwidth]{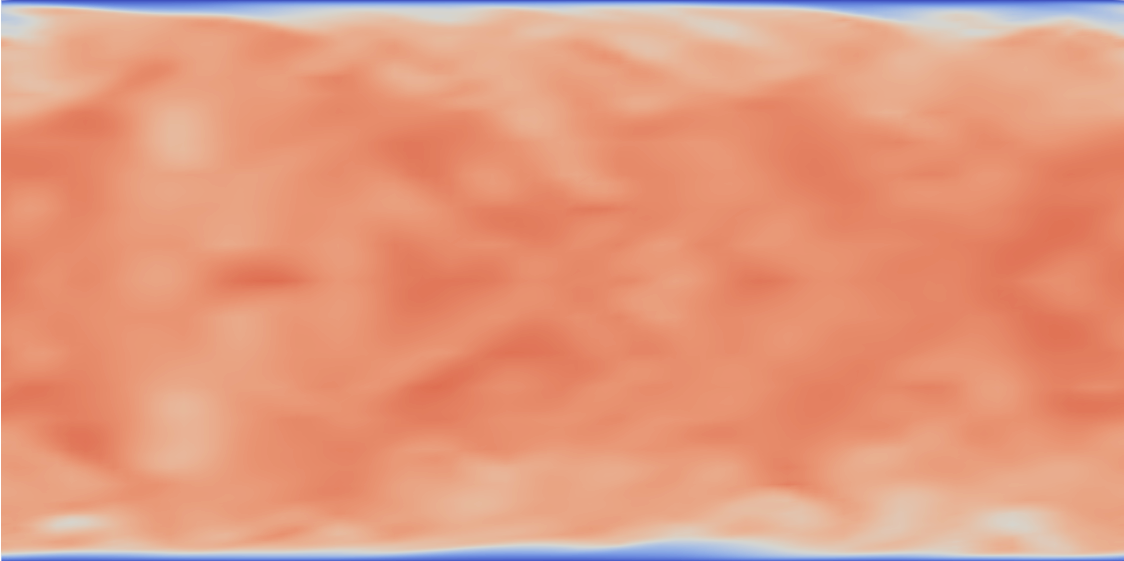}
         \caption{$t=25$}
     \end{subfigure}

     \begin{subfigure}[b]{\textwidth}
         \centering
         \includegraphics[width=0.3\textwidth]{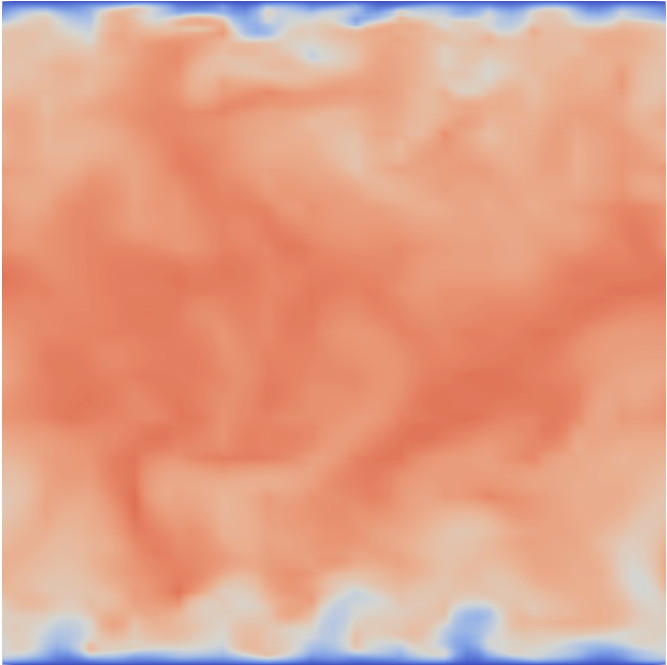}
         \hfill
         \includegraphics[width=0.6\textwidth]{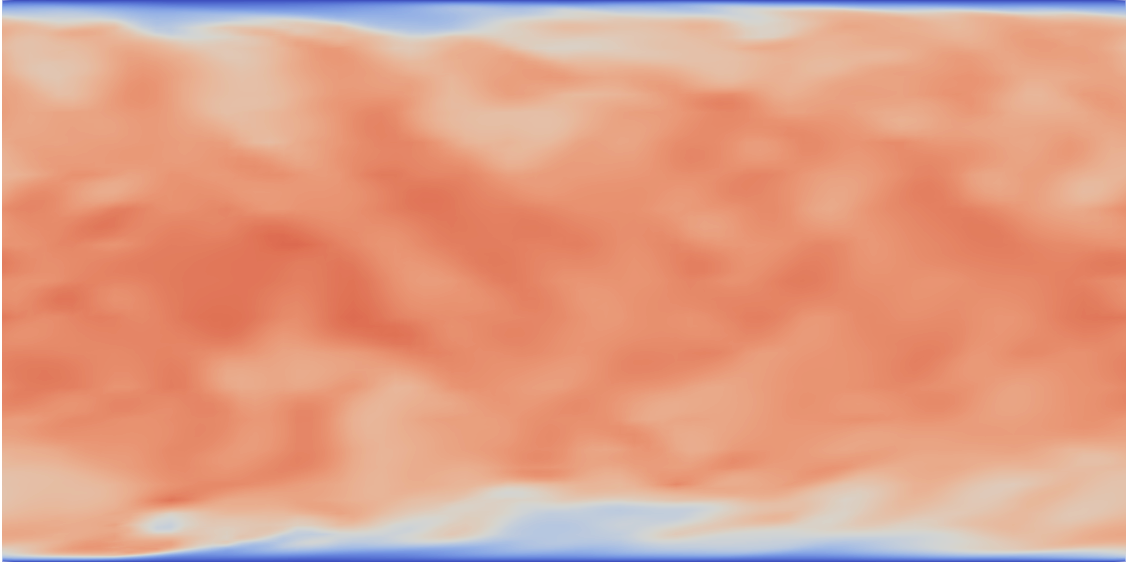}
         \caption{$t=50$}
     \end{subfigure}

     \begin{subfigure}[b]{0.8\textwidth}
         \centering
         \includegraphics[width=\textwidth]{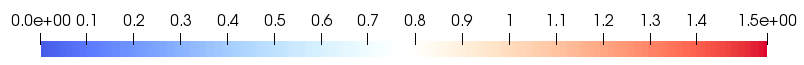}
     \end{subfigure}

     \caption{Velocity magnitude at different times and planes. Left figures show $x_3,x_2$-plane at $x_1=1$ and right figures show $x_1,x_2$-plane at $x_3=0.5$.}
     \label{transition_figure}
\end{figure}

\clearpage
\section{RANS}
Three different two-equation eddy-viscosity models from Chapter~\ref{chapter:simulation} are used for calculating the RANS equations. The $K-\epsilon$ model has been left out here, since we did not obtain very good results in this case with this model and the superiority of the $K-\omega$ models is unsurpassable in wall-bounded flows. Additionally, a second version of the $K-\omega$ model by Peng \cite{Peng} was introduced. This model incoporates a weakend form of the cross-diffusion term and different damping functions. The following values for the respective constants and the corresponding damping functions have been used throughout all computations.
\begin{itemize}
\item $K-\omega$ version from 1998:
\begin{table}[!htbp]
	\centering
	\begin{tabular}{|c|l|l|}
		\hline
		Description & Value/Function   \\ \hline
		$C_{\mu}$ & $\frac{0.024+\frac{Re^*}{6}}{1+\frac{Re^*}{6}}$ \\
		$\beta^*$ & $0.09\big(\frac{\frac{4}{15}+(\frac{Re^*}{8})^4}{1+(\frac{Re^*}{8})^4}\big)$    \\
		$C_{\omega 1}$ &  $0.52\big(\frac{\frac{1}{9}+\frac{Re^*}{2.95}}{1+\frac{Re^*}{2.95}}\big)$    \\
		$C_{\omega 2}$ & $0.072$           \\
		$Re^*$ & $\frac{K}{\nu\omega}$ \\
		$\sigma_K$ & $2$       \\
		$\sigma_{\omega}$ & $2$       \\
		\hline
	\end{tabular}
\end{table}
\item $K-\omega$ version from Peng \cite{Peng}:
\begin{table}[!htbp]
	\centering
	\begin{tabular}{|c|l|l|}
		\hline
		Description & Value/Function    \\ \hline
		$C_{\mu}$ & $0.025+(1-e^{-(\frac{Re^*}{10})^{\frac{3}{4}}})(0.975+0.001e^{-(\frac{Re^*}{200})^2})/Re^*$ \\
		$\beta^*$ & $0.09\big(1-0.722e^{-(\frac{Re^*}{10})^4}\big)$    \\
		$C_{\omega 1}$ &  $0.42\big(1+4.3e^{-(\frac{Re^*}{1.5})^{\frac{1}{2}}}\big)$    \\
		$C_{\omega 2}$ & $0.075$           \\
		$Re^*$ & $\frac{K}{\nu\omega}$ \\
		$\sigma_K$ & $0.8$       \\
		$\sigma_{\omega}$ & $1.35$       \\
		\hline
	\end{tabular}
\end{table}
\\The cross-diffusion term $\frac{3\nu_T}{4K}\nabla K\,\cdot\,\nabla\omega$ is added to the right hand-side of the $\omega$ equation.
\item $K-\omega$ SST:
\begin{table}[H]
	\centering
	\begin{tabular}{|c|l|l|}
		\hline
		Description & Value/Function   \\ \hline
		$a_1$ & $0.31$ \\
		$\beta^*$ & $0.09$    \\
		$\Pi$ & $min(\Pi,0.9K\omega)$ \\
		$C_{\omega 1}$ &  $\chi(\frac{5}{9},0.44)$    \\
		$C_{\omega 2}$ & $\chi(0.075,0.0828)$           \\
		$C_{\omega 3}$ & $0.856$           \\
		$\sigma_K$ & $\chi(0.85,1)$       \\
		$\sigma_{\omega}$ & $\chi(0.5,0.856)$       \\
		\hline
	\end{tabular}
\end{table}
\end{itemize}

The $K-\omega$ SST is not using any damping-functions and the constants are calculated via the blending function Equation~\eqref{blendingfunction}.
\\[1ex]
The initial field for $K$ and $\omega$ is given as
\begin{align*}
&K_0=1.5(T_{int})^2, \\
&\omega_0 = \frac{\sqrt{K_0}}{l_T},
\end{align*}
where the turbulent intensity is initially set to $T_{int}=0.05$ and the turbulent length scale to $l_T=0.076\delta$.\\
The Dirichlet boundary conditions at the solid wall are
\begin{align*}
&K_D=0, \\
&\omega_D = 10^5.
\end{align*}
The high value of $\omega$ at the wall demands fine meshing and small time stepping, especially in the first phase of the flow development. Strictly speaking, $\omega$ is infinity directly at the wall but this is actually not feasible.
\begin{figure}[H]
	\centering

	\begin{subfigure}[b]{0.6\textwidth}
		\centering
		\includegraphics[width=\textwidth]{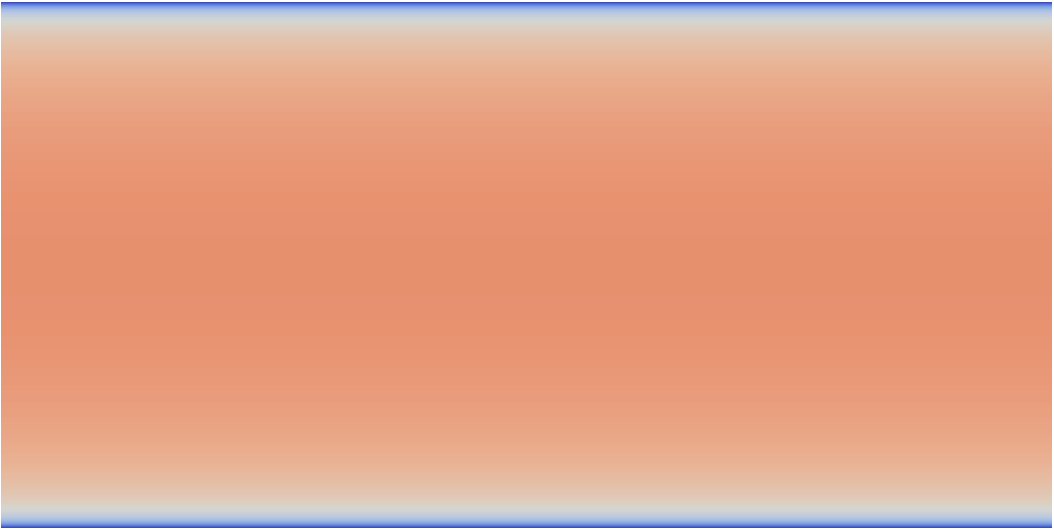}
     \end{subfigure}

     \begin{subfigure}[b]{0.6\textwidth}
		\centering
		\includegraphics[width=\textwidth]{images/results/bar.png}
     \end{subfigure}
     \caption{Mean velocity magnitude field.}
     \label{RANS_field}
\end{figure}
In the case of the channel flow with streamwise and spanwise periodic boundaries, the Reynolds-averaged equations would even allow a one-dimensional consideration of the problem. However, we choose a domain with dimensions $2\delta\times \delta$ in $x_1$ and $x_2$ direction, since it is more represantive for this setup. The use of stretched uniform mesh in $x_2$-direction allows a very high grid density in the vicinity of the walls with nearest-wall cell spacing of $\Delta y^+=1$.\\
The polynomial order of the Taylor-Hood elements is $Q_3-Q_2$. For the HDG version, $BDM_2$ elements with tangential finite element space of the same order and the finite space of piecewise linear discontinuous functions for the pressure.
\\[1ex]
Once a steady-state is achieved, the corresponding quantities are obtained and the results are seen below. In Figure~\ref{RANS_field}, the magnitude of the mean velocity field is shown.\\
The maintained mean streamwise velocity profiles compare relatively well with the DNS data, as shown in Figure~\ref{RANS_vel}. The agreement in the viscous sublayer is very good. Although, all models significantly suffer in the buffer layer region, since this area is of greatest difficulty to correctly model. As expected, the $K-\omega$ SST model gives the best result. Furthermore, in Figure~\ref{RANS_other} the normalized turbulent kinetic energy $K$ and the shear stress component of the modeled RST are shown. Especially, the Reynolds stress term shows a suprisingly good match with the DNS data for all model approaches. In the last figure, the viscosity ratio is given. It can be seen, that the different curves of the models considerably diverge after about $y^+\approx 100$, leading to very high ratios. In the case of the $K-\omega$ SST even higher than 50.
\\[1ex]
One remarkable notice is that the results obtained from the standard CG discretization with Taylor-Hood elements (marked as TH in the figures) and from the HDG approach do not differ at all. This is mainly due to two reasons. Firstly, the mean velocity field is still very uniform and any form of vortical motions and eddies are missing. The absence of such turbulent structures allows the standard CG method to perform quite well since the conservation error might be relatively small. Secondly, as the turbulent diffusion is comparatively much higher than the viscous one, the modeled eddy-viscosity is more than one order of magnitude higher than the kinematic viscosity as it can be seen in Figure~\ref{RANS_nut}. Therefore, the highly increased total viscosity gives (locally) an appearance of a laminar flow.
\\[1ex]
Overall, the outcome of the various RANS models has a reasonably good agreement with the DNS data and all models work very well for both types of discretization techniques.

\begin{figure}
	\centering

	 \begin{subfigure}[b]{0.9\textwidth}
         \centering
         \includegraphics[width=\textwidth]{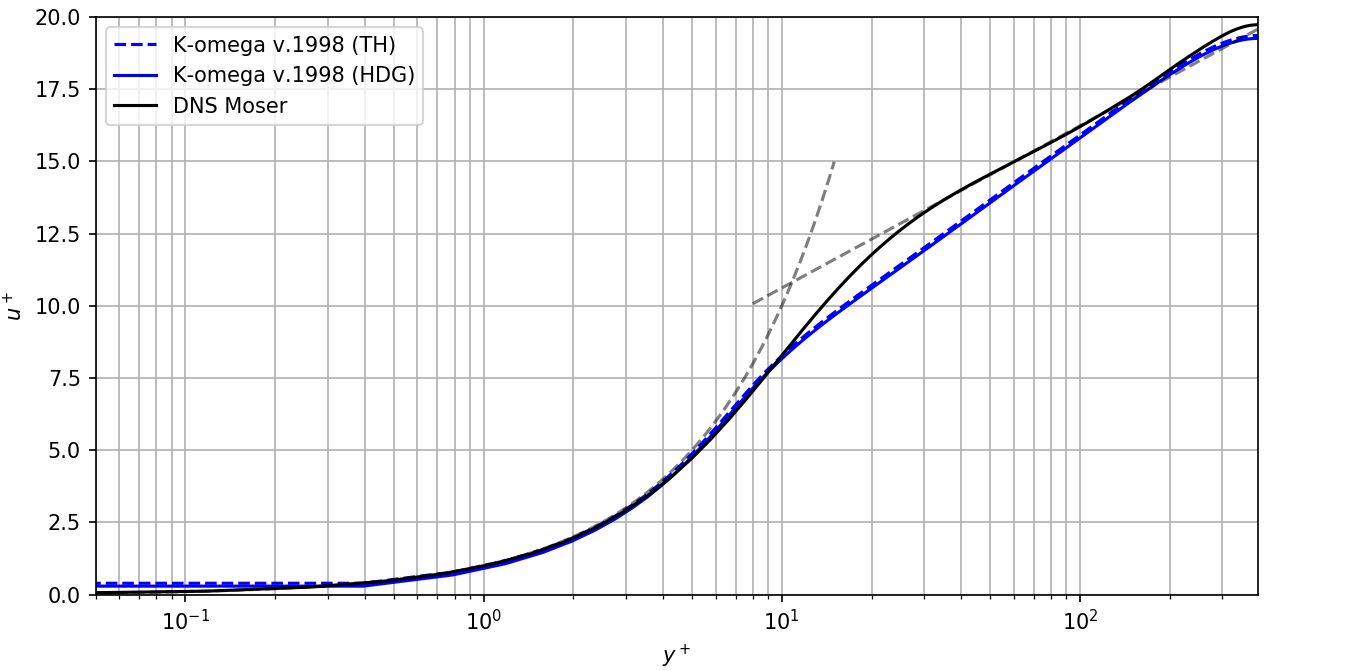}
         \caption{}
     \end{subfigure}

	 \begin{subfigure}[b]{0.9\textwidth}
         \centering
         \includegraphics[width=\textwidth]{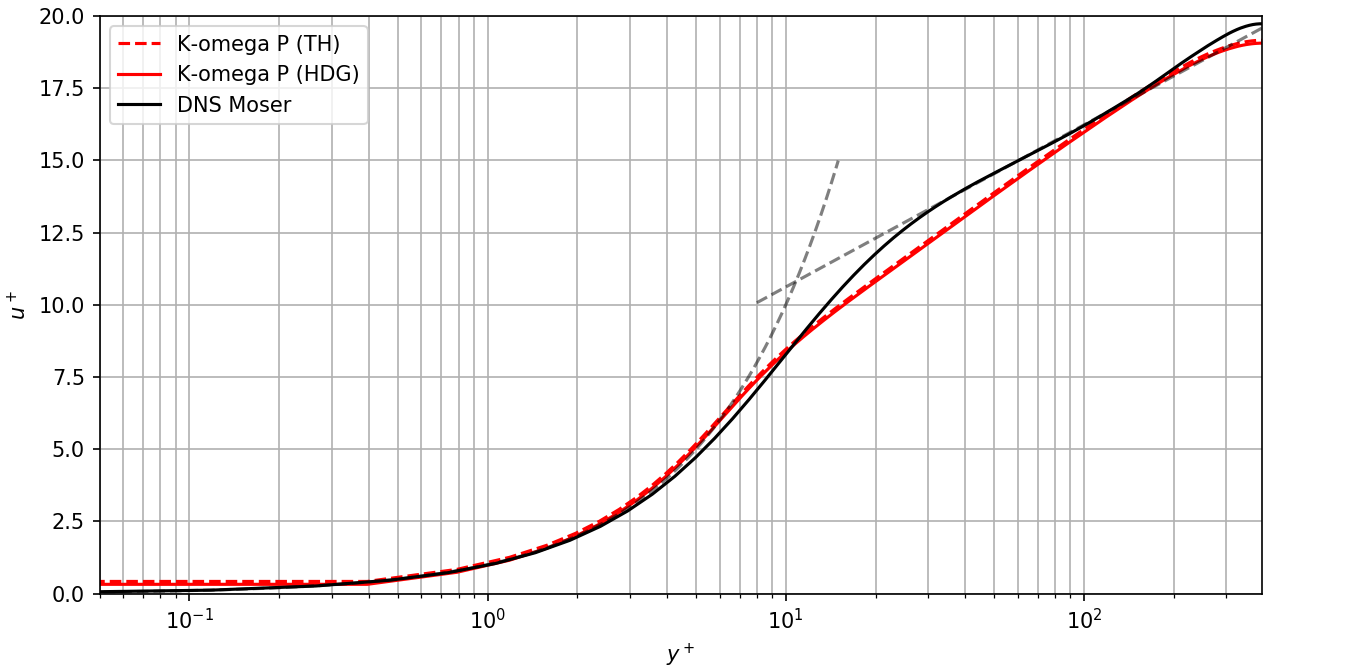}
         \caption{}
     \end{subfigure}

	 \begin{subfigure}[b]{0.9\textwidth}
         \centering
         \includegraphics[width=\textwidth]{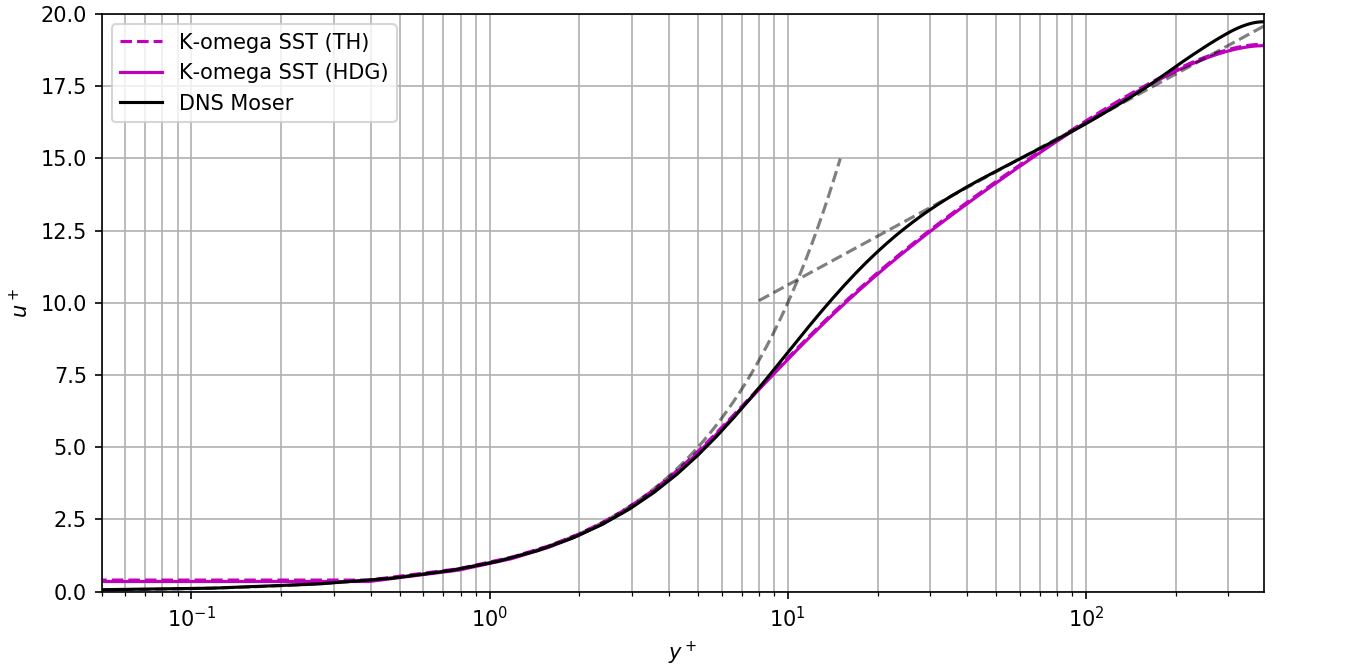}
         \caption{}
     \end{subfigure}
	 \caption{Normalized mean streamwise velocity profiles calculated with (a): $K-\omega$ v.1998, (b): $K-\omega$ Peng \cite{Peng}, (c): $K-\omega$ SST model and compared with DNS data from Moser \cite{DNSmoser}.}
	 \label{RANS_vel}
\end{figure}

\begin{figure}
	\centering

	 \begin{subfigure}[b]{0.85\textwidth}
         \centering
         \includegraphics[width=\textwidth]{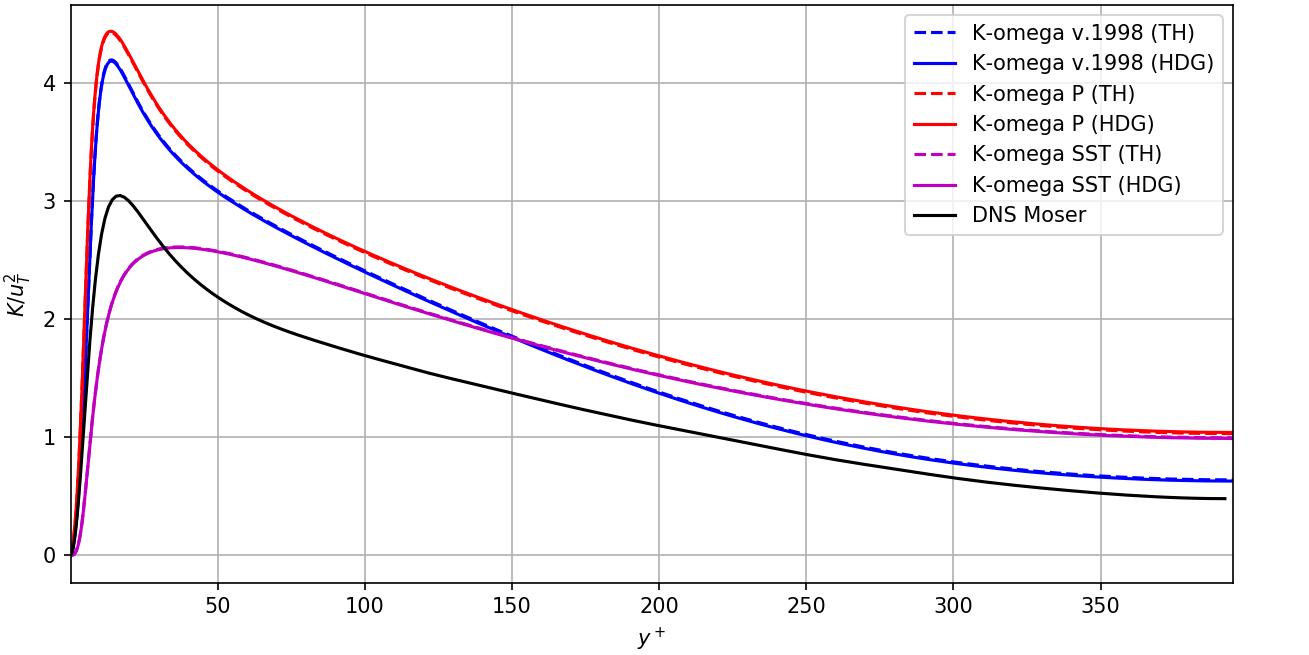}
         \caption{}
     \end{subfigure}

	 \begin{subfigure}[b]{0.9\textwidth}
         \centering
         \includegraphics[width=\textwidth]{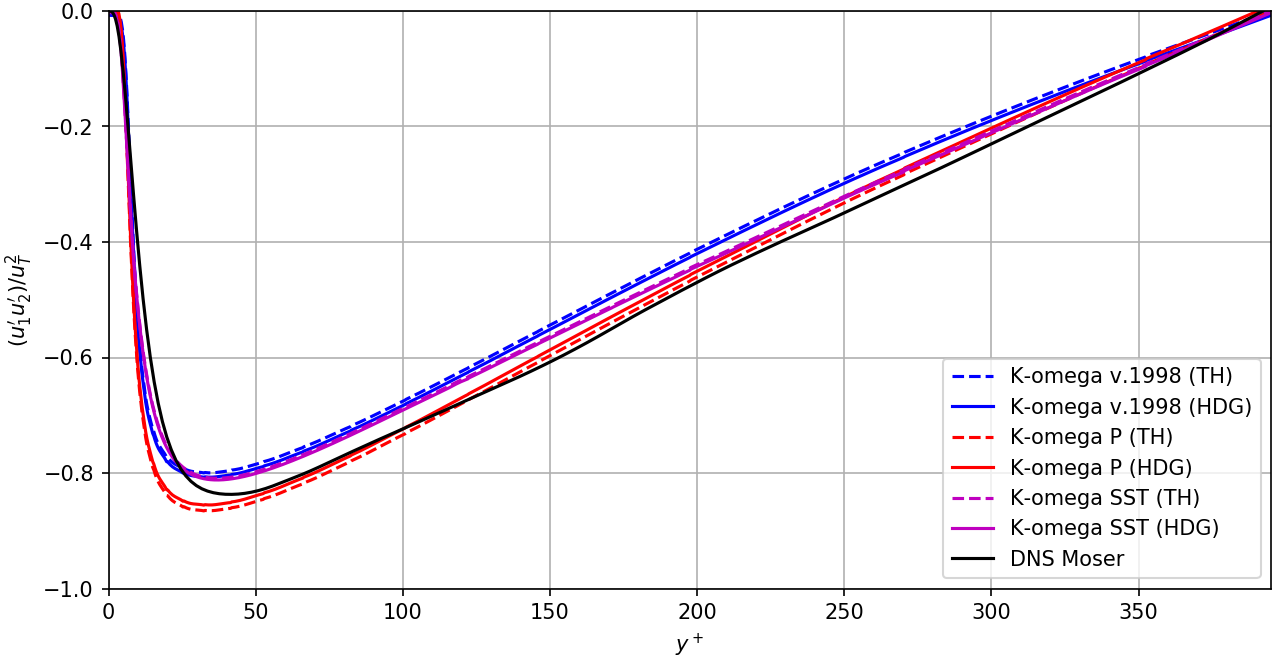}
         \caption{}
     \end{subfigure}

	 \begin{subfigure}[b]{0.85\textwidth}
         \centering
         \includegraphics[width=\textwidth]{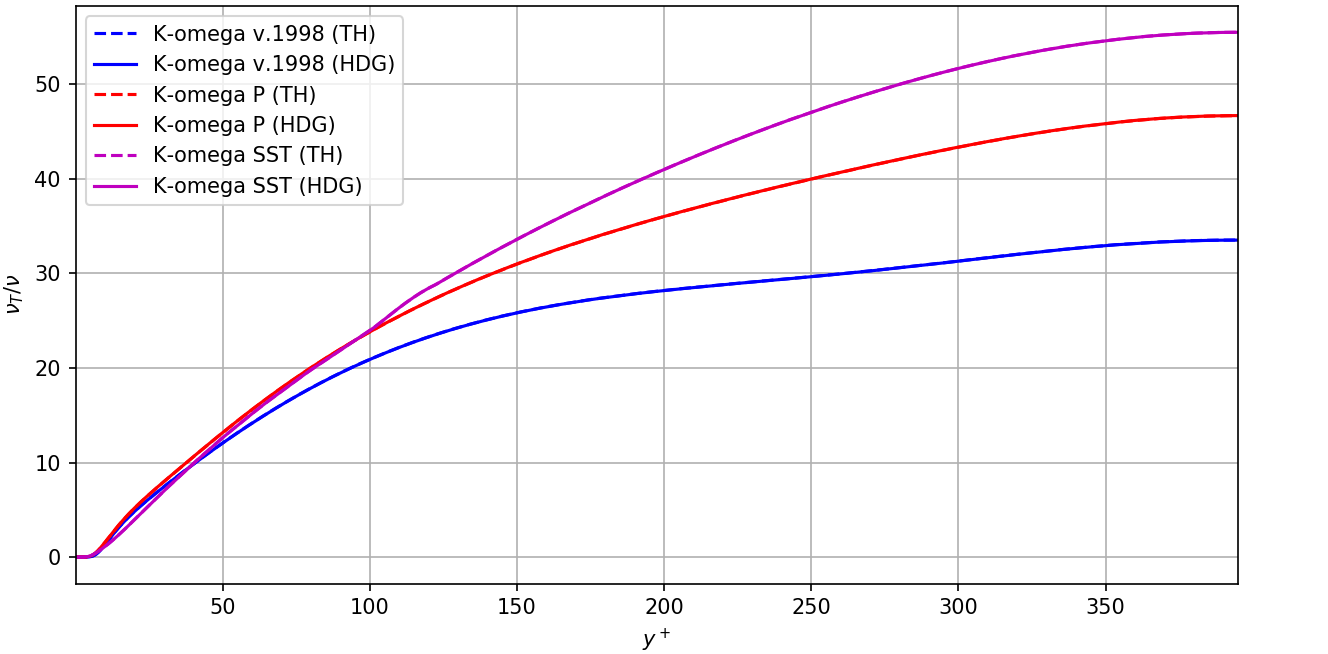}
         \caption{}
         \label{RANS_nut}
     \end{subfigure}
	 \caption{(a): Turbulent kinetic energy, (b): shear stress component of the modeled RST and (c): eddy-viscosity, normalized by the squared friction velocity or kinematic viscosity, calculated with RANS turbulence models and compared with DNS data from Moser \cite{DNSmoser}.}
	 \label{RANS_other}
\end{figure}

\clearpage
\section{DNS}
Within the scope of this thesis, a fully-resolved DNS of the turbulent channel case for $Re_{\tau}=395$ would have exceeded the computational effort. Nevertheless, a so called quasi DNS or implict LES of sufficiently well resolution gives surprisingly good results and therefore the outcome is discussed in this work.
\\[1ex]
As initially expected, a DNS calculation with the $H^1$-conforming method (without any additional stabilization technique) early becomes very unstable and obtaining a stable steady-state solution is not possible. The forming of highly three-dimensional vortical structures leads to a strong mixing of the flow quantities. The property of only discrete divergence-free velocity is not sufficient in such highly turbulent diffusive processes. It is observed, that the divergence of the velocity is far from zero, reaching one order of magnitude at maxmimum. From a physical point of view, it seems that the conservation of momentum is kind of "lost" because of the insufficiently satisfied incompressibility constraint.\\
However, the HDG discretization method works very well for DNS of the channel flow and its outcome is shown in the next figures. Again, a stretched uniform hexahedral mesh with $BDM_2$ elements is used and the dimensions of the grid are $20\times 40 \times 15$. The wall-adjacent cell height is $\Delta y^+=2$.
\\[1ex]
By the time it reaches a statistical steady-state of the fully turbulent flow, time and spatial averaging is done and the following mean streamwise velocity, as seen in Figure~\ref{DNS_vel}, is obtained. The velocity profile agrees very well with the DNS benchmark within all regions of the turbulent boundary layer.\\
In Figure~\ref{DNS_stresses}, each component of the RST is given, which compares relatively well with the reference data. The normal Reynolds stresses $v'v'$ and $w'w'$ give slightly overpredicted values in the whole $y^+$ range. This effect is consistent with published results for under-resolved meshes, and is primarily due to excessive resolved scale motion. Especially, the shear stress component shows good conformity. The averaged trace of the RST is seen via the turbulent kinetic energy in Figure~\ref{DNS_k}.
\\[1ex]
By looking at the energy spectrum of the resolved fluctuations in Figure~\ref{DNS_E}, a better idea of the effect of the resolution of the turbulent scales on the frequency range can be observed. The plots are a one-dimensional spectral representation of the turbulent kinetic energy at different planes in the flow, corresponding to its $y^+$ values. The dark dotted line is the $k^{-\frac{5}{3}}$ power curve, which corresponds to the gradient of the inertial range by Kolmogorov hypothesis, while the light dotted line represents the $k^{-1}$ curve associated with the inverse energy cascade.\\
Generally, it is clear that most of the turbulent energy is expressed as lower frequency eddies. Up to $k<10^1$, the obtained energy spectrum from the quasi DNS is in good agreement with the benchmark spectra for almost all positions. None of the profiles exhibits a well developed inertial range, mostly because the region where isotropic turbulence dominates is fairly small in a channel flow. In the streamwise direction, however, all different $y^+$ positions tend toward the predicted $k^{-1}$ slope. As seen in all plots, after about $k>10^1$ a significant drop in the energy spectrum is observed. The fact that the coarser mesh does not reproduce the higher frequency eddies very well is in line with the drop in the corresponding curve. In spanwise direction, the drop seems less pronounced as in the streamwise direction, meaning that the turbulent structures are slightly better resolved there.

\begin{figure}
	\centering

	 \begin{subfigure}[b]{0.83\textwidth}
         \centering
         \includegraphics[width=\textwidth]{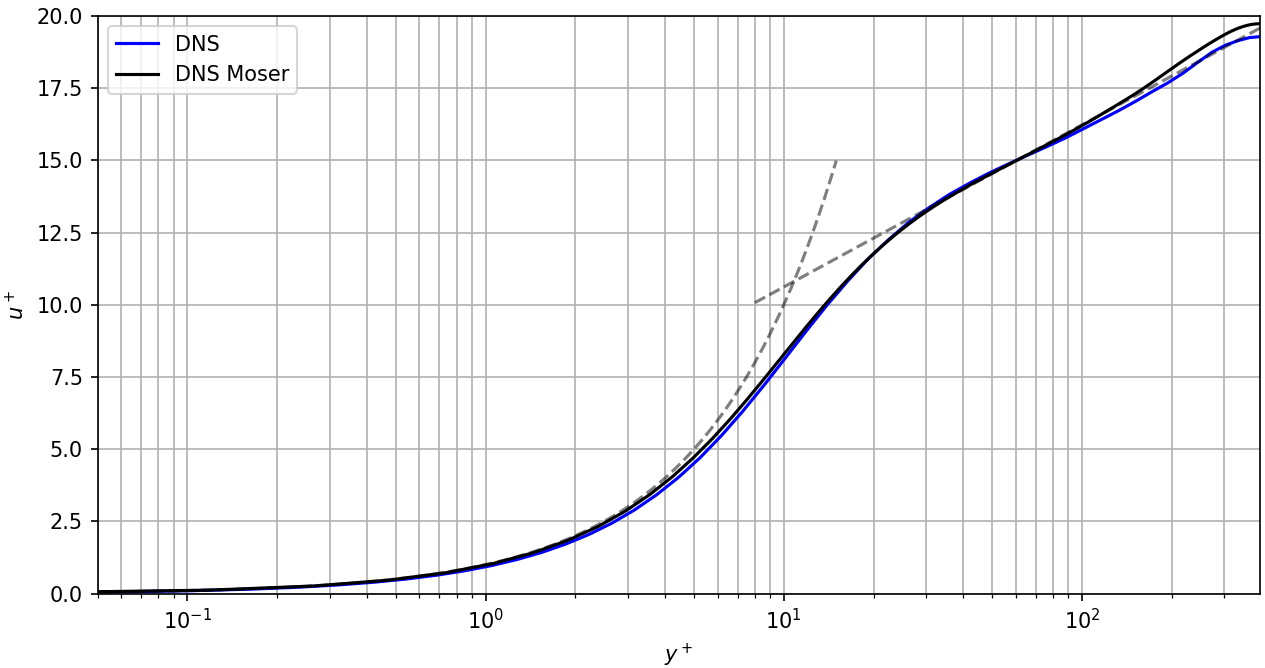}
         \caption{}
         \label{DNS_vel}
     \end{subfigure}

	 \begin{subfigure}[b]{0.83\textwidth}
         \centering
         \includegraphics[width=\textwidth]{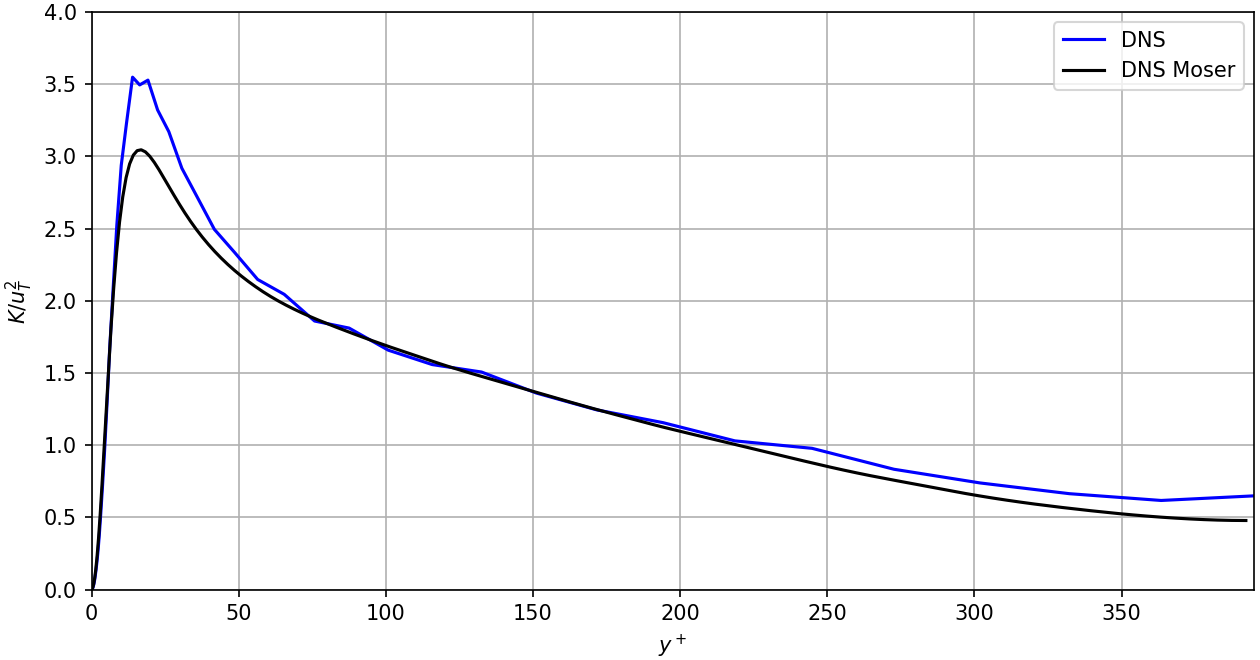}
         \caption{}
         \label{DNS_k}
     \end{subfigure}

	 \begin{subfigure}[b]{0.83\textwidth}
         \centering
         \includegraphics[width=\textwidth]{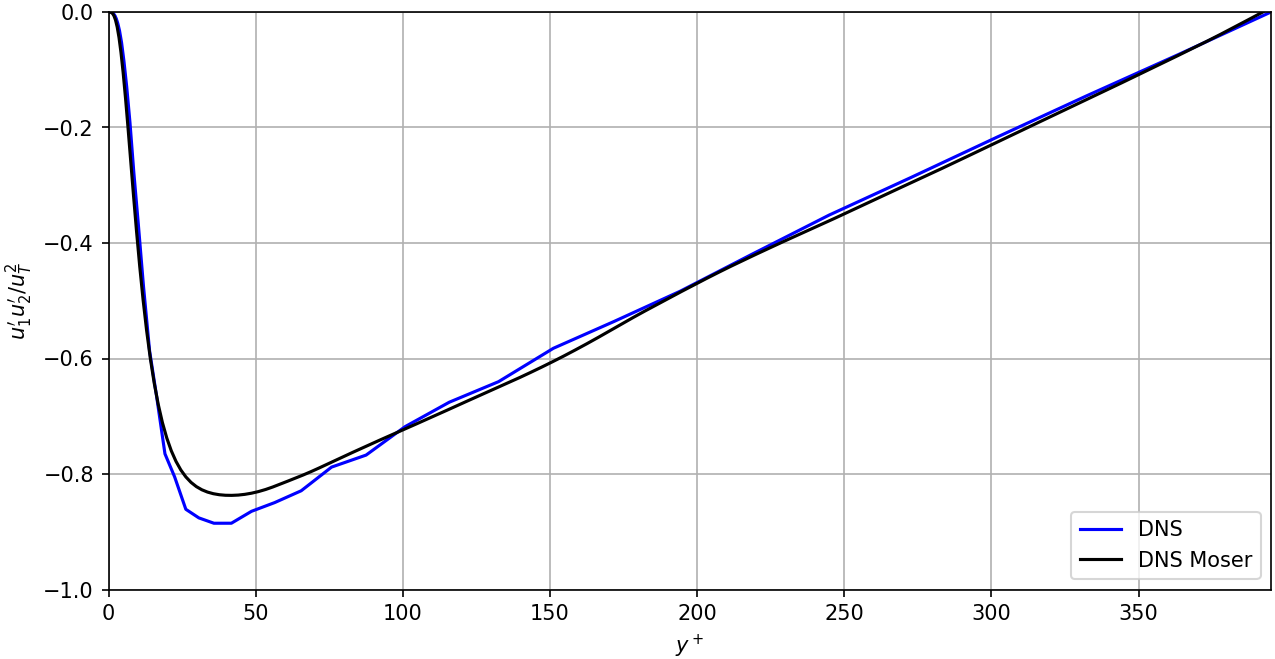}
         \caption{}
         \label{DNS_uv}
     \end{subfigure}

	 \caption{(a): Mean streamwise velocity, (b): turbulent kinetic energy and (c): shear stress component of the RST, normalized by the squared friction velocity, calculated with DNS and compared with DNS data from Moser \cite{DNSmoser}.}
\end{figure}

\begin{figure}
	\centering

	 \begin{subfigure}[b]{0.49\textwidth}
         \centering
         \includegraphics[width=\textwidth]{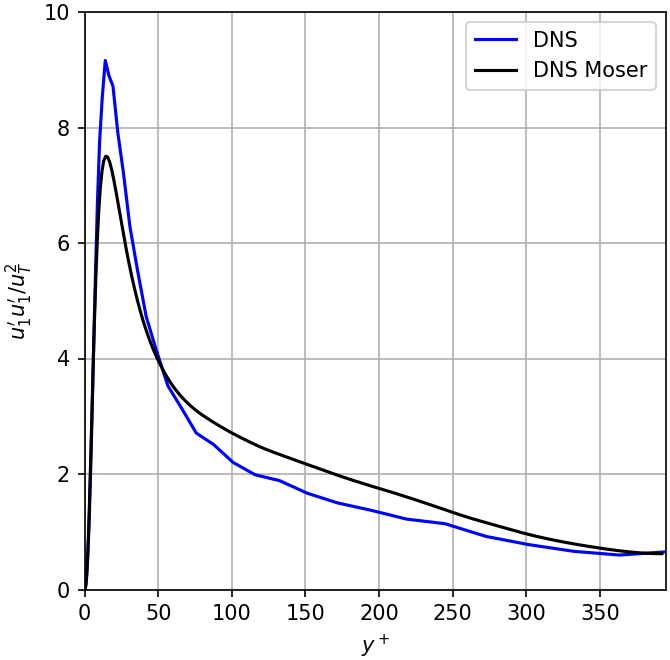}
         \caption{}
     \end{subfigure}
     \hfill
	 \begin{subfigure}[b]{0.49\textwidth}
         \centering
         \includegraphics[width=\textwidth]{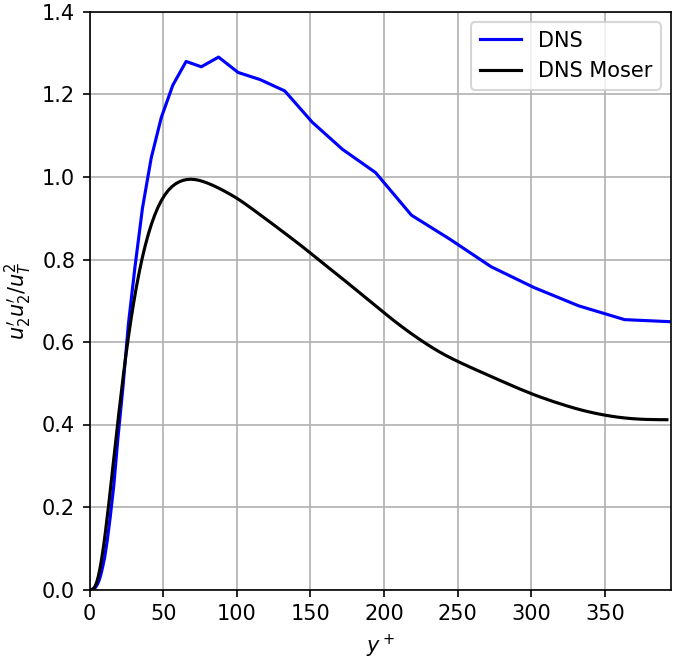}
         \caption{}
     \end{subfigure}

	 \begin{subfigure}[b]{0.49\textwidth}
         \centering
         \includegraphics[width=\textwidth]{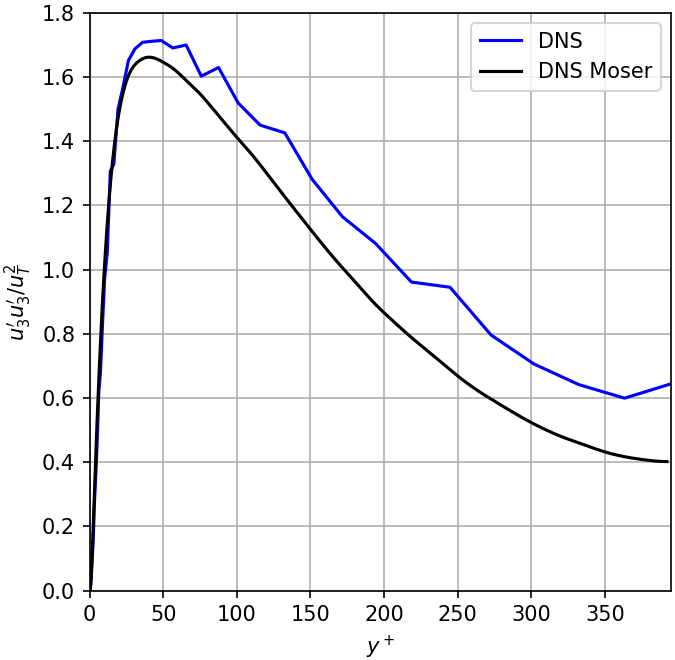}
         \caption{}
     \end{subfigure}

	 \caption{The different diagonal components of the RST normalized by the squared friction velocity, calculated with DNS and compared with DNS data from Moser \cite{DNSmoser}.}
	 \label{DNS_stresses}
\end{figure}

\begin{figure}
	\centering

	 \begin{subfigure}{0.44\textwidth}
         \centering
         \includegraphics[width=\textwidth]{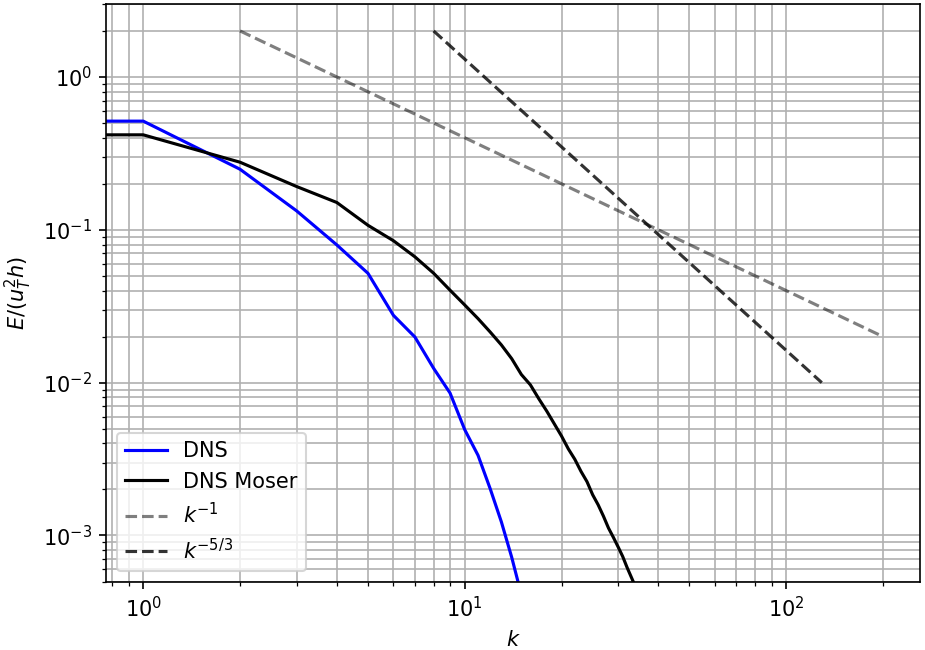}
         \caption{$y^+=5$}
     \end{subfigure}
     \hfill
	 \begin{subfigure}{0.44\textwidth}
         \centering
         \includegraphics[width=\textwidth]{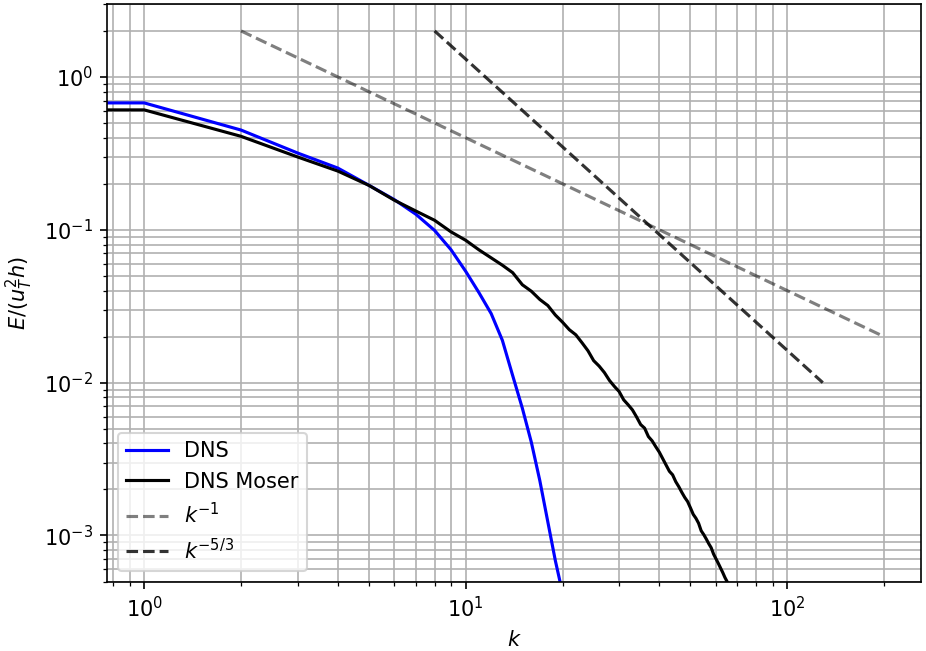}
         \caption{$y^+=40$}
     \end{subfigure}
     \medbreak
	 \begin{subfigure}{0.44\textwidth}
	     \centering
         \includegraphics[width=\textwidth]{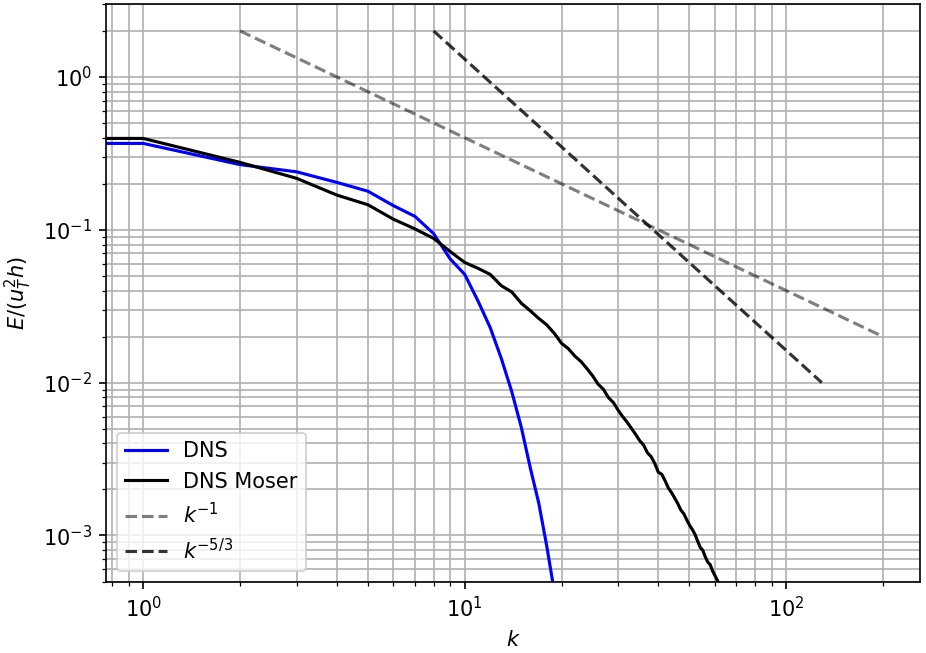}
         \caption{$y^+=100$}
     \end{subfigure}

	 \begin{subfigure}{0.44\textwidth}
         \centering
         \includegraphics[width=\textwidth]{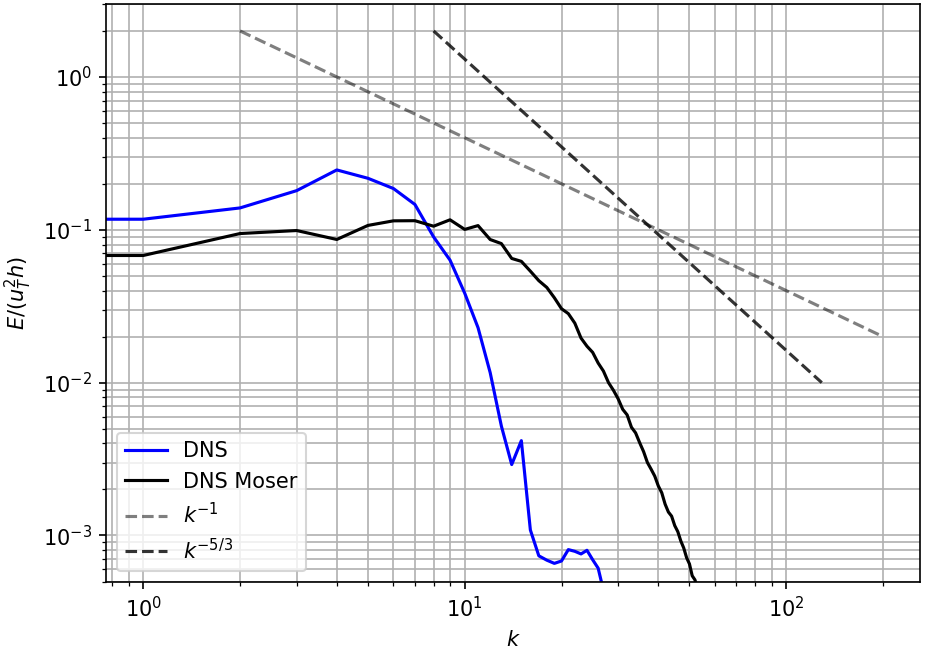}
         \caption{$y^+=5$}
     \end{subfigure}
     \hfill
	 \begin{subfigure}{0.44\textwidth}
         \centering
         \includegraphics[width=\textwidth]{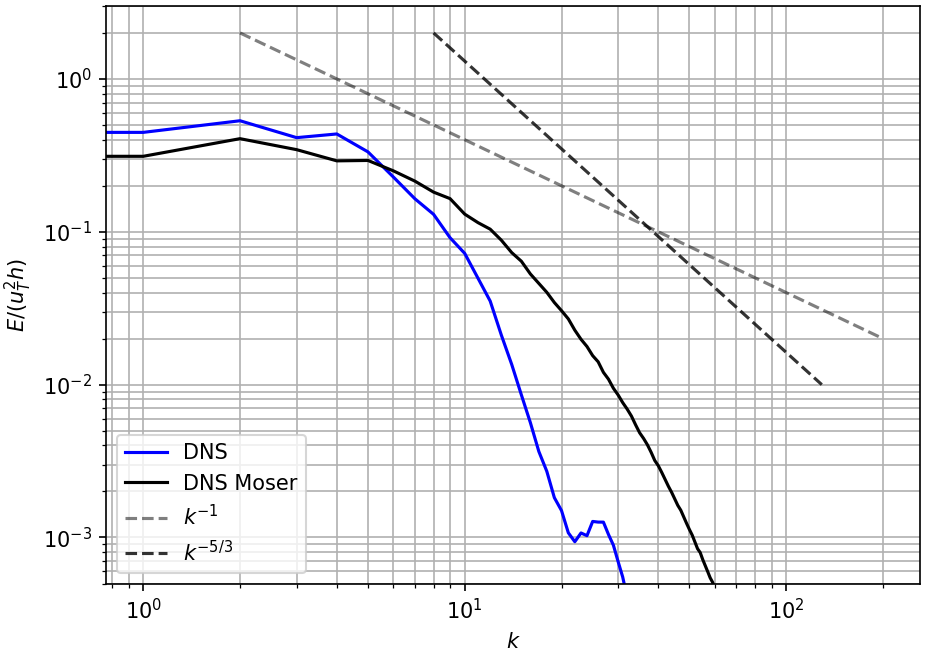}
         \caption{$y^+=40$}
     \end{subfigure}

	 \begin{subfigure}{0.44\textwidth}
         \centering
         \includegraphics[width=\textwidth]{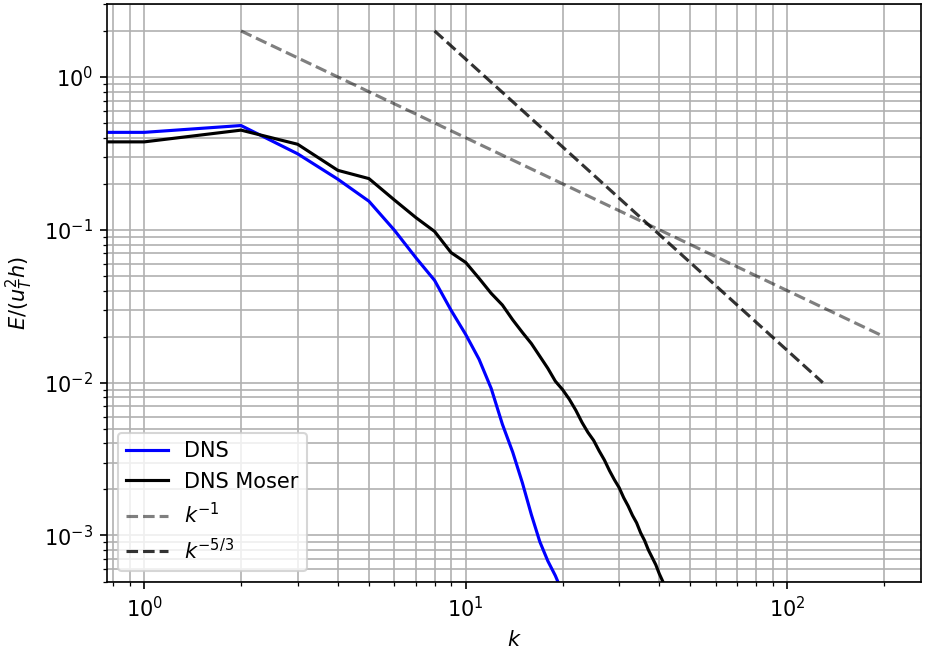}
         \caption{$y^+=100$}
     \end{subfigure}
	 \caption{(a), (b), (c): Total energy spectrum in streamwise direction, (d), (e), (f): total energy spectrum in spanwise direction, normalized by the squared friction velocity and channel half width, calculated with DNS and compared with DNS data from Moser \cite{DNSmoser}.}
	 \label{DNS_E}
\end{figure}

\clearpage
\section{LES}
LES computations are conducted using the Smagorinksy model from Equation~\eqref{Smagorinsky} combined with the Van Driest damping function. The rightful choice of the Smagorinksy constant $C_S$ seemed to be not trivial and different values have been suggested in literature. By using the value of $C_S=0.1$, the filtered flow appears to be very diffusive and strongly damping any fluctuations. Therefore, the parameter is decreased to $C_S=0.05$ and applied to all LES and VMS simulations.
\\[1ex]
Since we still resolve large eddies and coherent structures in LES, the use of the standard CG discretization method with Taylor-Hood elements again shows very bad results and numerical instabilities are highly likely to occur. Thus, no steady-state in statistical sense was achieved and any results are omitted here. The $H(\mathrm{div})$-conforming HDG method performes very well for LES.
\\[1ex]
Compared to the DNS case from the previous section, the same type of grid and finite elments of same order are used but the mesh resolution is reduced to $10\times 24\times 10$ with a nearest-wall cell spacing of $\Delta y^+=4$. This mesh resolution is fairly coarse even for a LES, but despite that the final results are reasonably fine.
\\[1ex]
The mean velocity profile given in Figure~\ref{LES_vel} deviates slightly in the viscous and logarithmic layer region with the DNS benchmark. The reasons are mainly that the excessive turbulent eddy-viscosity damps the near-wall eddies and the coarse near-wall resolution is incapable of carrying the fine turbulence producing features.\\
The turbulent kinetic energy produced by LES (Figure~\ref{LES_k}) overestimates in the region of the viscous and buffer layer and underrates in higher regions. This outcome verifies the small deviations in the streamwise mean flow. The total Reynolds shear stress consists of the sum of the shear stress component of the resolved RST and averaged modeled SGS tensor as given in Figure~\ref{LES_nut}. It can be clearly seen that the resolved part makes up the largest percentage of the total Reynolds stress. Further on, each diagonal component of the resolved RST is shown in Figure~\ref{LES_stresses}. The low values of the $v'v'$ and $w'w'$ components are mainly due to the excessive damping and shortcomings of assuming isotropy turbulence of the SGS turbulence model.
\\[1ex]
As it was expected, the energy spectrum of the coarser LES mesh resolves less vortical structures than the DNS from the previous section. The cut-off wave number is set to a lower value therefore and the drop in the curve starts at smaller wave numbers. This is consistent with the obtained spectral representation.

\begin{figure}
	\centering

	 \begin{subfigure}[b]{0.83\textwidth}
         \centering
         \includegraphics[width=\textwidth]{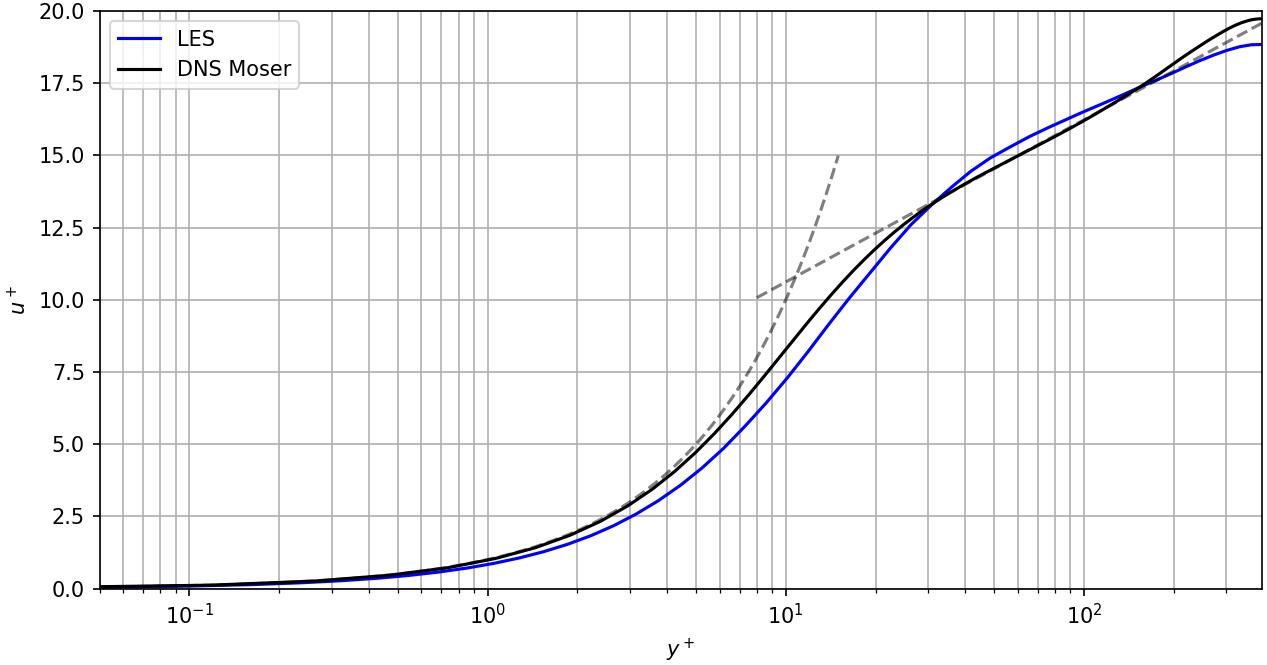}
         \caption{}
         \label{LES_vel}
     \end{subfigure}

	 \begin{subfigure}[b]{0.83\textwidth}
         \centering
         \includegraphics[width=\textwidth]{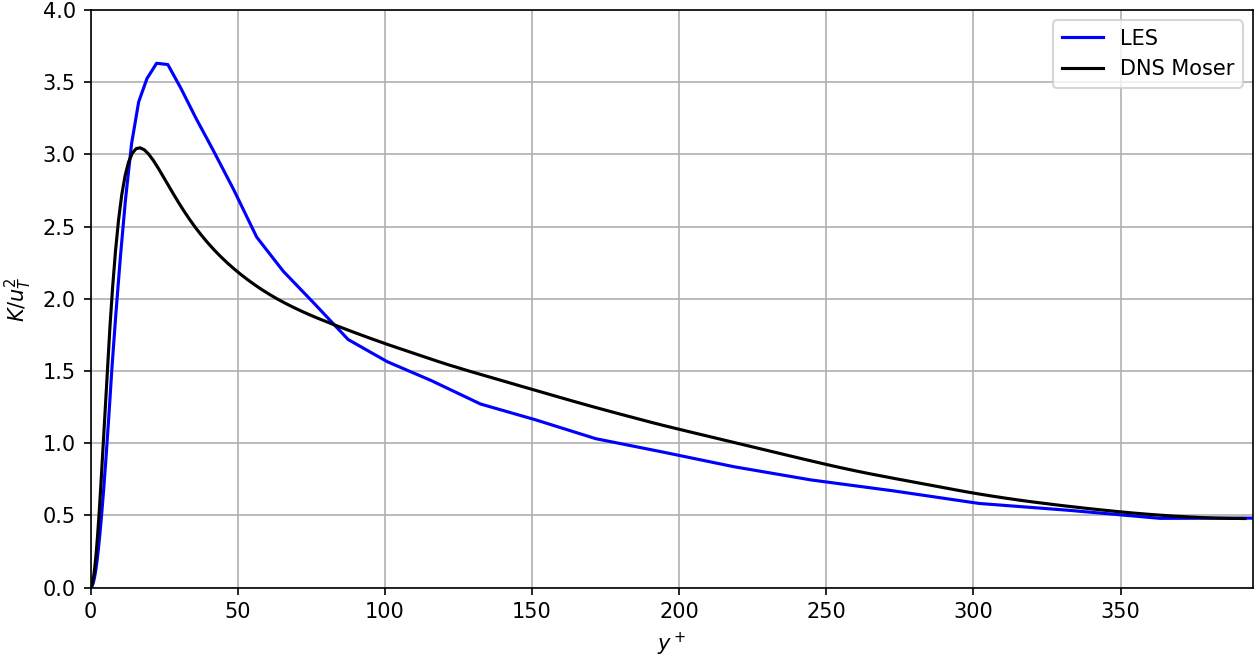}
         \caption{}
         \label{LES_k}
     \end{subfigure}

	 \begin{subfigure}[b]{0.83\textwidth}
         \centering
         \includegraphics[width=\textwidth]{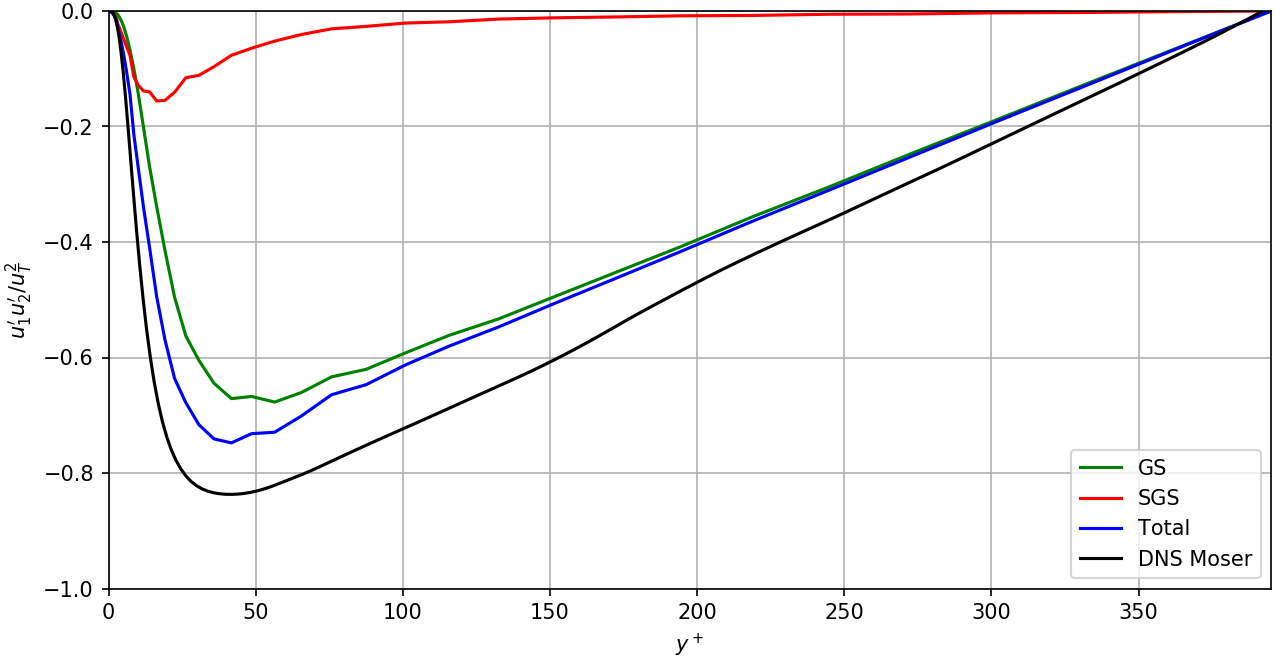}
         \caption{}
         \label{LES_nut}
     \end{subfigure}

	 \caption{(a): Mean streamwise velocity, (b): turbulent kinetic energy and (c): resolved and modeled shear stress components, normalized by the squared friction velocity, calculated with LES and compared with DNS data from Moser \cite{DNSmoser}.}
\end{figure}

\begin{figure}
	\centering

	 \begin{subfigure}[b]{0.49\textwidth}
         \centering
         \includegraphics[width=\textwidth]{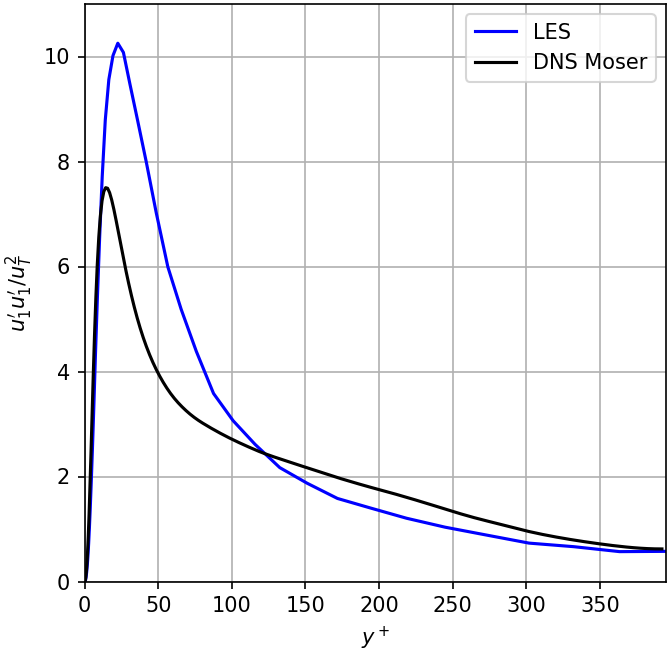}
         \caption{}
     \end{subfigure}
     \hfill
	 \begin{subfigure}[b]{0.49\textwidth}
         \centering
         \includegraphics[width=\textwidth]{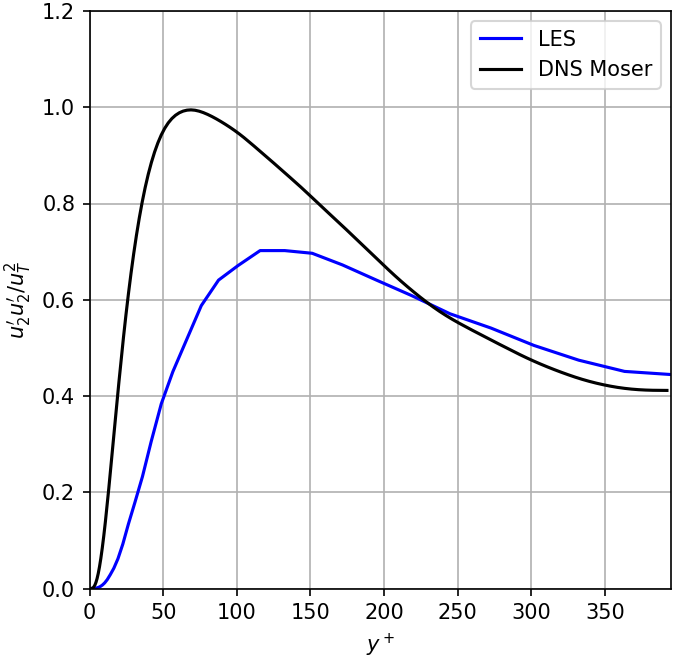}
         \caption{}
     \end{subfigure}

	 \begin{subfigure}[b]{0.49\textwidth}
         \centering
         \includegraphics[width=\textwidth]{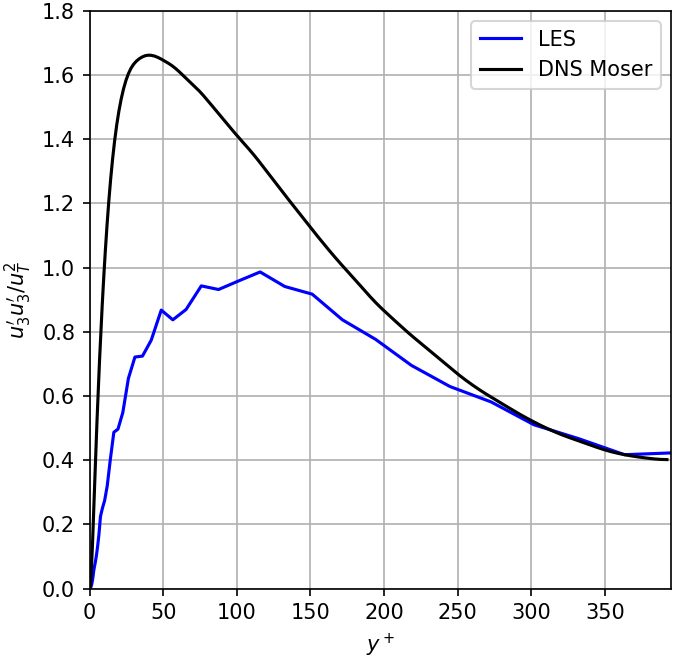}
         \caption{}
     \end{subfigure}

	 \caption{The different diagonal components of the RST normalized by the squared friction velocity, calculated with LES and compared with DNS data from Moser \cite{DNSmoser}.}
	 \label{LES_stresses}
\end{figure}

\begin{figure}
	\centering

	 \begin{subfigure}{0.44\textwidth}
         \centering
         \includegraphics[width=\textwidth]{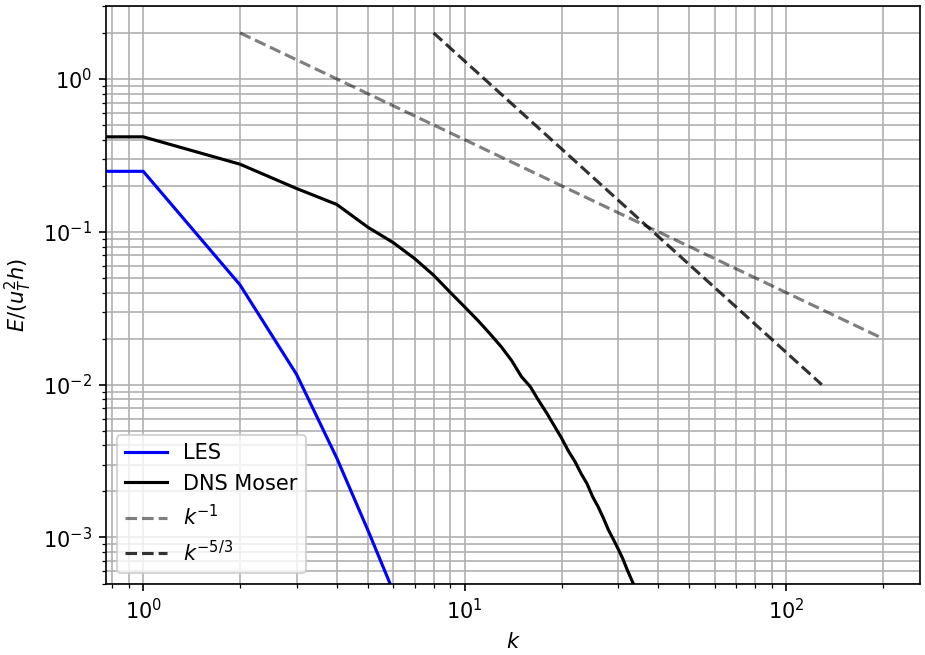}
         \caption{$y^+=5$}
     \end{subfigure}
     \hfill
	 \begin{subfigure}{0.44\textwidth}
         \centering
         \includegraphics[width=\textwidth]{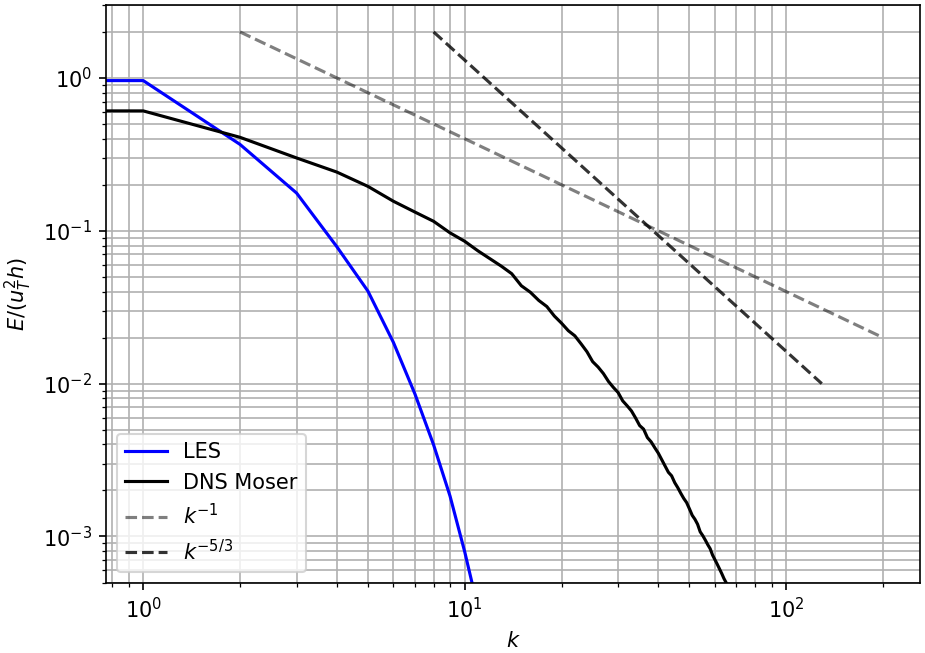}
         \caption{$y^+=40$}
     \end{subfigure}
     \medbreak
	 \begin{subfigure}{0.44\textwidth}
	     \centering
         \includegraphics[width=\textwidth]{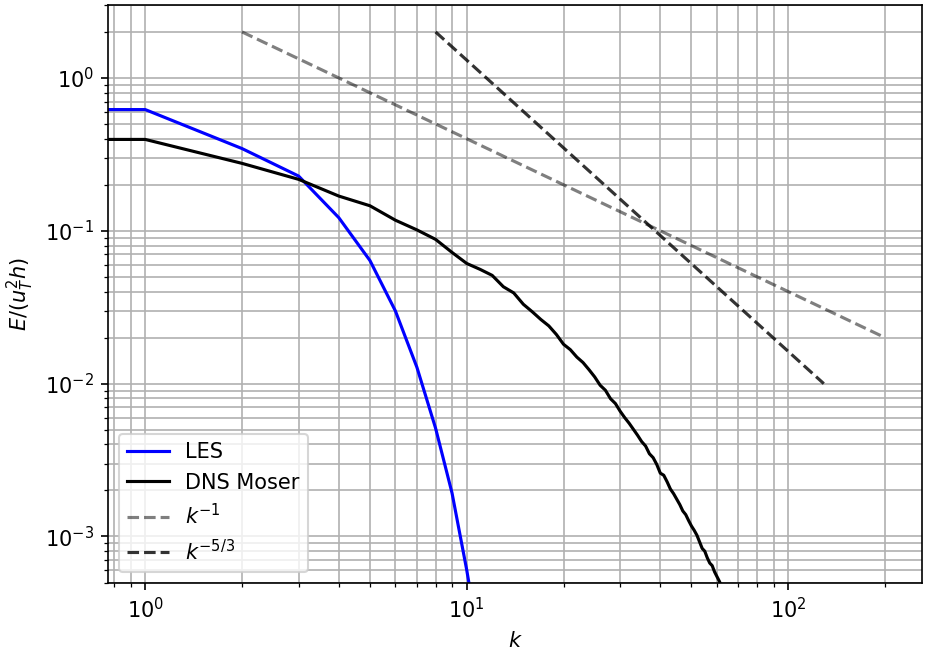}
         \caption{$y^+=100$}
     \end{subfigure}

	 \begin{subfigure}{0.44\textwidth}
         \centering
         \includegraphics[width=\textwidth]{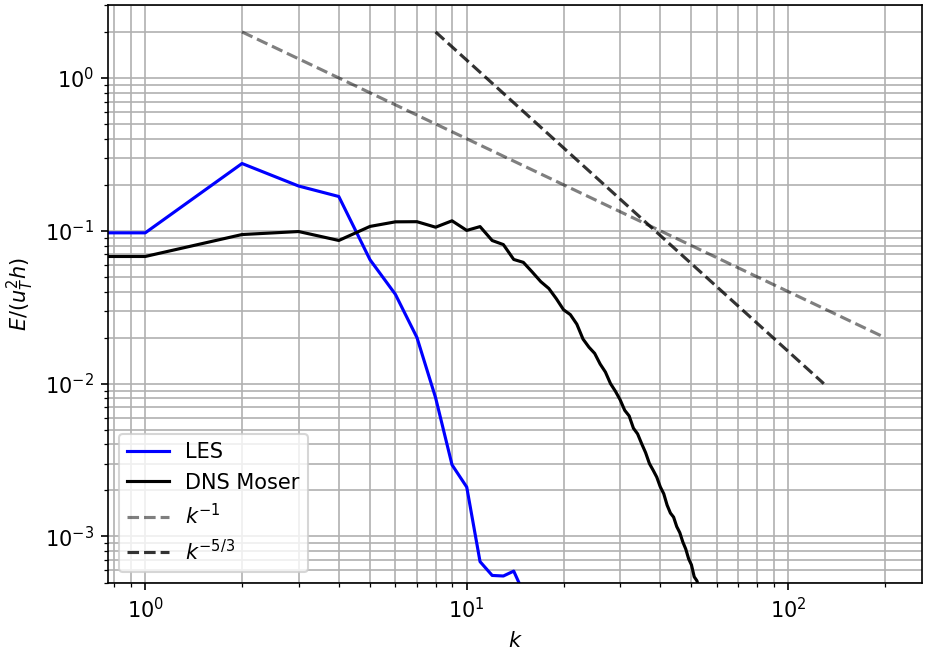}
         \caption{$y^+=5$}
     \end{subfigure}
     \hfill
	 \begin{subfigure}{0.44\textwidth}
         \centering
         \includegraphics[width=\textwidth]{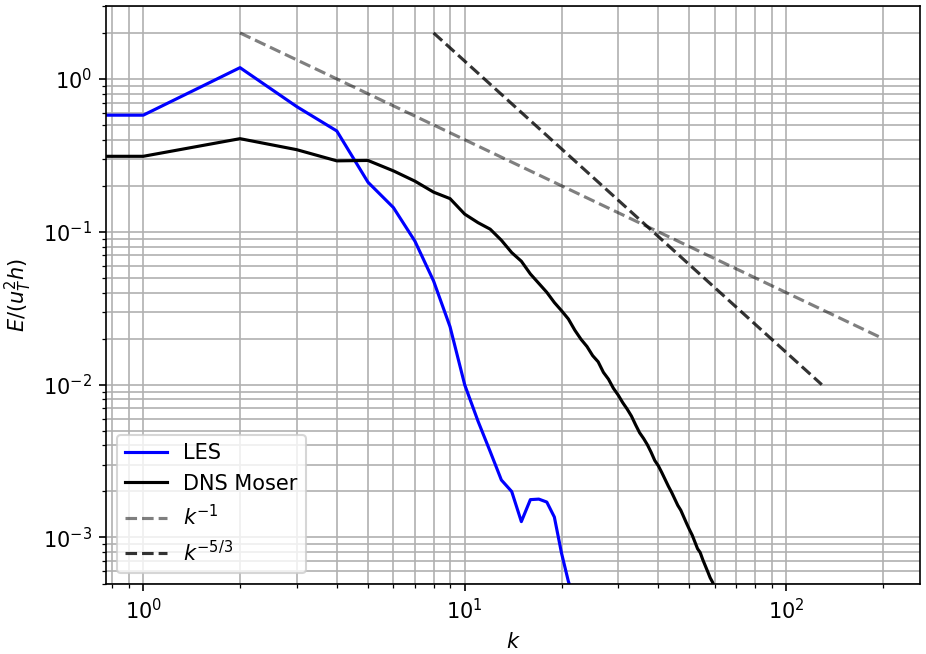}
         \caption{$y^+=40$}
     \end{subfigure}

	 \begin{subfigure}{0.44\textwidth}
         \centering
         \includegraphics[width=\textwidth]{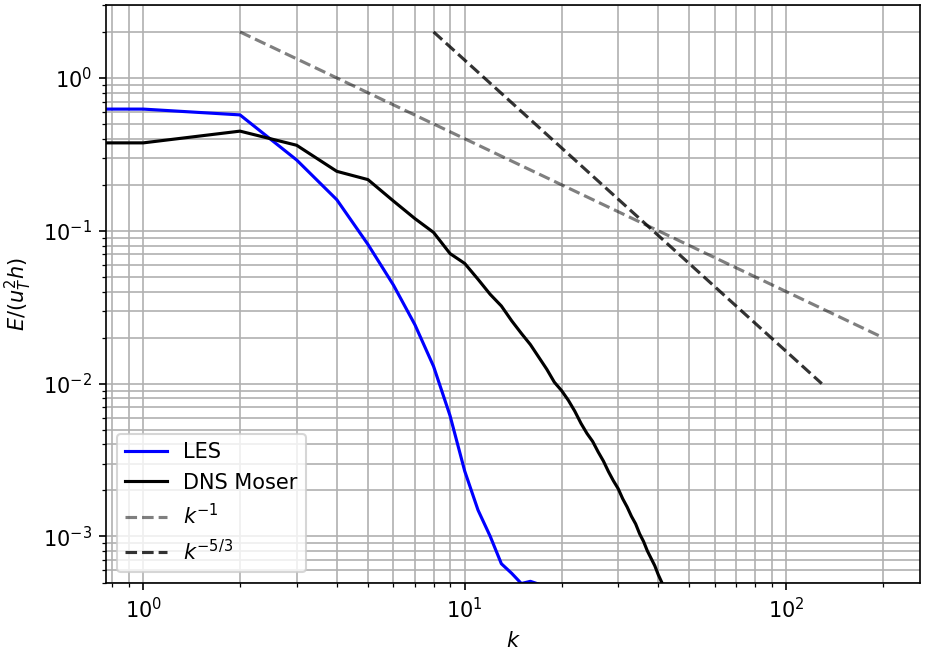}
         \caption{$y^+=100$}
     \end{subfigure}
	 \caption{(a), (b), (c): Total energy spectrum in streamwise direction, (d), (e), (f): total energy spectrum in spanwise direction, normalized by the squared friction velocity and channel half width, calculated with LES and compared with DNS data from Moser \cite{DNSmoser}.}
	 \label{LES_E}
\end{figure}

\clearpage
\section{VMS}
For the VMS method, the same mesh and eddy-viscosity model for the unresolved scales as for LES is used. The polynomial order of the $L^h$ is chosen as $p=1$ and therefore assuming less turbulent activities in the whole computational domain. In this case as well, the standard discretization method with the described finite element pairing from Section~\ref{section:TH} clearly failed to correctly predict a stable solution and the HDG approach is in all aspects ahead.
\\[1ex]
For this choice of the polynomial order for $L^h$, the VMS method performes significantly better than the LES with the same setup. The profile of the mean velocity is shown in Figure~\ref{VMS_vel} and basically agrees pretty good with the DNS reference in all regions.\\
The second order statistics also show better conformity with the benchmark than with the traditional LES method. Still, in the vicinity of the wall the $u'u'$ component of the RST is significantly overestimated as seen in all turbulence simulation principles before. A better resolution in the vicinity of the walls would improve this behavior. However, the normal stresses $v'v'$ and $w'w'$ shows astonishing good agreement with the DNS reference.\\
As we know, in VMS the large and small scales are resolved and the impact of the modeled unresolved scales only influences the small scales. The choice of the large scale deformation tensor space determines the amount of the resolved small scales among all resolved scales and therefore restricts the influence of the modeled unresolved structures. For $p=1$, the effect of the modeled scales can be observed in Figure~\ref{VMS_nut} for the Reynolds shear stress. There it can be clearly seen, that on average the impact of the unresolved to small scales is diminishing for such higher order. Although, except to the small peak in the buffer region, the total $u'v'$ stress coincides with the reference solution.
\\[1ex]
The streamwise and spanwise energy spectrum is given in Figure~\ref{VMS_E}. As obvious, the VMS method adequately resolves less scales than the quasi DNS, but shows little bit better results than the LES. The reason for that is mainly the less excessive damping of the model and therefore the better approximation of the statistics.
\\[1ex]
A comparison of the normalized mean streamwise velocity of all simulation principles is given in Figure~\ref{Comparison_vel}.

\begin{figure}
	\centering

	 \begin{subfigure}[b]{0.83\textwidth}
         \centering
         \includegraphics[width=\textwidth]{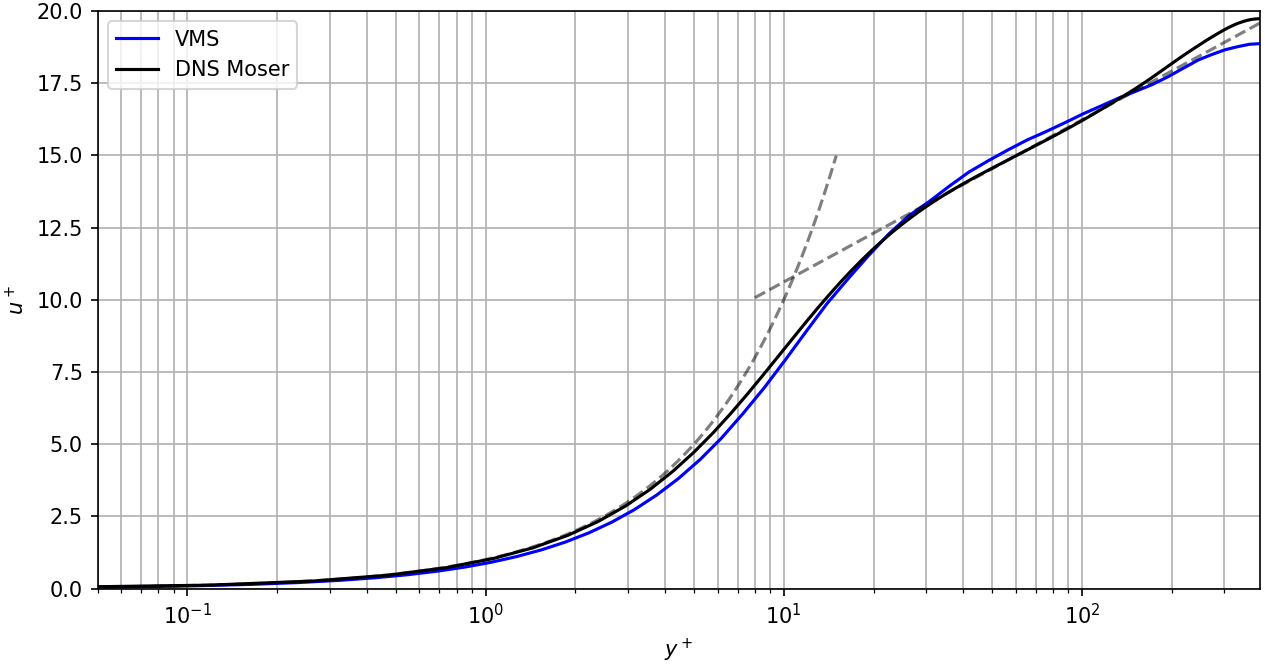}
         \caption{}
         \label{VMS_vel}
     \end{subfigure}

	 \begin{subfigure}[b]{0.83\textwidth}
         \centering
         \includegraphics[width=\textwidth]{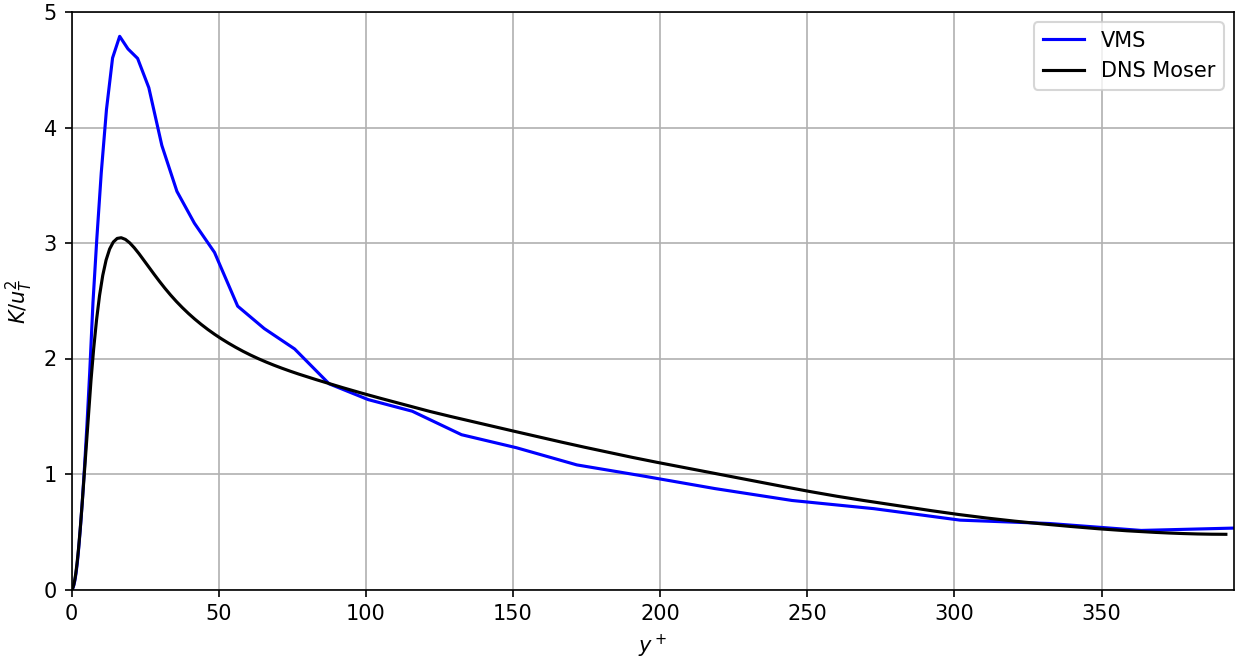}
         \caption{}
         \label{VMS_k}
     \end{subfigure}

	 \begin{subfigure}[b]{0.83\textwidth}
         \centering
         \includegraphics[width=\textwidth]{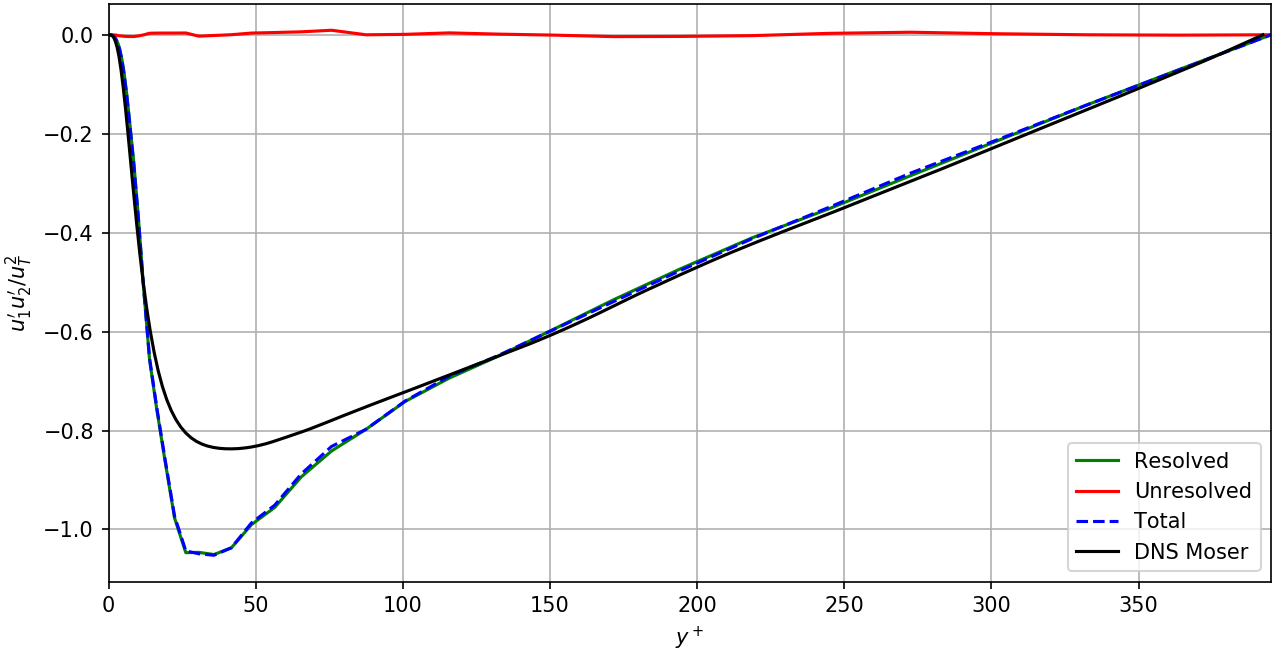}
         \caption{}
         \label{VMS_nut}
     \end{subfigure}

	 \caption{(a): Mean streamwise velocity, (b): turbulent kinetic energy and (c): resolved and modeled shear stress components, normalized by the squared friction velocity, calculated with VMS and compared with DNS data from Moser \cite{DNSmoser}.}
\end{figure}

\begin{figure}
	\centering

	 \begin{subfigure}[b]{0.49\textwidth}
         \centering
         \includegraphics[width=\textwidth]{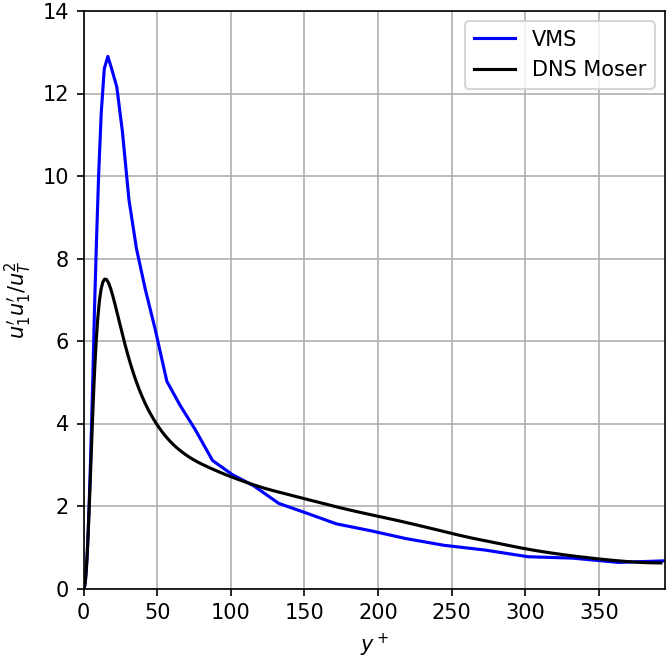}
         \caption{}
     \end{subfigure}
     \hfill
	 \begin{subfigure}[b]{0.49\textwidth}
         \centering
         \includegraphics[width=\textwidth]{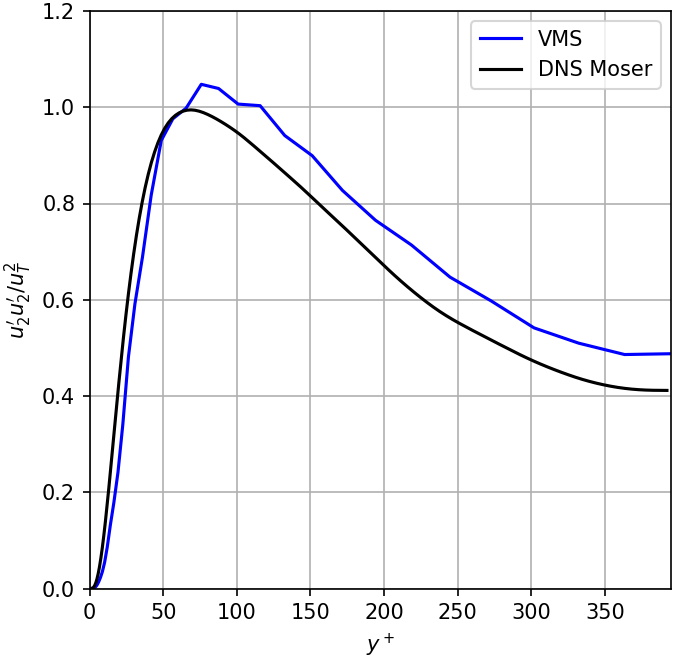}
         \caption{}
     \end{subfigure}

	 \begin{subfigure}[b]{0.49\textwidth}
         \centering
         \includegraphics[width=\textwidth]{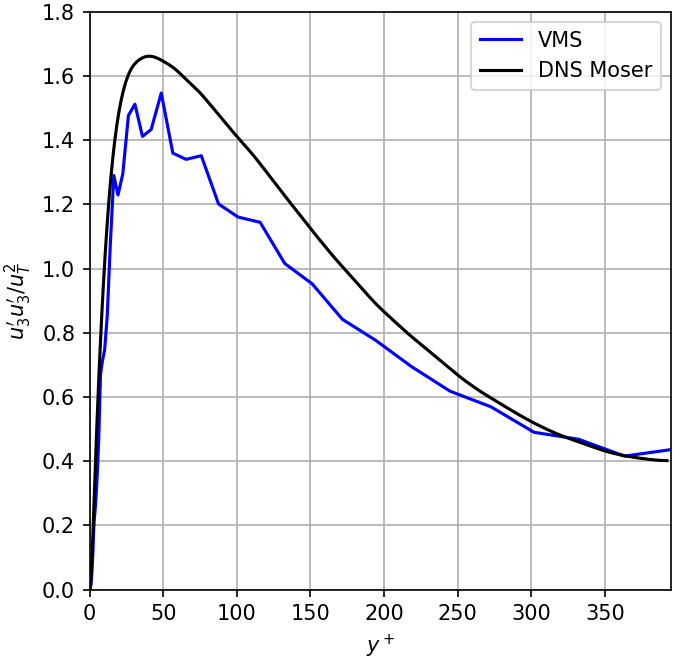}
         \caption{}
     \end{subfigure}

	 \caption{The different diagonal components of the RST normalized by the squared friction velocity, calculated with VMS and compared with DNS data from Moser \cite{DNSmoser}.}
	 \label{VMS_stresses}
\end{figure}

\begin{figure}
	\centering

	 \begin{subfigure}{0.44\textwidth}
         \centering
         \includegraphics[width=\textwidth]{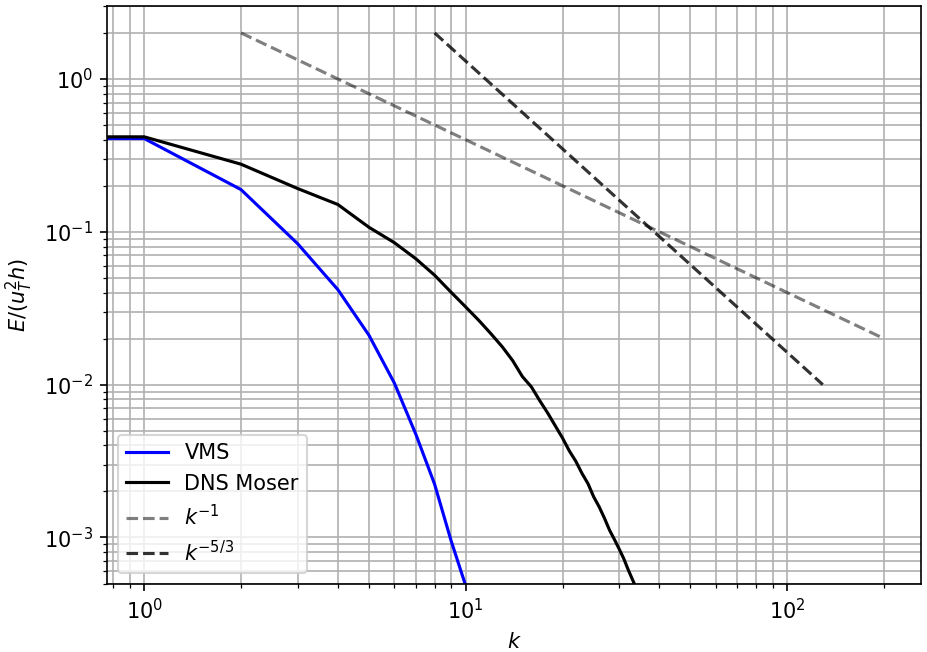}
         \caption{$y^+=5$}
     \end{subfigure}
     \hfill
	 \begin{subfigure}{0.44\textwidth}
         \centering
         \includegraphics[width=\textwidth]{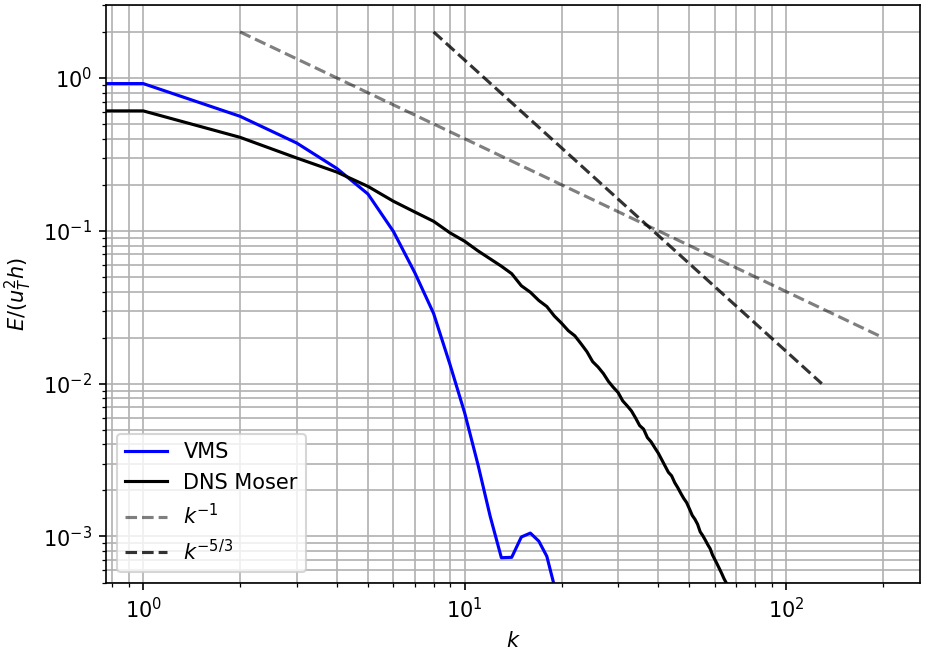}
         \caption{$y^+=40$}
     \end{subfigure}
     \medbreak
	 \begin{subfigure}{0.44\textwidth}
	     \centering
         \includegraphics[width=\textwidth]{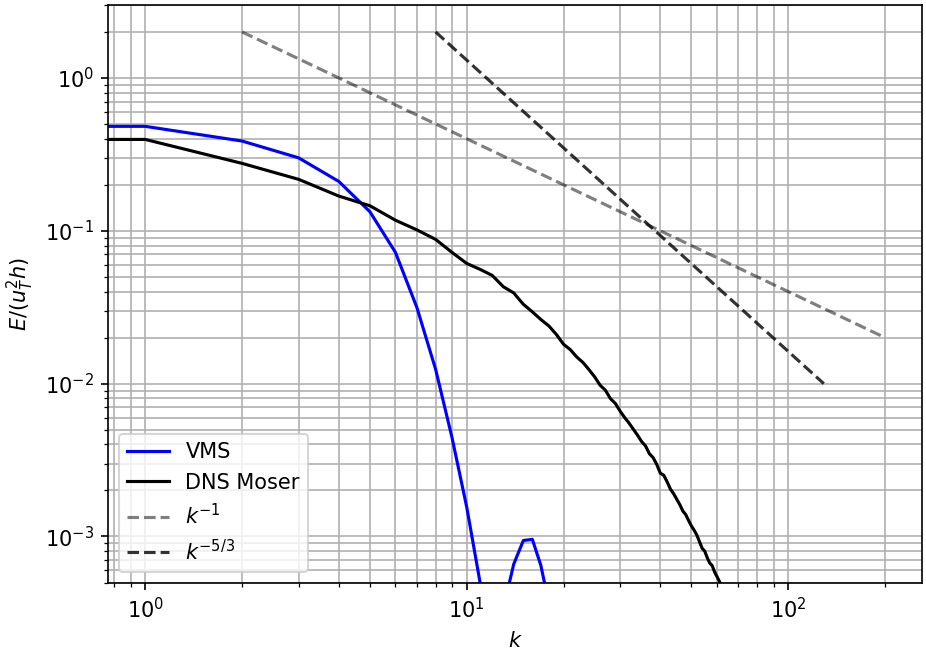}
         \caption{$y^+=100$}
     \end{subfigure}

	 \begin{subfigure}{0.44\textwidth}
         \centering
         \includegraphics[width=\textwidth]{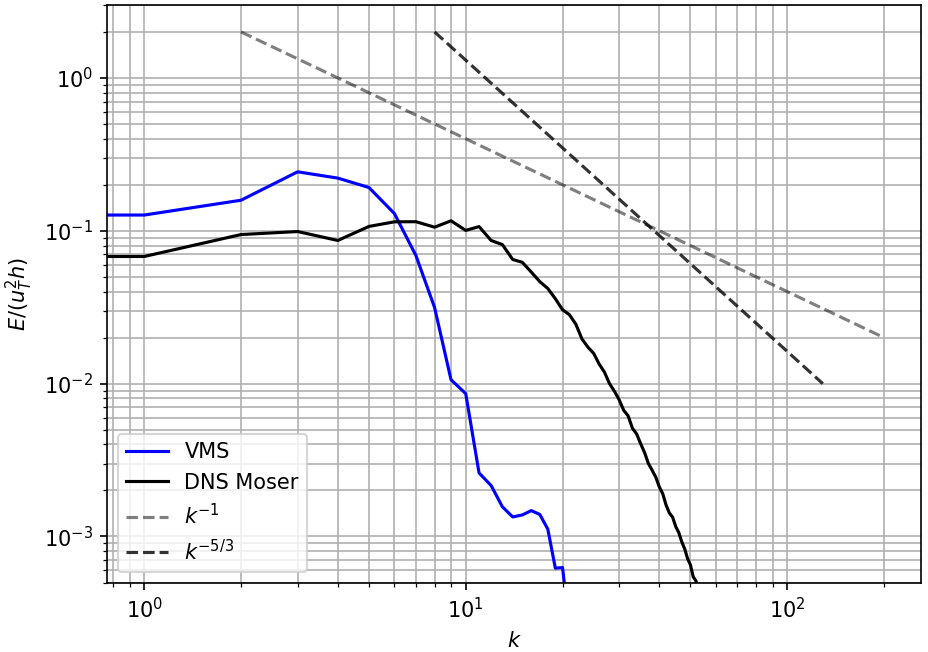}
         \caption{$y^+=5$}
     \end{subfigure}
     \hfill
	 \begin{subfigure}{0.44\textwidth}
         \centering
         \includegraphics[width=\textwidth]{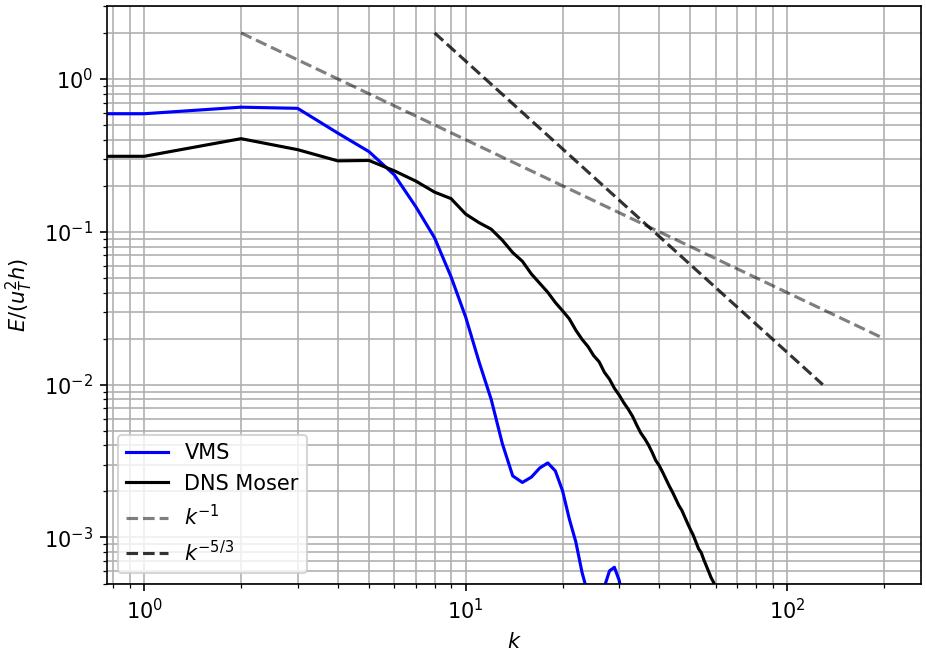}
         \caption{$y^+=40$}
     \end{subfigure}

	 \begin{subfigure}{0.44\textwidth}
         \centering
         \includegraphics[width=\textwidth]{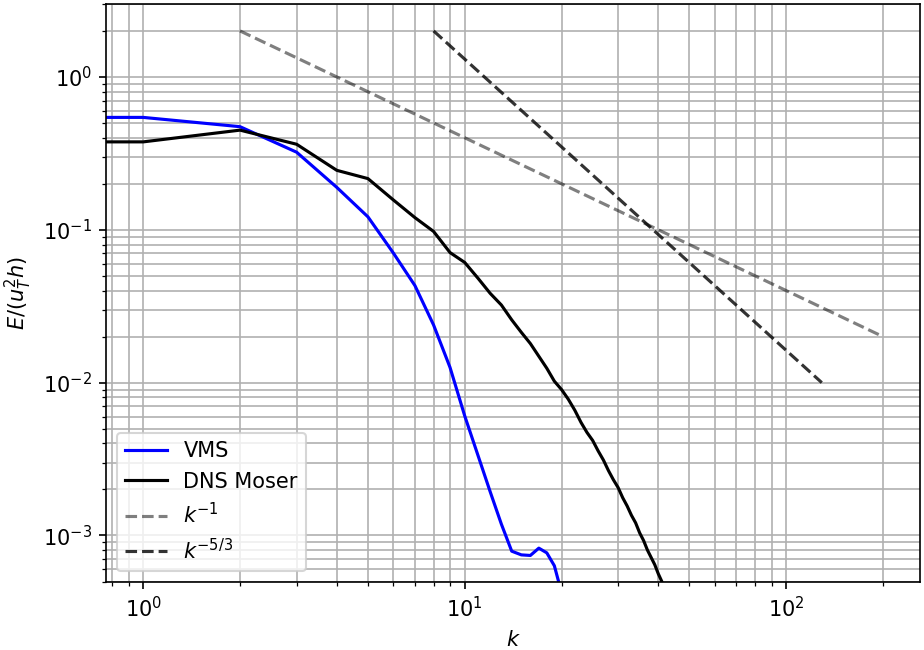}
         \caption{$y^+=100$}
     \end{subfigure}
	 \caption{(a), (b), (c): Total energy spectrum in streamwise direction, (d), (e), (f): total energy spectrum in spanwise direction, normalized by the squared friction velocity and channel half width, calculated with VMS and compared with DNS data from Moser \cite{DNSmoser}.}
	 \label{VMS_E}
\end{figure}

\begin{figure}

    \includegraphics[width=\textwidth]{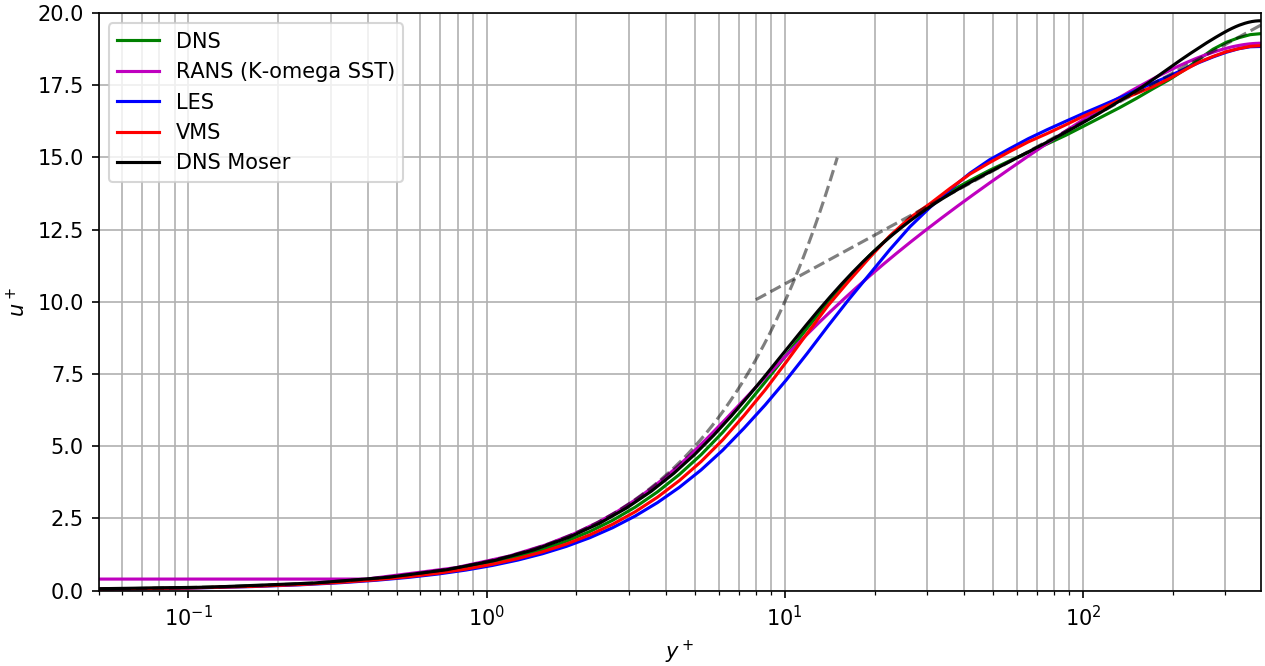}

	\caption{Comparison of the normalized mean streamwise velocity of all simulation principles and the DNS data from Moser \cite{DNSmoser}.}
	\label{Comparison_vel}
\end{figure}

%% file: chapters/conclusion.tex
\chapter{Conclusion}
\label{chapter:conclusion}
%%%%%%%%%%%%%%%%%%%%%%%%%%%%%%%%%%%%%%%%%%%%%%%%%%%%%%%%%%%%%%%%%%%%%%%%%%%%%%%%%%%%%%%%%%%%%%%%%%%%%%%%%%%%%%
%%%%%%%%%%%%%%%%%%%%%%%%%%%%%%%%%%%%%%%%%%%%%%%%%%%%%%%%%%%%%%%%%%%%%%%%%%%%%%%%%%%%%%%%%%%%%%%%%%%%%%%%%%%%%%

{
\section{Summary}
In this thesis, simulations of the turbulent plane channel flow with the $H(\mathrm{div})$-conforming hybrid discontinuous Galerkin method have provided insight into the capabilities and efficiency of this relatively new discretization method in order to predict wall-bounded turbulent flows. Basically, this was done by attempting to compare its result with a standard $H^1$-conforming method with the well-known Taylor-Hood pairing.
\\[1ex]
These discretization techniques have been applied to the four main principles of simulating incompressible turbulent flow. Numerical experiments clearly showed the supremacy of the HDG scheme in resolving turbulent coherent structures and noisy eddies. A good coincidence of both types has been observed for the RANS case. Herein, no improvement of performance with respect to the new method have been noticed and the standard method worked faultless. For LES/VMS and DNS, computations with the conventional method with the Taylor-Hood elements have arised stability issues and the property of only discrete divergence-free velocity has been shown to be not sufficient. The quasi DNS simulation has provided surprisingly good agreement with the reference data, even though the smaller scales have not been adequately resolved. A separation of the turbulent scales leads to the LES/VMS approach. Both modelling principles have proven their abilities of providing good approximations in this numerical test case. Within this comparison, the VMS slightly outperformed the traditional LES method through the different definiton of the scale separation.
\\[1ex]
On this basis, we conclude that relatively to its computational effort, the $H(\mathrm{div})$-conforming HDG method produces qualitatively very good results compared to the given benchmark case for all principles.

\section{Future work}
Since VMS methods are quite new in the field of turbulence simulation, future research could continue in investigating this method combined with the HDG discretization technique. The used eddy-viscosity models for modelling the unresolved scales allows further research and modifications. As well as, to give more insight into the correlation between the used model and the space of the strain rate tensor of the large scales. 
\\[1ex]
Further on, interesting questions for future research could examine more advanced and improved HDG methods for incompressible turbulent flow. In the work of Lederer \cite{lederer_diss}, a new formulation of the Navier-Stokes equations within a HDG scheme was posited. Examination of the application of the new method to turbulence would possibly bring further improvements.
\\[1ex]
}